\documentclass[11pt]{article}
\usepackage[utf8]{inputenc}
\usepackage[T1]{fontenc}
\usepackage{geometry}
\usepackage{cite}
\usepackage{ulem}
\usepackage[T1]{fontenc}
\usepackage{lmodern}
\usepackage{verbatim}
\usepackage{amsmath}

\usepackage{authblk}
\usepackage{natbib}
\usepackage{float}
\usepackage{graphicx}
\usepackage{caption}
\usepackage{subcaption}
\usepackage{url}
\usepackage{fancyhdr}
\newcommand{\JELClassification}[1]{}
\usepackage{cleveref}
\usepackage[toc,page]{appendix}

\usepackage{epstopdf}
\AppendGraphicsExtensions{.tif}

\title{Systemic liquidity contagion\\ in the European interbank market}
\author{\large  ~~~~~~~V. Macchiati\textsuperscript{1\thanks{Corresponding author. Email: valentina.macchiati@sns.it}} , G. Brandi\textsuperscript{2}, G. Cimini\textsuperscript{3,4,5}, G. Caldarelli\textsuperscript{3,4,6}, D. Paolotti\textsuperscript{7}, \\T. Di Matteo \textsuperscript{2,8} }

\affil{\small
	\textsuperscript{1} Scuola Normale Superiore, 56126 Pisa, Italy \\
	\textsuperscript{2} Department of Mathematics, King's College London, WC2R 2LS London, UK\\	 
	\textsuperscript{3} IMT School for Advanced Studies, 55100 Lucca, Italy\\
	\textsuperscript{4} Institute for Complex Systems (ISC-CNR) UoS Sapienza, 00185 Roma, Italy\\
	\textsuperscript{5} Dipartimento di Fisica, Università degli Studi di Roma Tor Vergata, 00133 Roma, Italy\\
	\textsuperscript{6} European Centre for Living Technology, 30124 Venice, Italy\\
	\textsuperscript{7}	 Digital and Computational Epidemiology group, ISI Foundation, 10126 Turin, Italy\\
	\textsuperscript{8} Complexity Science Hub Vienna, 1080 Vienna, Austria}

\date{}

\begin{document}
	
\maketitle

\begin{abstract}
	Systemic liquidity risk, defined by the IMF as "the risk of simultaneous liquidity difficulties at multiple financial institutions", is a key topic in macroprudential policy and financial stress analysis. Specialized models to simulate funding liquidity risk and contagion are available but they require not only banks' bilateral exposures data but also balance sheet data with sufficient granularity, which are hardly available. 
	Alternatively, risk analyses on interbank networks have been done via centrality measures of the underlying graph capturing the most interconnected and hence more prone to risk spreading banks. In this paper, we propose a model which relies on an epidemic model which simulate a contagion on the interbank market using the funding liquidity shortage mechanism as contagion process. The model is enriched with country and bank risk features which take into account the heterogeneity of the interbank market. The proposed model is particularly useful when full set of data necessary to run specialized models is not available. Since the interbank network is not fully available, an economic driven reconstruction method is also proposed to retrieve the interbank network by constraining the standard reconstruction methodology to real financial indicators. We show that the contagion model is able to reproduce systemic liquidity risk across different years and countries. This result suggests that the proposed model can be successfully used as a valid alternative to more complex ones. 	
\end{abstract}

\section*{Introduction}

\begin{figure}[h!]
	\centering
	\includegraphics[width=0.75\linewidth]{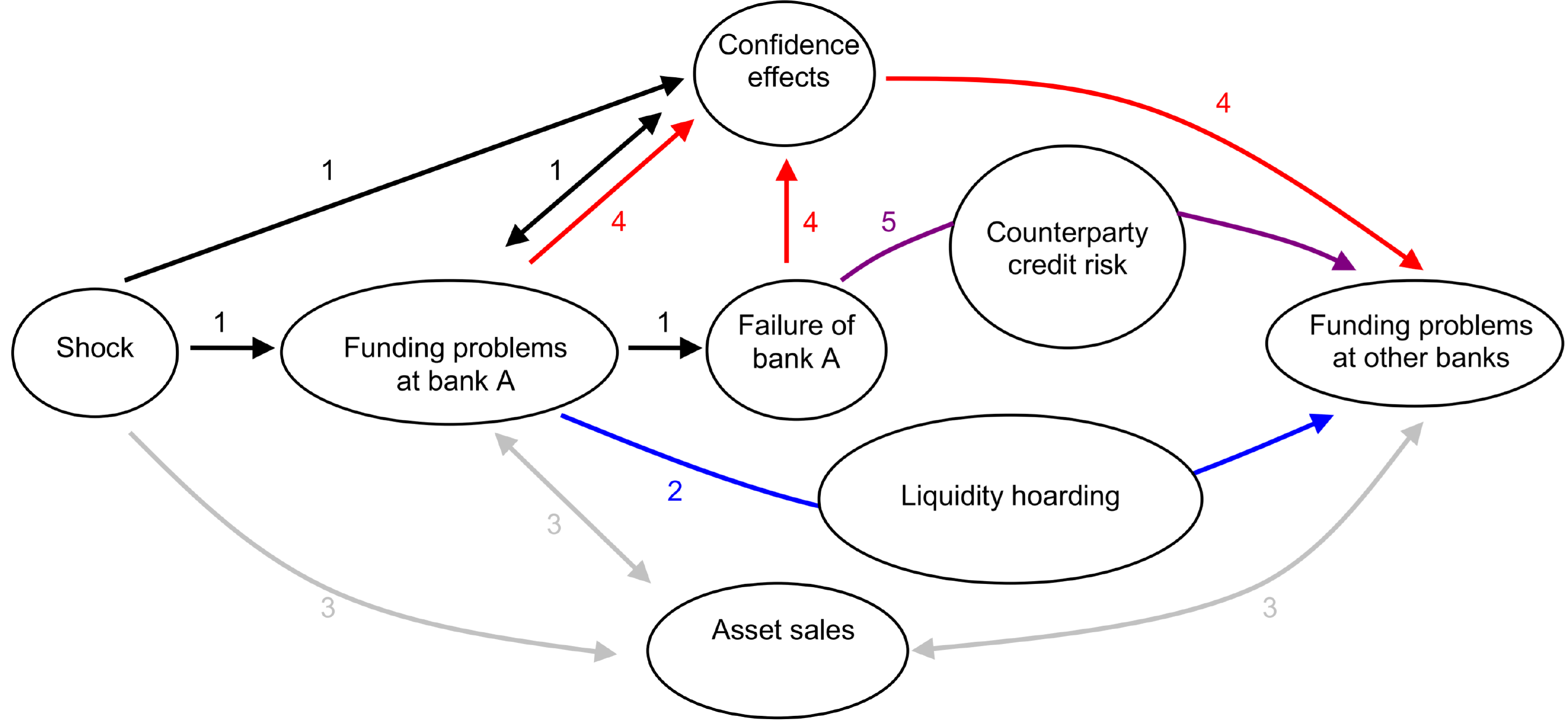}
	\caption{Channels of liquidity risk in the funding market. Source: Bank of England \citep{kapadia2012liquidity}}
	\label{fig:fig_BoE}
\end{figure}

The global financial crisis has shown how fundamental is the role of liquidity risk in the stability of the financial system. Recently, an IMF Working Paper on macroprudential stress testing has pointed out that liquidity shocks can rapidly manifest and impact the whole financial system, while bank solvency concerns tend to take more time to build up \citep{jobst2017macroprudential}. On the same line, the Bank of England \citep{kapadia2012liquidity} draws attention on the fact that "although the failure of a financial institution may reflect solvency concerns, it often manifests itself through crystallisation of funding liquidity risk". Likewise, the ECB \citep{liquidity2018systemic} has recently released a paper on the monitoring framework for systemic liquidity and the use of macroprudential liquidity tools in the banking system of the European Union. However, despite its fundamental relevance in macroprudential policy and financial stress analysis, funding liquidity contagion has only received limited attention in the literature \citep{kapadia2012liquidity}. Recent research papers \citep{gai2010contagion,gai2011complexity} go in this direction, employing a theoretical economic model supported by an interbank network to show how the fragility of the funding interconnectedness of the system can generate a financial collapse. The 2012 case of Northern Rock in the UK made clear that liquidity shocks may well come from the wholesale interbank funding market rather than the traditional depositor run. In fact, in that situation, the banks in the financial systems stopped lending money to the UK bank, generating a liquidity shock which hit the bank, making it default.
Liquidity risk can be differentiated in two main types, funding liquidity risk and market liquidity risk. The former is related to the fact that the bank is not able to meet its liquidity needs in case of a funding shock while the latter is referred to the case in which an institution is not able to buy or sell securities without a huge price impact, usually measured by the bid-ask spread. The literature has shown that while distinct in nature, the increased reliance on wholesale bank funding has made the two liquidity risks strongly connected, with possible feedback and spiral effect between the two, especially during distress periods \citep{jobst2017macroprudential,brunnermeier2008market,brunnermeier2009deciphering,cai2008liquidity,drehmann2013funding,bonfim2012liquidity,bonfim2012systemic}. Recently, the research community started to adopt complex networks as the underlying measure of interconnectedness between financial institution \citep{gai2010contagion,gai2011complexity,glasserman2015likely,aldasoro2017bank}. In fact, the global crisis has clearly shown that the complex structure of the financial system plays a pivotal role in the amplification of liquidity risk propagation \citep{gai2010contagion,haldane2011systemic,philippas2015insights}. Even if the literature has flourished after the financial crisis, many studies on the interbank market structure had already been conducted even before the most recent crisis in 2007-2008 \citep{boss2004network,inaoka2004fractal,soramaki2007topology,bech2010topology,iori2008network,de2006fitness}. The inability to completely capture and quantify systemic financial risk has led to immense social and economic costs \citep{lewis2010financial,poledna2015multi}. The whole financial system, including the interbank market, has been increasingly accepted by both regulators and the academic community to be a complex system \citep{cont2010network,haldane2009banks,haldane2013rethinking}. In this context, it has become more and more natural and accepted to represent the interbank market as a directed weighted network with links' directionality and heterogeneities. Study of financial systems has greatly benefited by networks science, because network representation allows to clearly and conveniently figure out all the institutions involved and the topology structure of their connections \citep{battiston2010structure,nier2007network}.
Several sources and dynamics of contagion have been identified and analysed in order to assess systemic risk and resilience. Many works \citep{allen2000financial,nier2007network,cocco2009lending,huser2015too,gai2010contagion} have highlighted the crucial role played by the topology structure of the interbank market. This has lead to the development of new models of contagion spreading \citep{eisenberg2001systemic,furfine2003interbank,gai2011complexity,mistrulli2011assessing,battiston2012debtrank,bardoscia2015debtrank,thurner2013debtrank} and more attention has been devoted to the interplay between contagion dynamics and the underlying topology \citep{glasserman2016contagion,di2018network,battiston2012liaisons}. This work is part of this research line and the goal is to model the propagation of liquidity shocks as an epidemic disease spreading over the interbank market network. This gives us the possibility to adapt a contagion-like model to the framework of systemic liquidity risk on the interbank network \citep{brandi2018epidemics,philippas2015insights,toivanen2013contagion}. We build the model on the observation made by the Bank of England \citep{kapadia2012liquidity} on the channels of liquidity risk spreading. Figure \ref{fig:fig_BoE} represents the channels indicated by the Bank of England for which the funding shock can propagate. 

According to the Bank of England, there are 5 possible channels:
\begin{enumerate}
	\item Reputational shocks and negative feedback loops.
	\item Liquidity hoarding.
	\item Asset prices depreciation (market liquidity risk).
	\item Confidence contagion (run on banks similar to defaulted ones).
	\item Counterparty credit risk.
\end{enumerate}
In this paper, we will base the model mostly on funding liquidity risk, so we will not touch channel 5 and we will consider market liquidity as an amplification mechanism rather than a contagion channel. Based on this argument, we use compartmental models that allow the simplification of the contagion dynamics so that few assumptions on the contagion and transmission parameters are required. Those models are particularly useful when the full set of data necessary to run more specialized and sophisticated models is not available \citep{greenwood2015vulnerable,manna2012externalities,cont2019liquidity,smaga2015can}. In fact, effective contagion models require not only the real underlying network but also their complexity demands balance sheet data with sufficient granularity and other economic data which are hardly available. When the real topology of the interbank network is not available, these missing data have to be compensated using reconstruction techniques, based on the principles of statistical mechanics. We have thus based the reconstruction of the interbank market structure on the only available information for the network, i.e. aggregated data of banks' exposure provided by Bankscope database  \citep{battiston2016leveraging}, implementing techniques already developed in previous works \citep{cimini2015systemic,squartini2017network,squartini2018reconstruction}. We further develop a new reconstruction technique by imposing macroeconomic constraints to the standard reconstruction methodology. In particular, we used as constraints the volumes of inter and intra-country exposures provided by the BIS database \footnote{BIS Statistics Warehouse, \url{https://www.bis.org/statistics/index.htm}}. In this way, the lending relationship is not only driven by the amount of the aggregate interbank assets and liabilities but it also takes into account the real preferential relationship between countries. It is in fact observed in the data that volumes exchanged by banks of the same country account for most of the interbank lending volume, and this preferential lending is retrieved by the proposed reconstruction method.

The paper aims to assess systemic liquidity risk defined by the IMF \citep{international2011global} as "the risk of simultaneous liquidity difficulties at multiple financial institutions", which asks for both a contagion mechanism and an interbank topology. We tackle the problem by proposing a tool which is easy to implement and use but still being flexible and containing financial information, a contagion mechanism and underlying market topology. In particular, the model is built in order to reflect both the funding and market liquidity in the mechanics and amplification of the contagion. This is done via considering also node's properties, such as the banks' country, its size, the liquidity resilience in addition to the heterogeneities due to the different volumes of money exchanged in the interbank market. The model builds on the epidemiological-like models discussed in \citet{brandi2018epidemics} and improves on it adding node-specific information. We show that the differentiation of banks' susceptibility and contagiousness, as well as the network topology, play a non-negligible role in the spread of liquidity distress. In the European interbank market, systemic funding liquidity risk started to set up during the global financial crisis but particularly escalated during the European sovereign debt crisis. Our dataset covers the period 2006-2013 so it is not possible to understand if the systemic liquidity risk we found falls back to pre-crisis levels.
The rest of the paper is structured as follows. Section \ref{cap1} introduces the epidemic-like contagion processes implemented, section \ref{cap2} presents the data and the network reconstruction technique used to build the network while section \ref{cap3} reports the results of the network reconstruction and of the liquidity-driven dynamics. Section \ref{conclusions} is devoted to final remarks and conclusions.

\section{Contagion process}\label{cap1}
The recent global crisis has shed light on the importance of the interconnection among players in the financial systems to assess their systemic risk. This led to the new paradigm "too-interconnected-to-fail" to emerge \citep{battiston2012debtrank,haldane2009banks,toivanen2013contagion} alongside the "too-big-to-fail". Network representation \citep{haldane2013rethinking,catanzaro2013network} allows to clearly depict and model the different interactions between the financial institutions to deeply understand complex phenomena as financial crisis \citep{huser2015too}. Many studies \citep{freixas2000systemic,iori2006systemic,rochet1996interbank} focus on assessing the systemic risk in the interbank market network, i.e. on banks' direct exposures and connections. The interbank market network is, in fact, the first line of defence against liquidity and credit shocks \citep{iori2006systemic}. Its role is to quickly provide liquidity to banks through interbank lending. During the crisis, however, this market has proved its fragility \citep{fricke2015core} and inability to give to the system the liquidity needed \citep{allen2014transmission,brunnermeier2009deciphering,gabbi2015financial}, because of liquidity hoarding caused by financial turmoil and by fear of contagion \citep{arinaminpathy2012size,brandi2018epidemics,may2009systemic,smaga2015can,toivanen2013contagion}. It is also shown \citep{cifuentes2005liquidity,upper2011simulation} as bank failures due only to losses in the interbank market have not been observed, maybe because liquidity risk can generate solvency issues and then be masked or because of government interventions.\footnote{An exception is the case of Northern Rock that due to funding liquidity cuts in the interbank market and bank run defaulted in 2012.} However, exploring and assessing the systemic risk via the interbank market is important to take into consideration its interaction with other possible channels of contagion \citep{huser2015too}. 

Several sources and dynamics of contagion have been identified and analysed in order to assess systemic risk and resilience. A seminal work \citep{gai2010contagion} has shown the robust yet fragile tendency of the financial network whereby even a low probability of contagion could cause a disruptive crisis. Besides, this study has pointed out as identical (same magnitude) financial shocks may lead to very different outcomes because of the crucial role played in the shock spreading by the complex topology of the connections between financial institutions. Much attention is devoted by both regulators and by academics to assess if such a complex and interconnected interbank structure lessens or enhances the contagion spreading \citep{halaj2013assessing,nier2007network}. From the point of view of a single bank, the aim is to diversify and then to increase the number counterparties in order to dampen its risk of contagion. However, considering all the financial institutions and their risk containment strategies, the systems turn to show increased complexity and interconnectivity \citep{glasserman2016contagion}. Network connections have thus both positive effects (risks' diversification) but also they create channels through which shocks can spread. In addition, before the crisis, banks have shown the tendency to increase their debt levels and their exposure to derivatives and securitizations. The interplay of all of these factors has affected the stability of the financial system \citep{glasserman2016contagion,upper2011simulation} that have become so much prone to shock propagation and then to unexpected failure knock-on effects. Starting from the pioneering models of \citet{eisenberg2001systemic,furfine2003interbank}, an increasing stream of literature have focused on the development of new dynamics and metrics to describe the spread of financial distress over the network \citep{battiston2012debtrank,bardoscia2015debtrank,elsinger2013network,nier2007network,krause2012interbank,amini2016resilience,bardoscia2017pathways,serri2016interbank,cifuentes2005liquidity,diamond2001liquidity,kapadia2012liquidity,hurd2016double,cimini2016entangling,brandi2018epidemics,cimini2015estimating}. The aim of these models is not only to assess the systemic risk but also understand which features and how their interplays are involved in the system robustness and resilience against shock spreading. Our work is part of this context as its aim is to model the spread of financial distress over a network representing the interbank market, by means of an epidemiological-like model. 
\subsection{Epidemic contagion processes}
Epidemiologist and financial policymakers, in a way, share a common interest in studying spreading phenomena and rely on models for the description of the contagion diffusion over the population considered. So, epidemic modelling, with its methods and results, could be adapted to describe the domino effect that could arise in the interbank market subject to financial shocks \citep{keeling2005networks,pastor2015epidemic,barrat2008dynamical}. The main idea is to adapt the well-known Susceptible-Infected compartmental model (SI model) in the framework of contagion so that liquidity and credit shocks propagate as an epidemic disease over the interbank market.  

In the literature, epidemic-like contagion processes have been adapted to the interbank market in order to study and quantify the liquidity-driven \citep{brandi2018epidemics} and the credit-driven \citep{philippas2015insights,toivanen2013contagion} spread of financial distress. In this work, we focus on the infection due to banks' liquidity hoarding and the consequent counterparties' liquidity shortages, further enhancing such model by considering not only the links' heterogeneity but also nodes' features. In fact, each bank has different financial structure and operate in a specific country which also plays a central role in the contagion spreading and then systemic risk. As aforementioned, banks have not complete information on the financial structure of their counterparties, so, a contagion, through such direct exposures, is also driven by psychological aspects as the perceived counterparty risk and expectations. In order to overwhelm further this extended uncertainty, we have implemented stochastic compartmental models that add randomness to the possible contagion channels and to the propagation dynamics and outcomes.
\subsubsection{Liquidity driven model}
The interbank market allows banks temporarily short on liquidity to borrow money from other banks. The major part of these loans has an overnight duration so the borrower must repay the lender at the start of business the next day. These loans let the bank to temporarily meet reserve requirements and find additional liquidity without having to sell their assets. Usually, these liquidity shortcomings are longer than a night so the overnight loans need to be replaced by new debt. The system is thus subject to roll-over risk: distressed banks might not find in the interbank market the liquidity they need and then they are forced to sell assets, eventually also illiquid ones whose sale may trigger further losses (fire sales) \citep{brandi2018epidemics}. Besides, if other banks perceive these troubles, they will hoard liquidity and this will induce another wave of liquidity shocks and then lead to a shrinkage of the volume exchanged in the interbank market. 

Our contagion process is an adaption of the SIR (Susceptible-Infected-Recovered) model without considering the recovery state but instead two different infectious compartments, $I_1$ and $I_2$. This is because the time-scale of the contagion spreading is supposed to be too fast with respect to the period required for an infected bank to recover from their liquidity problems.\footnote{If such a mechanism would be of economic relevance, the model would be easily adapted to an SIS (Susceptible-Infected-Susceptible) model, with a stochastic rate of recovery.} Indeed, an infected bank remains contagious because the removal procedure needs longer times than the dynamics considered. In epidemiology, the system of equations describing the dynamics of this Susceptible-Infected model is given by:
\begin{equation}
\begin{cases}
\frac{ds}{dt}=-\lambda(\epsilon~i_1(t)+i_2(t)) s\\
\frac{di_1}{dt}=+\lambda(\epsilon~i_1(t)+i_2(t)) s  -\mu i_1(t)\\
\frac{di_2}{dt}= \mu i_1(t)
\end{cases} 
\label{eq:si ode}
\end{equation}
where $s(t)$, $i_1(t)$, $i_2(t)$ are the fraction of individuals in the susceptible and infectious states at time $t$ and $\epsilon$ is a scale parameter. The transition from a group to another is regulated by the contagion rate $\lambda$ and the default rate $\mu$.
In the liquidity-driven model, we adapt this SI model into the EDB (Exposed-Distressed-Bankrupted) model.
Exposed compartment corresponds to the susceptible one while both distressed and bankrupted correspond to two different infectious compartments. Exposed banks (hereinafter $S$) are those that have obtained the needed liquidity in the interbank market but they are also susceptible to contagion because their lenders may be distressed or bankrupted banks. They are thus healthy at the inception but the liquidity shortages of their counterparties could deprive them of the liquidity provision. When that happens, the contagion spreads to the exposed bank that goes into the distressed state and in turn becomes contagious for its counterparties. Distressed banks (hereinafter $I_1$) are those infectious due to their liquidity troubles and they could turn into the bankrupted state (hereinafter $I_2$) when these losses become overwhelming.  In this work, we set $\epsilon=1$ so that the contagion rate of the distressed and bankrupted banks has the same value.\footnote{It is easy to adapt this to more complex situation in which the economic environments suggests how the two should differ.} We further improve the SI model (eq. \ref{eq:si ode}) considering node and link dependant contagion ($\lambda$) and bankruptcy ($\mu$) rate, i.e. driven by lending relationship and lenders' state. In a previous work \citep{brandi2018epidemics}, the rate of the contagion from a distressed/bankrupted bank $i$ to an exposed $j$ is defined as follows:
\begin{equation}
\lambda_{ij}=\frac{w_{ij}}{\sum_jw_{ij}}
\label{lambdabrandi}
\end{equation}
where $w_{ij}$ is the transaction volume between the lender $i$ and the borrower $j$. The economic rationale behind this contagion rate specification is that the funding shock is proportional to the amount of money $i$ lend to $j$ with respect to its total interbank lending volume. 
The bankruptcy rate $\mu$, instead depends on the real-time state of the banks and it is defined as follows:
\begin{equation}
\mu_{i}(t) =\frac{\sum_{j\in H(t)}w_{ji}}{\sum_{j}w_{ji}}
\label{eq:bankrupt rate}
\end{equation}
where $H(t)$ is the set of distressed and bankrupted lenders of $i$ at time $t$.
As defined, the bankruptcy rate is the fraction of liquidity that bank $i$ needs that were previously lent by infectious banks. The idea is that bank's losses become overwhelming when they are so high that bank has no possibility to reallocate its assets to absorb quickly the losses. As it is defined (eq.\ref{lambdabrandi}), the contagion rate does not take into account any nodes' feature. This would imply that the contagion rate is independent of bank features and the country in which they operate. 

Unlike previous models \citep{brandi2018epidemics}, we take into account lenders' features as how risky they are and how suffering and indebted are the economies of their countries. Indeed, liquidity shocks are more probable for banks with higher liquidity risk operating in countries with a more fragile economy. To incorporate this feature on the contagion model, we redefine the contagion rate as follows:
\begin{equation}
\lambda_{ij}^{*}=\lambda_{ij}^{(1-\gamma_i)}
\label{lambdafinale}
\end{equation}
where $\gamma_i \in [-1,1]$ is the node-term of the infected bank $i$. This functional form ensures that the contagion rate is always bounded between $0$ and $1$. In fact, as $\gamma_i$ tends to $1$, the bank $i$ contagion rate tends to $1$ whilst in the case $\gamma_i$ tends to $-1$, the contagion rate diminishes\footnote{It is important to notice that the effect is not symmetric and that $\gamma_i>0$ has a higher impact.}. In the neutral case with $\gamma_i=0$, we restore the functional form of the contagion rate already discussed in literature \citep{brandi2018epidemics}, which we use as a benchmark model (BM).
The node-term variable $\gamma_i$ that we take is a combination of a bank liquidity risk measure and a country-specific risk measure. Regarding the bank-specific liquidity risk, we compute the following liquidity indicator:
\begin{equation}\label{liq_ratio}
Liq_i=\frac{T_i}{F_i}
\end{equation}
where $T_i$ are the values of the $i$-th bank total assets and $F_i$ is a proxy of banks' liquid assets (hereafter $LA$).\footnote{ We will use the total deposits, money market and short-term funding as a proxy.} The more a bank is fragile and much higher is the value of the liquidity indicator and the bank’s infectiousness. In fact, when systemic risk comes into being, the $LA$ plays the role of the first defence. If the liquidity indicator is high, the bank is likely to be more severely stricken than other banks in the market.
Instead, with regards to the country-specific liquidity risk, we use as a proxy of it the bid-ask spread $\delta_P$ computed on the 10-year government bonds, which is a popular measure of market liquidity given by:
\begin{equation}\delta_P=B_P-A_P,
\label{bid_ask}
\end{equation}

where $B_P$ is the bid price of the 10-year government bond, which is defined as the highest price that a buyer is willing to pay for the bond while $A_P$ is the ask price, that is the lowest price that a seller is willing to accept in order to sell the bond. A high value of this measure indicates lower liquidity, while a lower value signals higher liquidity in the market. The European Central Bank has recently released a paper on systemic liquidity \citep{liquidity2018systemic} where they state that such a measure "shows a clear distinction between periods of stress and periods of more benign market conditions". For the current research, we computed the measure from the bid and ask prices\footnote{Bloomberg L.P., "Bid and ask price of the 10-year government bonds 12/30/06 to 12/30/13." } for government bonds for every country at the end of each year.

As mentioned in eq. \ref{lambdafinale}, the node-term variable $\gamma_i$ should be in the interval $[-1,1]$. In order to achieve this, different transformations are available. In this paper, we use the following transformation $f(\cdot)$:
\begin{equation}
\underbar{x}=f(x)=\frac{x-m(x)}{x+m(x)}
\label{mapping}
\end{equation}
where $x$ is a generic variable, $m(x)$ is the median of $x$ and $\underbar{x}$ is the transformed variable. This transformation has the feature to be in the prescribed range and additionally it is symmetric around the median. The rationale behind the choice of such transformation is to discriminate between banks increasing the contagion effect from the ones which decrease it. 
We define the node-term  $\gamma_i$ of the bank $i$ as the rescaled version of the product of its country and bank liquidity risk. The motivation for using the product between the two terms is to have an amplification mechanism between the country liquidity risk and the bank liquidity structure so that a solid bank liquidity structure would imply a contraction of the country risk while a fragile structure would magnify it.\footnote{Other possibilities arise in this context. One can sum the two variables and then apply the $f(\cdot)$ transformation or can firstly transform each variable and then use the geometric or arithmetic mean between the rescaled variables, taking into account the interval in which $\gamma_i$ should be defined. We tested for such alternatives and the results remain qualitatively equal.} The bank $i$ node-term $\gamma_i$ is therefore defined as:
\begin{equation}
\gamma_i=f(Liq_i \cdot {\delta_P}_i)
\label{eq:nodeterm}
\end{equation}      
where ${\delta_P}_i$ is the bid-ask spread of the 10-year government bonds and $Liq_i$ is the liquidity indicator. This model could be further enhanced by considering a bankruptcy rate $\mu^{*}$ which depends on nodes liquidity resilience. As we introduced the node-term $\gamma$ to take into account the fragility of the infected banks, we follow the same reasoning for the bankruptcy rate. In particular, it is clear that the transition from the distressed to the bankrupted state does not depend only on banks' connectivity but also on bank's capabilities to absorb liquidity shocks, so we introduce a bank (node) specific indicator $\nu$ which the liquidity resilience of the distressed bank. The bankruptcy rate is so defined as follows:
\begin{equation}
\mu_{i}^{*}(t)=\mu_{i}(t)^{(1-\nu_i)}
\label{dopomuu}
\end{equation}
where the $\nu_i \in [-1,1]$ is the rescaled liquidity resilience indicator of the bank $i$ that is defined as:
\begin{equation}
\nu_{i}=f(\frac{L_i}{F_i})
\label{liqindicator}
\end{equation}
where $L_i$ are the interbank liabilities and $F_i$ is a liquidity proxy of the bank $i$. This indicator is a proxy of the capabilities of a bank to absorb the liquidity shock in the interbank market using liquid assets. As for the contagion rate, the node-term $\nu_i$ amplifies the default rate when $\nu_i>0$ while decreases it when $\nu_i<0$. This model specification is mostly relevant for regulatory activities since the asymptotic stage remains unchanged but what changes is the speed of the systemic default risk. In fact, knowing such resilience proxy for the whole system and a networked contagion mechanism, it can be used to tackle timely the systemic liquidity risk. However, as noted in the IMF Working Paper on macroprudential stress testing\citep{jobst2017macroprudential}, "liquidity crises are partly attributable to psychological factors or confidence effects". This feature of the market depends upon the state of health of the whole system and of specific banks which are in a state of distress. To account for this, we introduced a systemic-dependent variable which measures the entire health of the system, in order to capture the increasing pressure on the interbank market when most of the banks are infected or defaulted. Hence, we introduce a further variable $\theta(t)$, which we call systemic risk multiplier that enters the equation for the contagion rate. This modification makes the contagion rate dynamic and dependent on the state in which the whole market is at each point in time. The parameter $\theta(t)$ measures the health of the entire system and is defined as follows \footnote{In the appendix, we generalize the formulation of the systemic risk multiplier $\theta$ introducing a non-linear transformation $h(\theta,\phi)$, where $h(\cdot)$ is a non-linear function and $\phi$ represents the strength of non-linearity.}:
\begin{equation}
\theta(t)=\beta_0+\beta_1 s(t)-\beta_2 i_1(t)-\beta_3 i_2(t)
\label{labeltheta}
\end{equation}
where $s(t)$, $i_1(t)$ and $i_2(t)$ are respectively the fraction of exposed, distressed and bankrupted banks, $\beta_0$ is a normalizing parameter in order to have proper bounds and $\beta_1,\beta_2,\beta_3$ are weights attributed to the specific banks state. 
The $\beta$s serve to increase or decrease the importance of one segment or another and so the dynamic of the systemic risk multiplier. To avoid irregular patterns, we set $\beta_1$ to unity and the other $\beta$s are set to the same value ($\beta_0=\beta_2=\beta_3=\beta_*$) which relates to the importance attached to the distressed and bankrupted banks. Using this formulation, eq. \ref{labeltheta} simplifies to:
\begin{equation}
\theta(t)=(1+\beta_*) s(t)
\label{labeltheta2}
\end{equation}
In fact, if $\beta_*$ is high ($\geq1$) this translates to a financial system in which distressed and defaulted banks generate a negligible (or even positive) psychological effect to the other market participants. This will considerably slow down the contagion, which seems to be unrealistic from a behavioural economics point of view. On the other hand, if we set $\beta_*=0$, this generates a strong psychological effect even if only few banks are in distress or defaulted, which appears extreme. A reasonable value for $\beta_*$ should be in the range $(0,1)$, with small values having a stronger behavioural effect. 
This contagion specification, with the contribution of the parameter $\theta(t)$, results in:
\begin{equation}
\lambda_{ij}^{+}(t)=\lambda_{ij}^{(1-\gamma_i)(\theta(t))}=\lambda_{ij}^{(1-\gamma_i)(1+\beta_*)s(t)}
\label{lambdadinamicoo2}
\end{equation}
where $\gamma_i$ is the node-term introduced in eq. \ref{lambdafinale}. By definition, the parameter $\theta \in [0,1+\beta_*]$. This means that high (low) values $\theta>1$ ($\theta<1$) increase (decrease) the contagion rate and accelerate (slow down) the infection spreading. It is also possible to apply a transformation to $\theta$ such that the feedback effect is stronger or weaker with respect to the systemic risk. 

\section{Data}\label{cap2}
We use three different sources of data to our model. To take information on market liquidity, we use the data on bid-ask spread $\delta_P$ available on Bloomberg and which provides information on two main aspects: the country riskiness and liquidity tightness \citep{iachini2016systemic}. Secondly, we use the Bankscope dataset from which we compute the liquidity indicator $Liq_i$ and the liquidity resilience proxy $\nu$. We finally use the BIS dataset to extrapolate interconnectedness between banks operating in different European countries and use those as constraints for the network reconstruction. We now briefly present these datasets and the reconstruction technique used. 

\subsection{Bid-ask spread data}
\begin{figure}[h!]
	\centering
	\includegraphics[width=0.7\linewidth]{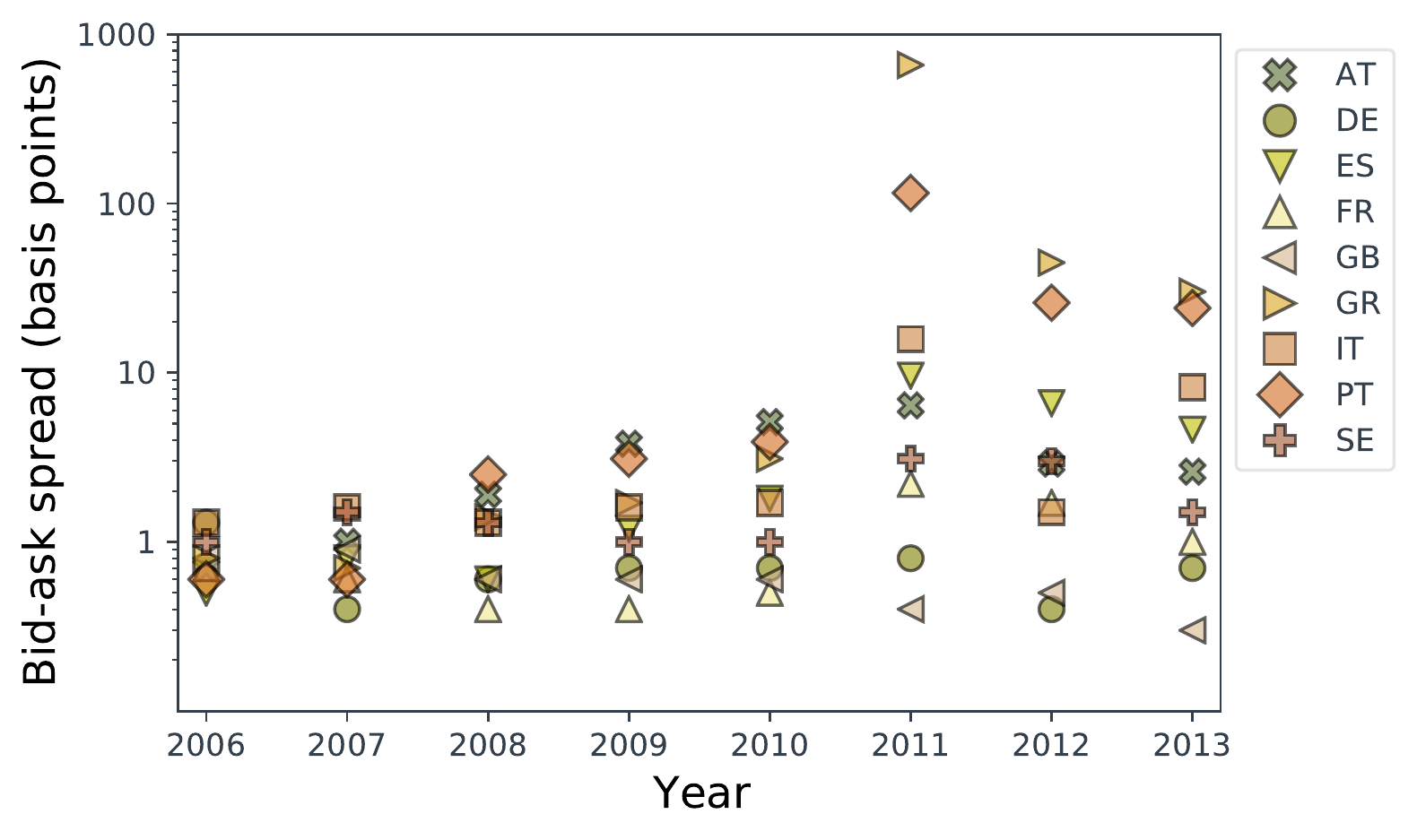}
	\caption{Bid-ask spread of the 10-year government bonds during the period 2006-2013. Semi-log plot. Source: Bloomberg L.P.}
	\label{fig:countryterm_bid_ask}
\end{figure}

The bid-ask spread is a widely adopted measure as a proxy of market liquidity in the marketplace and it is computed as in eq. \ref{bid_ask} using the 10 years maturity government bonds. In normal time, the bid-ask spreads mostly synthesize market structural features, while during turbulent periods it becomes a very responsive indicator of liquidity tightness (market liquidity risk) \citep{kyle1985continuous,iachini2016systemic}. In distress periods, a high spread indicates lower liquidity while a lower bid-ask spread indicates higher liquidity in the market. Since we have data from 2006 to 2013, we download the end of the year bid and ask prices from Bloomberg and compute the spread. Figure \ref{fig:countryterm_bid_ask} shows the dynamic (in semi-log scale) of the bid-ask spread for the countries in our dataset. It clearly shows that some countries, the so called PIGS, are at the top of the graph for the whole period of consideration. This indicates that such countries hugely suffer from liquidity risk and they can cause, through network contagion, a systemic event. 

\subsection{Bankscope dataset}
\begin{figure}[h!]
	\centering
	\includegraphics[width=0.55\linewidth]{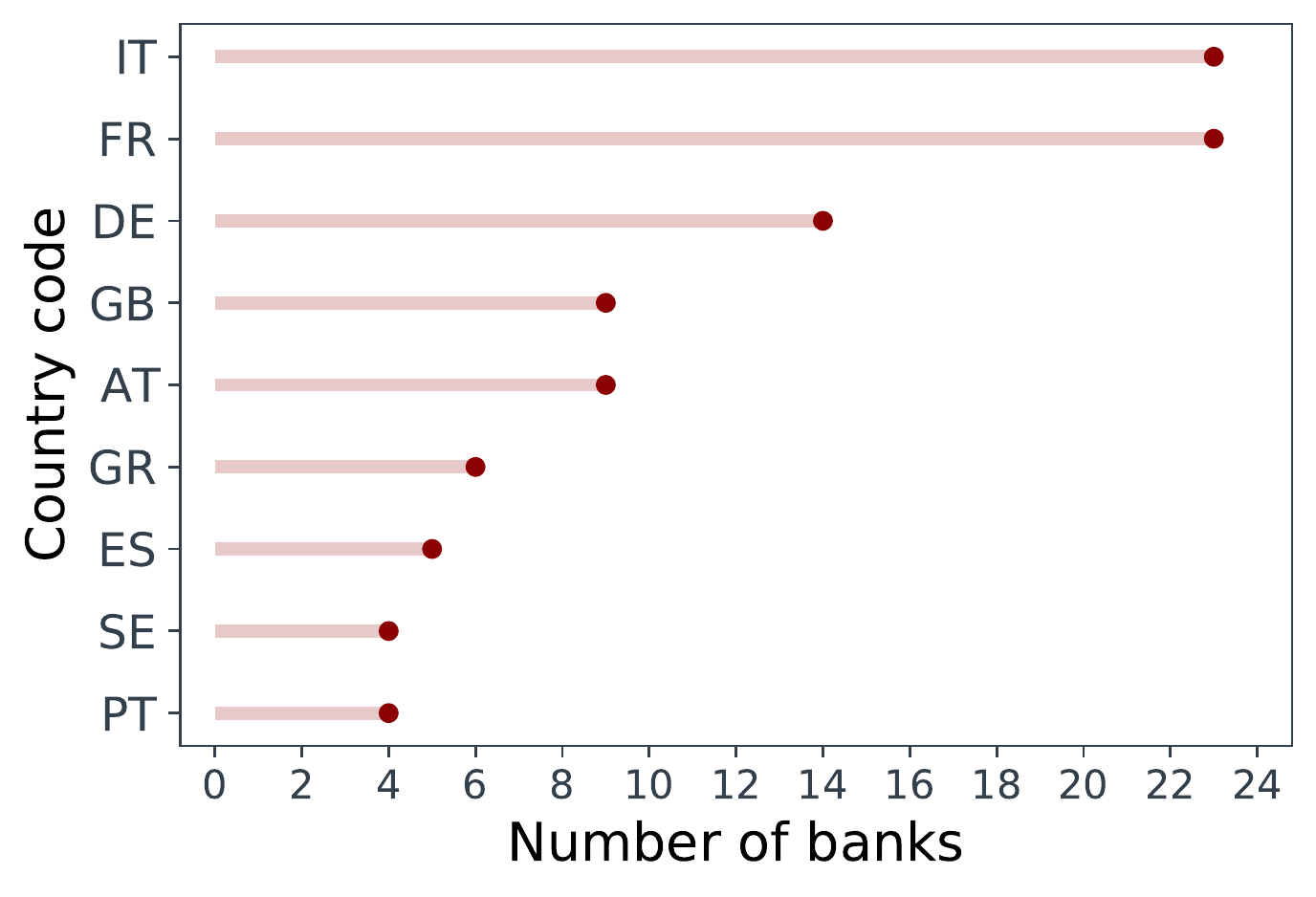}
	\caption{Number of banks per each country. Source: Bankscope database}
	\label{distr}
\end{figure}

\begin{figure}[h!]
	\centering
	\includegraphics[width=0.49\linewidth]{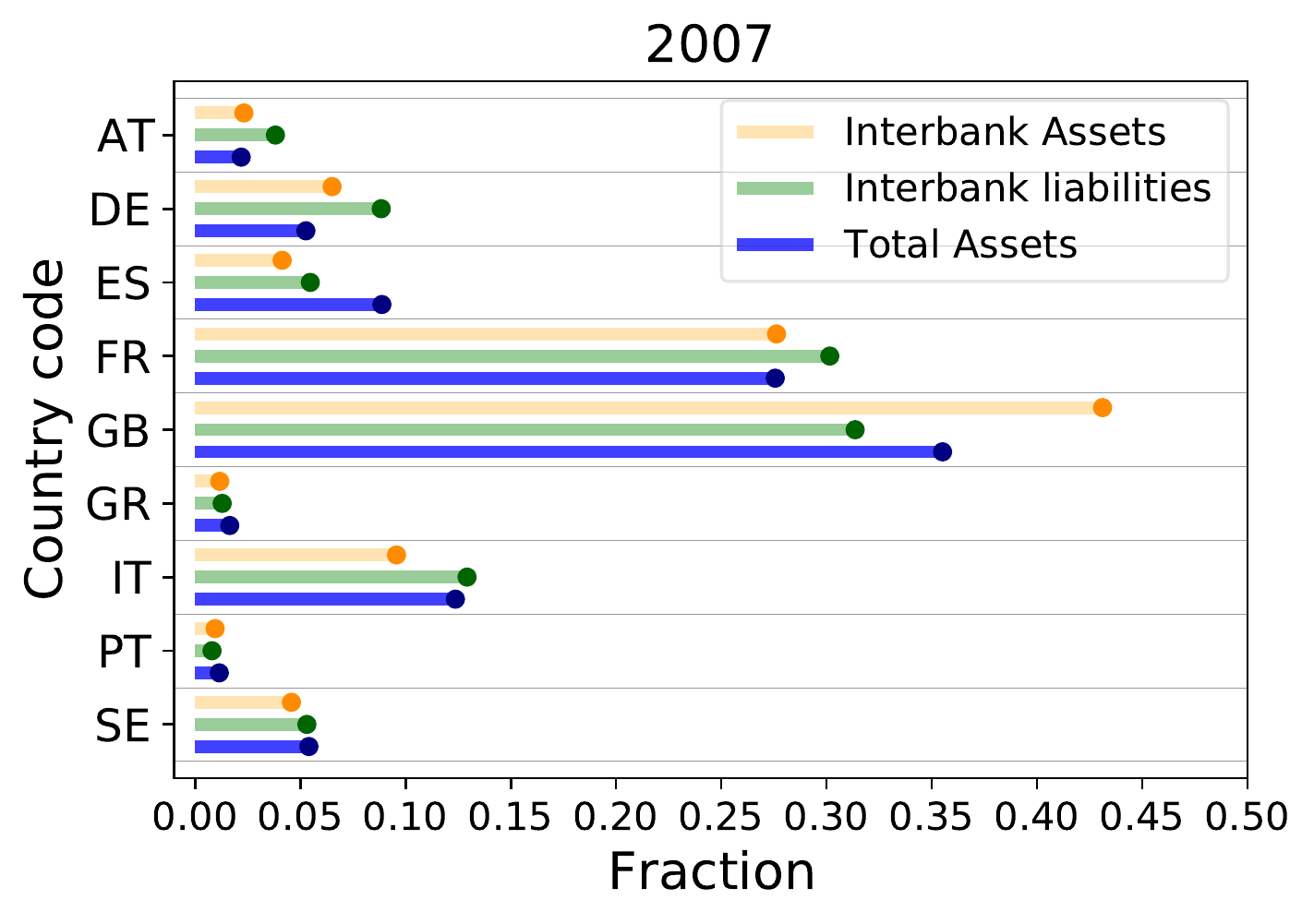}
	\includegraphics[width=0.49\linewidth]{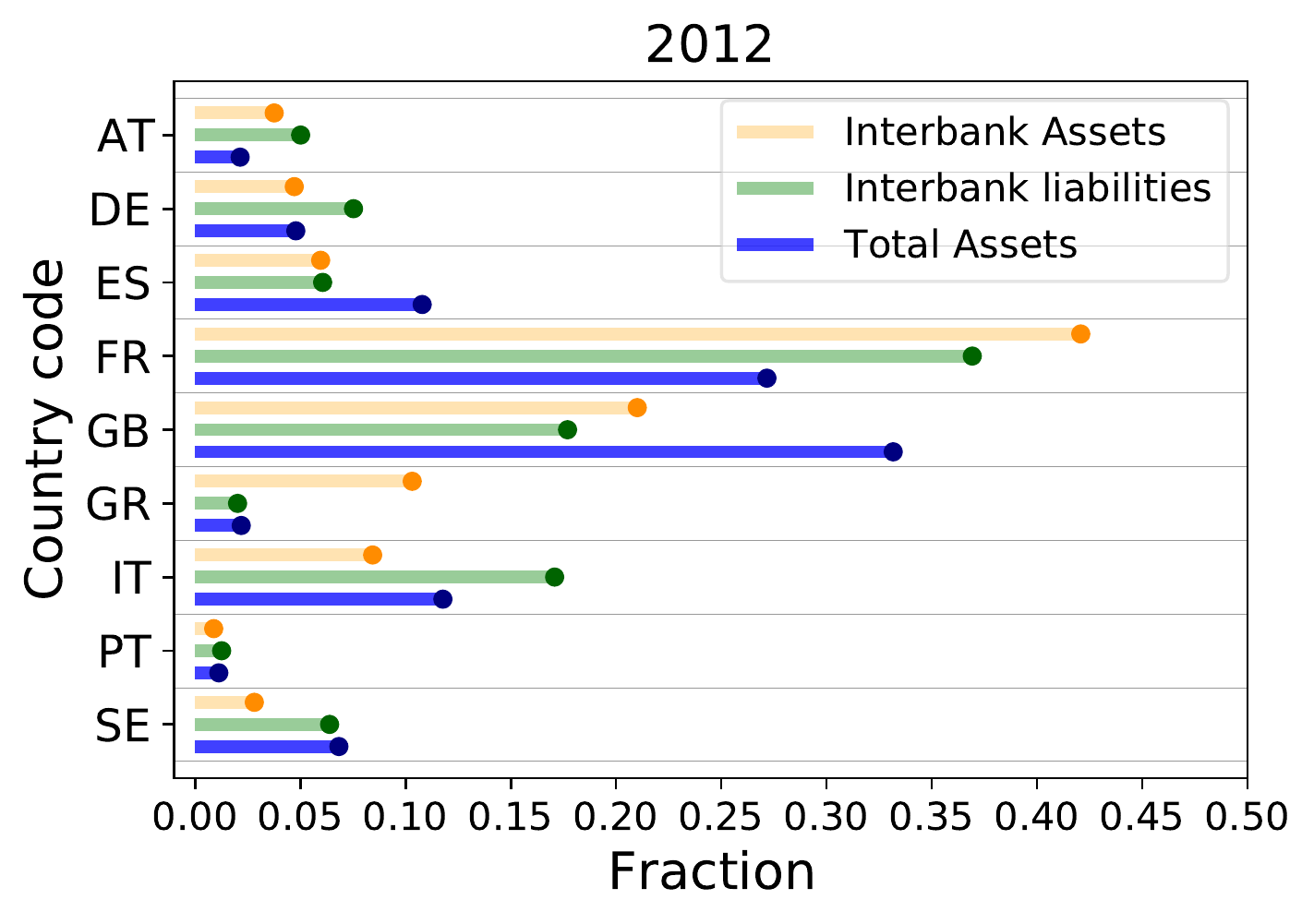}
	\caption{Fraction of total asset, interbank assets and liabilities per each country in the years 2007 and 2012. Source: Bankscope database}
	\label{distr1}
\end{figure}

Data about the bilateral interbank exposures are unfortunately not available but we have access to Bureau Van Dijk Bankscope database \citep{battiston2016leveraging} that contains information on banks’ aggregate exposures and on their balance sheets. We have access to the annual balance sheet of 97 anonymized European banks that were publicly traded between 2006 and 2013.\footnote{We use the subset of banks for which there was no missing information and whose country was represented by at least 4 banks.} In particular, we have information on the country in which each bank is based, their interbank assets, interbank liabilities, total assets, equities, total customer deposits, total deposits, money market, short-term funding and derivatives. The 97 banks of the dataset represent 9 different EU countries: Austria (AT), France (FR), Germany (DE), Great Britain (GB), Greece (GR), Italy (IT), Portugal (PT), Spain (ES) and Sweden(SE). The number of banks for each country is reported in figure \ref{distr} and it is possible to notice that there is heterogeneity between countries. 
The fraction of total asset, interbank assets and liabilities per each country for 2007 and 2012 is reported in figure \ref{distr1} and as is possible to notice, the higher number of banks of a specific country does not directly translate to an higher contribution of asset and liabilities in the market. In fact, Great Britain has higher figures relative to Italy, even if it is represented by less than half of banks. On the contrary, Germany seems to be under-represented in term of total assets. This is probably due to the fact that the biggest banks are not present in our dataset.

\begin{figure}[h!]
	\centering
	\includegraphics[width=0.7\linewidth]{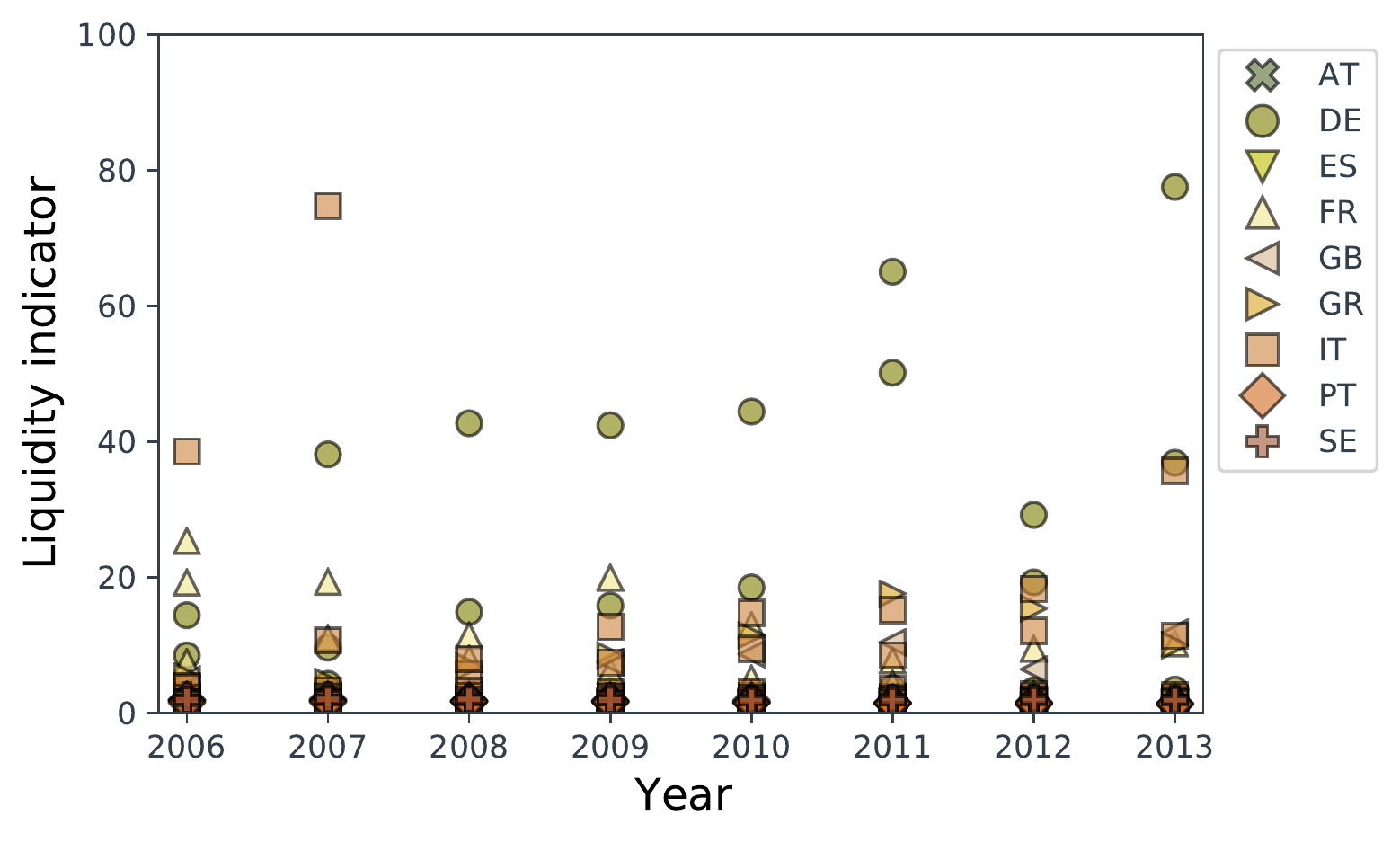}
	\caption{Banks' liquidity indicator during the period 2006-2013. Source: Bankscope database}
	\label{fig:bankterm}
\end{figure}

Total assets are used as the proxy for the bank size while the data on annual exposure in the interbank market (interbank assets and liabilities) defines the amount of money a bank lend or borrow from its counterparties. Since the actual interconnection between banks is not available, these data allow reconstructing the network topology using reconstruction methods that have been proved to have good performance in generating network close to the real one \citep{anand2018missing,cimini2015systemic,squartini2017network}.
The amount of money exchanged in the interbank market during the crisis shrank while the total volume of total assets and total deposits, money market and short-term funding increased during the same period. On one side, the banks increasingly hoard liquidity to counteract the crisis and on the other, several liquidity injections are made by Central Banks to support the banking system and to avoid the breakdown of the whole system. Thus, during the crisis, the interbank market freezes
\citep{fricke2015core,squartini2018reconstruction}. As explained in section \ref{cap1}, the contagion model takes into account some amplification mechanism. In particular, with the introduction of the node-term in both the contagion and default rates, we enrich the mechanical dynamic that the liquidity epidemics generate. Figure \ref{fig:bankterm} depicts the liquidity indicator of eq. \ref{liq_ratio}. Here it is possible to notice, most of the banks have a liquidity indicator of under 20, while a German bank has a value consistently over 40. This bank, if hit by a liquidity shock, could propagate the shock trough its neighbours because of the amplification mechanism the high liquidity indicator would have on the contagion rate.

\begin{figure}[h!]
	\centering
	\includegraphics[width=0.7\linewidth]{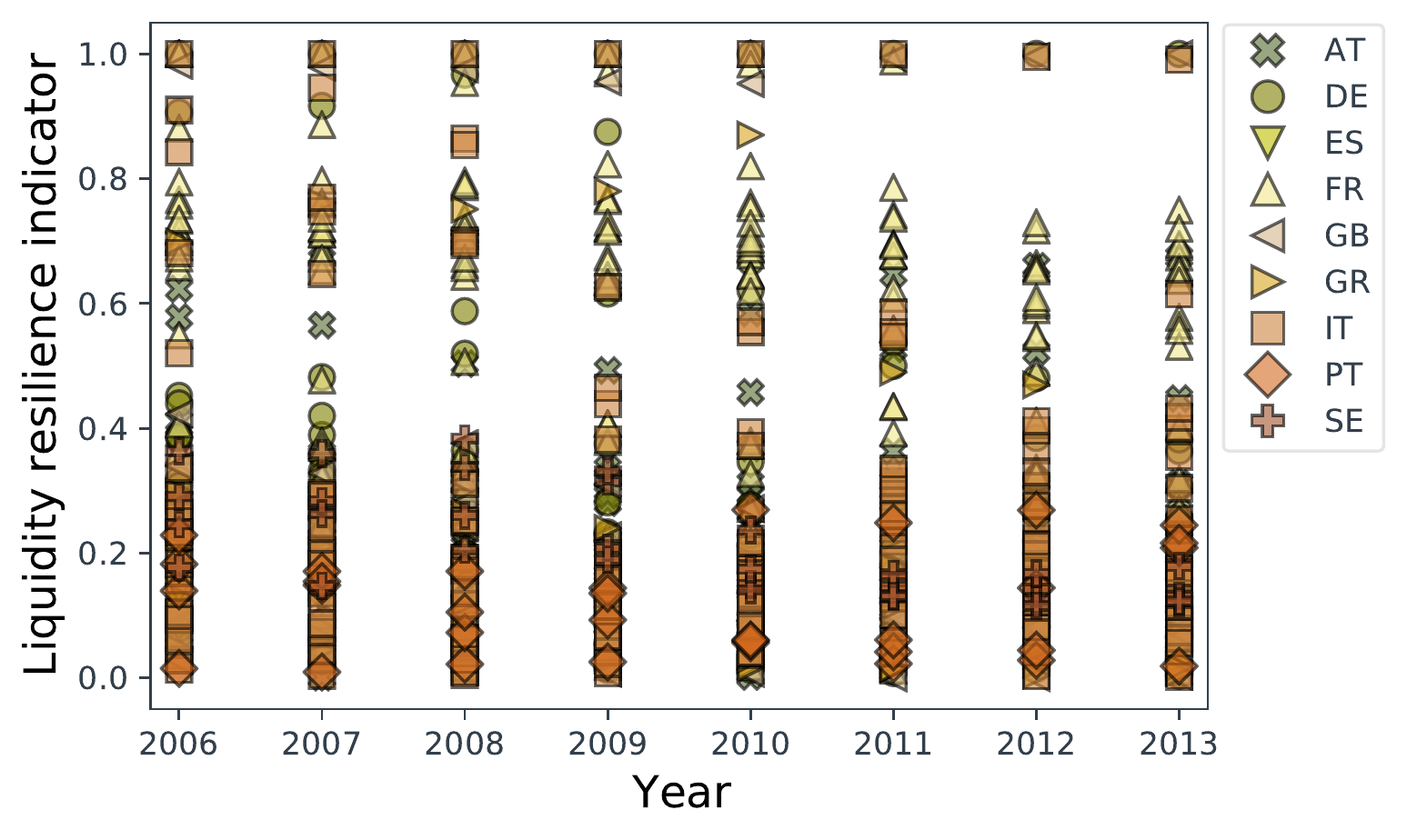}
	\caption{Banks' liquidity resilience indicator during the period 2006-2013. Source: Bankscope database}
	\label{fig:bankterm2}
\end{figure}

On the contrary, the liquidity resilience indicator introduced in eq. \ref{liqindicator} serves to accelerate or decelerate the default of a bank. In fact, if a bank has a liquidity buffer to sustain itself from an interbank liquidity shock, the probability of going bankrupted decreases. Figure \ref{fig:bankterm2} depicts the distribution for each year of the liquidity resilience indicator for all the banks in the dataset. 
As it is possible to notice, this indicator is always bounded between $0$ and $1$. Four banks have consistently high values so these ones are more inclined to default in case they are hit by a shock and consequently would be dangerous for the entire system.

\subsection{BIS data}
\begin{figure}[h!]
	\centering
	\includegraphics[scale=0.65]{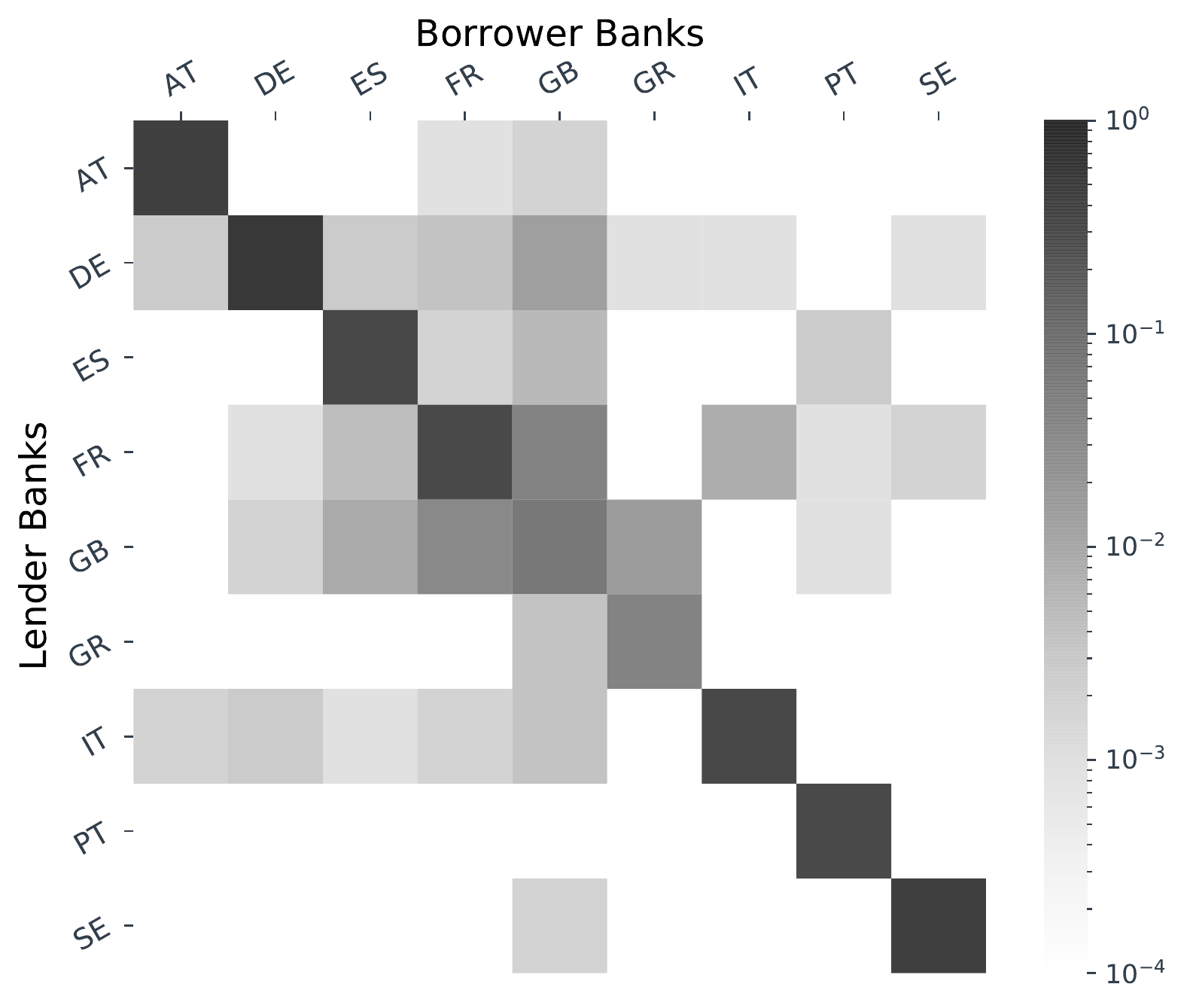}
	\caption{Rescaled inter-country interbank exposures (2013-Q4). Colorscale: log10. Source: BIS database}
	\label{bisriscalato}
\end{figure}

Cross-country interbank exposures are approximated by using consolidated banking statistics of the Bank for International Settlements (BIS). This statistical source provides information on banks’ financial claims, covering the cross-border bank exposures of a country’s banking sector with other countries. The idea is to incorporate countries' preferential lending relationship in the network reconstruction. This means that reconstructed networks have a block structure where each block represents the lending exposures between the two countries. BIS data give the annual amount of money that banks in country $U$ lend to banks in country $V$, hereinafter $exp_{UV}$. Considering the interbank market composed by the 9 countries in our dataset, the reconstruction takes into account the fraction of money borrowed and lent in each block when generates the links' weight between banks.\footnote{An alternative way to get the weights for the reconstruction could be to consider the whole bilateral exposure market and then restrict it to the set of countries available in the Bankscope dataset, renormalizing the weights. However, due to data availability, this is not possible in our case.}  Due to several missing values in the BIS dataset during 2006-2013, we use the bilateral exposures of 2013-Q4, which is the one with fewer missing values. In this respect, we need to impute the values for the pairs GB-GB, ES-ES, IT-AT and PT-PT. For the first three pairs we use the lending exposures GB-GB of 2014-Q1 and ES-ES and IT-AT of 2014-Q4 while for the PT-PT lending exposure, we imputed the value so that this exposure is proportional to the ES-ES, which has similar (in terms of proportion) bilateral exposures with the rest of the countries.\footnote{Alternatively, one could use the mean or median over the whole dataset or the PIGS countries (Portugal-Italy-Spain-Greece).} In figure \ref{bisriscalato} we report the rescaled BIS data defined as follows:
\begin{equation}
\centering
VOL_{UV}= \frac{(exp_{UV})^2}{\sum_V exp_{UV}\sum_U exp_{UV}} 
\label{volab}
\end{equation}
where $exp_{UV}$ is the BIS reported lending volume lent by banks of country $U$ to banks of the country $V$.
This figure shows two clear facts. First of all, it shows that there is a strong country preferential lending since banks tend to lend a higher amount of money to banks of the same country rather than to foreign banks. Secondly, there is a huge heterogeneity between foreign exposures among countries. Using this information in the reconstruction mechanism prevents the unconstrained reconstruction to be free of any economic structure. Having explained the issue of missing interbank market topology in our dataset, we now explain the state of the art in network reconstruction and our adaptation to our problem in which we impose financial specific constraints.   
\subsection{Network reconstruction}
The problem of network reconstruction is widely studied and several algorithms are proposed in order to overcome the limits of the information available. In the case of the interbank market networks, the data available are usually the aggregated exposures, usually taken from the banks' annual balance sheet, or in addition the total/partial number of bank's connections (nodes' degree). An overview of the reconstruction models is presented in the reference \citet{squartini2018reconstruction}.

Maximum-entropy reconstruction techniques are based on the maximum entropy principle. In fact, they consist of the maximization of the Shannon entropy that is defined using the banks' bilateral exposures ($w_{ij}$) as random variables. In fact, this reconstruction method is constrained only by the banks' aggregated positions (nodes' strength in and out). This is the most reasonable assumption if we do not have any further data on market structure. In this way, it is assumed that banks distribute their lending as evenly as possible given the assets and liabilities reported in the balance sheets of all other banks and then this leads to a maximization of the number of a bank's lending counterparties. This implies that the reconstructed networks have a fully-connected structure \citep{cimini2015systemic}. Maximum entropy (ME) techniques generate a fully connected network whilst real networks show rather a sparse topology. This disagreement leads then to a bad estimation of the resilience network properties. To overwhelm this limitation, it was developed other reconstruction techniques, the so-called fitness model. 

In many real networks, in fact, is observed that the links' probability can be expressed in term of intrinsic node-specific features, i.e. fitnesses. The idea is that nodes with stronger fitnesses are more likely to be network hubs, i.e. they have higher connectivity. 

\subsubsection{Fitness models}\label{finessformula}
Firstly introduced in \citet{bianconi2001competition,caldarelli2002scale}, fitness model is used to describe and reconstruct network topologies that are characterized by a strong correlation between the degree and the fitness of a node. Considering the interbank networks, the nodes' strengths themselves play the role of fitnesses since the total interbank assets and liabilities are considered as responsible for the nodes' connectivity distribution. These techniques allow reconstructing the networks' topological structure when only banks' aggregate exposures and link density are known. This approach consists of the two following steps: the reconstruction of the binary network topology and then the reconstruction of the weights to assign to those links \citep{cimini2015systemic,squartini2017network}. 

The binary topology is reconstructed from an ensemble of binary directed networks induced by the directed Configuration-Model \citep{cimini2015systemic,squartini2017network} assuming that nodes' fitnesses are linearly correlated to the out-degree and in-degree. In the case of interbank network reconstruction, the fitnesses are the in and out strength of the node $i$, $ s_i^{in}$ and $s_i^{out}$, i.e. the aggregated interbank liabilities and assets. We define $A_i$ and $L_i$ as the interbank assets and liabilities of the bank $i$, respectively, while $W=\sqrt{(\sum_iA_i)(\sum_i L_i)}$ is a normalization term that represents the expected induced total weight of the network. The outcomes are an ensemble of networks that meet the constraints given by the aggregated exposure on average. 

The adjacency matrix $a_{ij}$ is reconstructed from the probability matrix where the element $p_{ij}$ is the probability that the i-th bank lends to j-th bank:
\begin{equation}\label{pijold}
\centering
p_{i\rightarrow j}=\frac{z s_i^{out}s_j^{in}}{1+z s_i^{out}s_j^{in}}=\frac{zA_iL_j}{1+zA_iL_j}.
\end{equation}
The value of the free parameter $z$ is found imposing the value of the network density by mean of the following relation: 
\begin{equation}
\rho_{gen}=\sum_i\sum_j p_{ij}=\rho_{real}
\end{equation}
where $\rho_{gen}$ and $\rho_{real}$ are the density of the generated network and of the real one, respectively.

The weighted matrix $w_{ij}$ is then reconstructed using the degree-corrected gravity model. To respect the constraints given by the aggregated exposures in the balance sheets, the bilateral exposures are given by:
\begin{equation}\label{wijfitnessold}
\centering
w_{i\rightarrow j}=\frac{z^{-1}+s_i^{out}s_j^{in}}{W}{a}_{ij}
\end{equation}
where
\begin{equation}\label{aijfitnessold}
\centering
{a}_{ij}=
\begin{cases}
1&\text{\small with probability $p_{ij}$} \\
0&\text{\small otherwise.}
\end{cases}
\end{equation}

These reconstruction methods allow self-loops. However, in many real networks, self-loops are absent or even excluded. It is necessary to opportunely redistribute each diagonal term of the weight matrix across the columns and rows in order to still maintain the strengths constraints. For this reason, a procedure inspired to the Iterative Proportional Fitting algorithm (IPF) has to be implemented \citep{cimini2015estimating}. In a recent paper \citep{anand2018missing}, performances of different reconstruction models based only on the aggregated data are tested using the true bilateral data obtained from 25 markets spanning 13 jurisdictions. It is in fact postulated that the only information available is the aggregate asset and liability positions of each bank. Such datasets consist on interbank networks, payment networks and several other networks for different financial contracts: repurchase agreement, i.e. repos, foreign exchange derivatives (FX), credit default swaps (CDS), and equities. These networks' differences make the performance of the reconstruction algorithms depending on the real network considered since each method varies in term of the emphasis placed on the network features to reproduce. The purpose of the analysis in \citet{anand2018missing} is testing the different methods' performance in the reconstruction and it is shown as the aforementioned fitness model \citep{cimini2015systemic} is the clear winner between the ensemble methods across different markets. However, these methods are not anchored to any economic indicator, so potentially weakening their intrinsic information. In order to include reasonable macroeconomic constraints, we assume that the volume of the relative inter-country lending exposure is proportional to the one reported in the BIS Consolidated Banking Statistics dataset. First, we divide banks into different groups by their nationalities and then we divide the weight matrix into local blocks, where each block represents the lending relationship between banks of country $U$ to banks of country $V$. These blocks are reconstructed separately using the aforementioned method and imposing the following local fitnesses.

We define $exp_{UV}$ the interbank exposure of the banking sector of the country $U$ to the country $V$, where $U$ is the reporting country (lender) and $V$ is the counterparty (borrower).
We define the interbank assets and liabilities in the block $UV$ as:
\begin{equation}
\centering
\tilde{A_i}^{UV}=A_i \left( \frac{exp_{UV}}{\sum_Vexp_{UV}} \right)
\end{equation}
\begin{equation}
\centering
\tilde{L_j}^{UV}=L_j \left( \frac{exp_{UV}}{\sum_Uexp_{UV}} \right)
\end{equation}
from which it follows $\sum_V \tilde{A_i}^{UV}=A_i$ , $\sum_U \tilde{L_j}^{UV}=L_j$.

The probability matrix and the weight matrix are so defined as:
\begin{equation}\label{pij}
\centering
p_{i\rightarrow j}^{UV}=\frac{z\tilde{A_i}^{UV}\tilde{L_j}^{UV}}{1+z\tilde{A_i}^{UV}\tilde{L_j}^{UV}}.
\end{equation}
\begin{equation}\label{wijfitness}
\centering
w_{i\rightarrow j}^{UV}=\frac{\tilde{A_i}^{UV}\tilde{L_j}^{UV}}{W^{UV}p_{i\rightarrow j}^{UV}}a_{i\rightarrow j}^{UV},
\end{equation}
where $W^{}=\sqrt{(\sum_i\tilde{A_i}^{UV})(\sum_j \tilde{L_j}^{UV})}$.
We find the free parameter $z$ by solving the equation:
\begin{equation}\label{rhoo}
\centering
\rho_{gen}=
\sum_U\sum_V \left(\sum_{i\in U}\sum_{j\in V}p_{i\rightarrow j}^{UV}\right)=
\sum_U\sum_V \rho_{gen}^{UV}= \rho_{real}  .
\end{equation}
Imposing the same value of the free parameter $z$ in each block is the simplest assumption that requires only the overall density value of the interbank matrix. On the contrary, when the density values of each block are available, different block values of the free parameter $z_{UV}$ could be considered and found by imposing $\rho_{gen}^{UV}=\rho_{real}^{UV}~ \forall ~U,V$.

Results show as the imposition of such constraints preserves the network's core-periphery structure, typical of the interbank market network. However, it makes the network density not uniform over the country blocks, reflecting the preferential lending relationship, especially between banks of the same country.

\section{Results}\label{cap3}
Our work deals with two aspects: network reconstruction and financial contagion modelling. Firstly, we have implemented reconstruction techniques to generate, from aggregated data, the structure of banks' bilateral exposures. Such reconstructed networks constitute the underlying topology of the contagion spreading. Then, we have run our epidemic-like contagion models on the reconstructed interbank networks and we have tested models' sensitivity with respect to different scenarios. From now on, for the sake of readability, we report only plots related to 2007 and 2012.\footnote{ Plots of other years in the period 2006-2013 are available in the electronic supplementary materials.}

\subsection{Network reconstruction}
\begin{figure}[h!]
	\begin{subfigure}{1\textwidth}
		\centering
		
		\includegraphics[width=.48\linewidth]{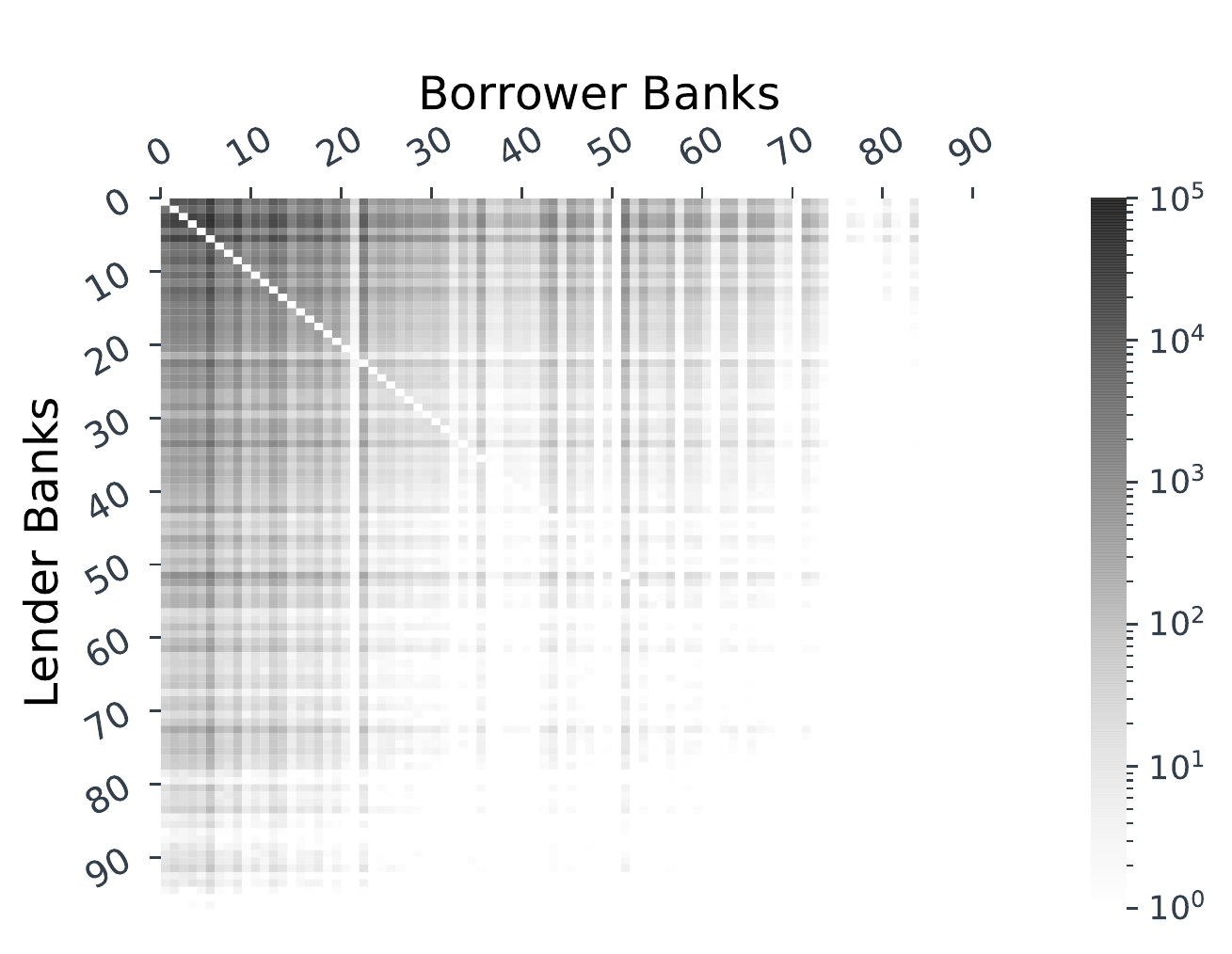}  
		\includegraphics[width=.48\linewidth]{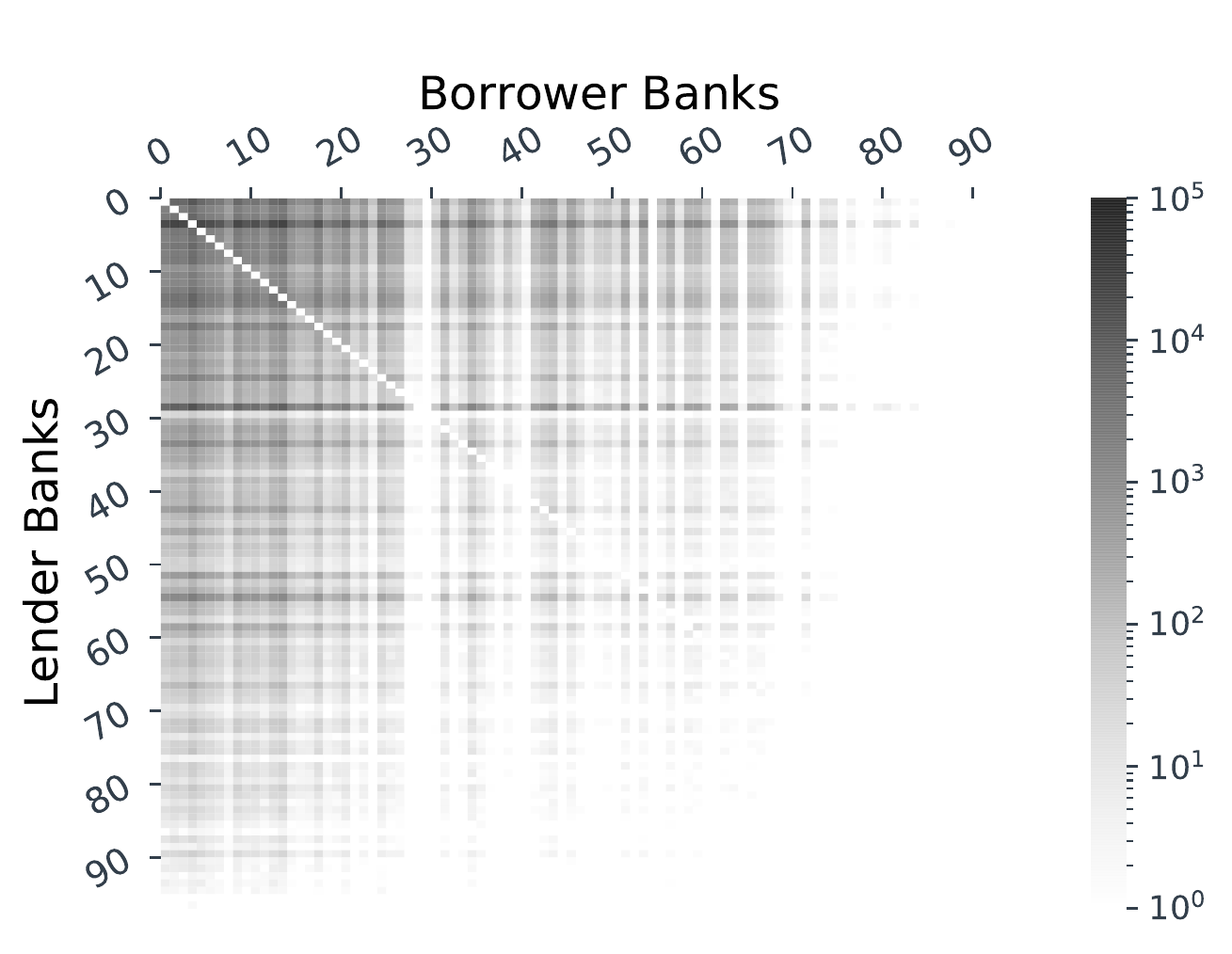}  
		\caption{Fitness model}
		\label{tatweight-giulio}
	\end{subfigure}	
	\begin{subfigure}{1\textwidth}
		\centering
		
		\includegraphics[width=.48\linewidth]{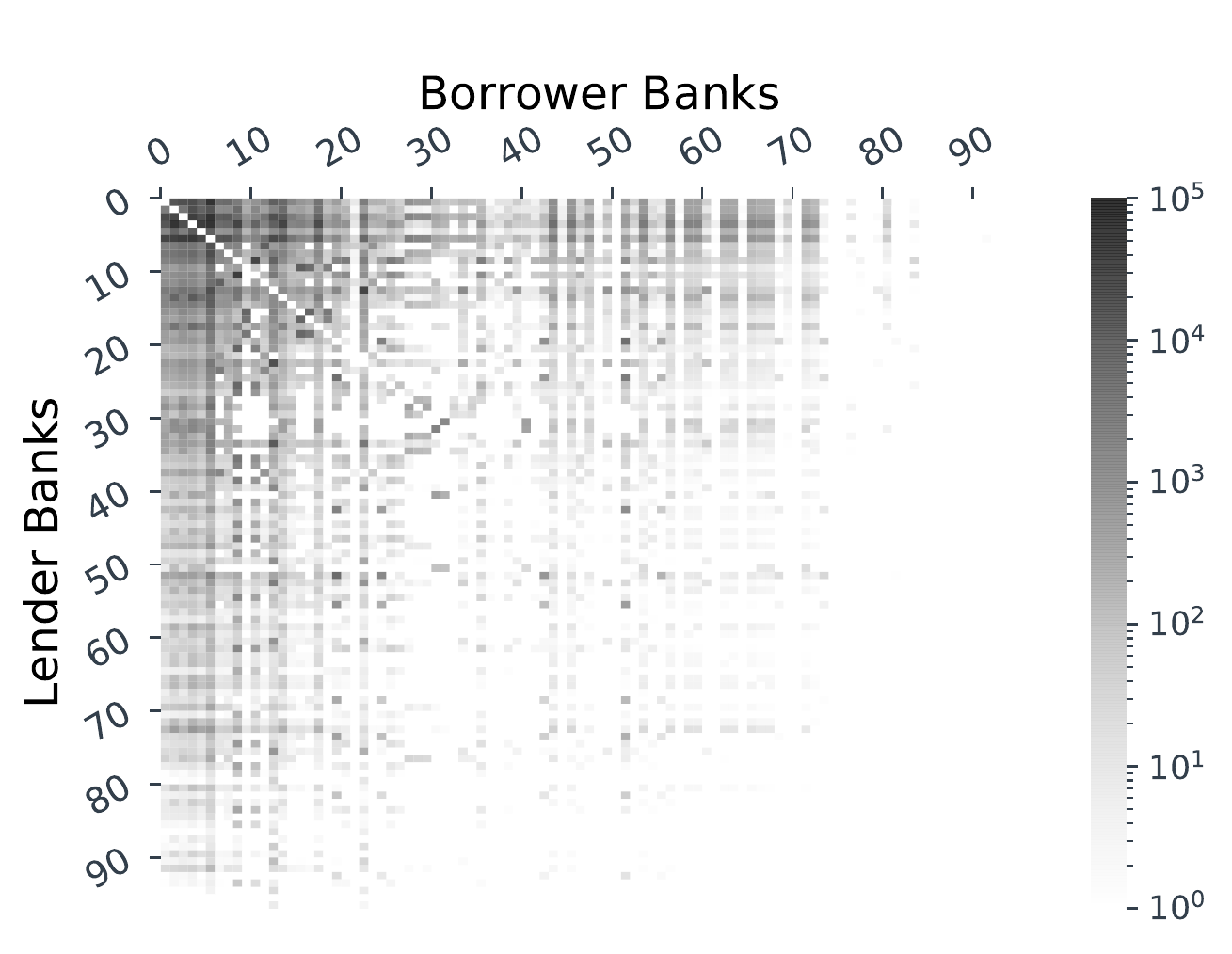}  
		\includegraphics[width=.48\linewidth]{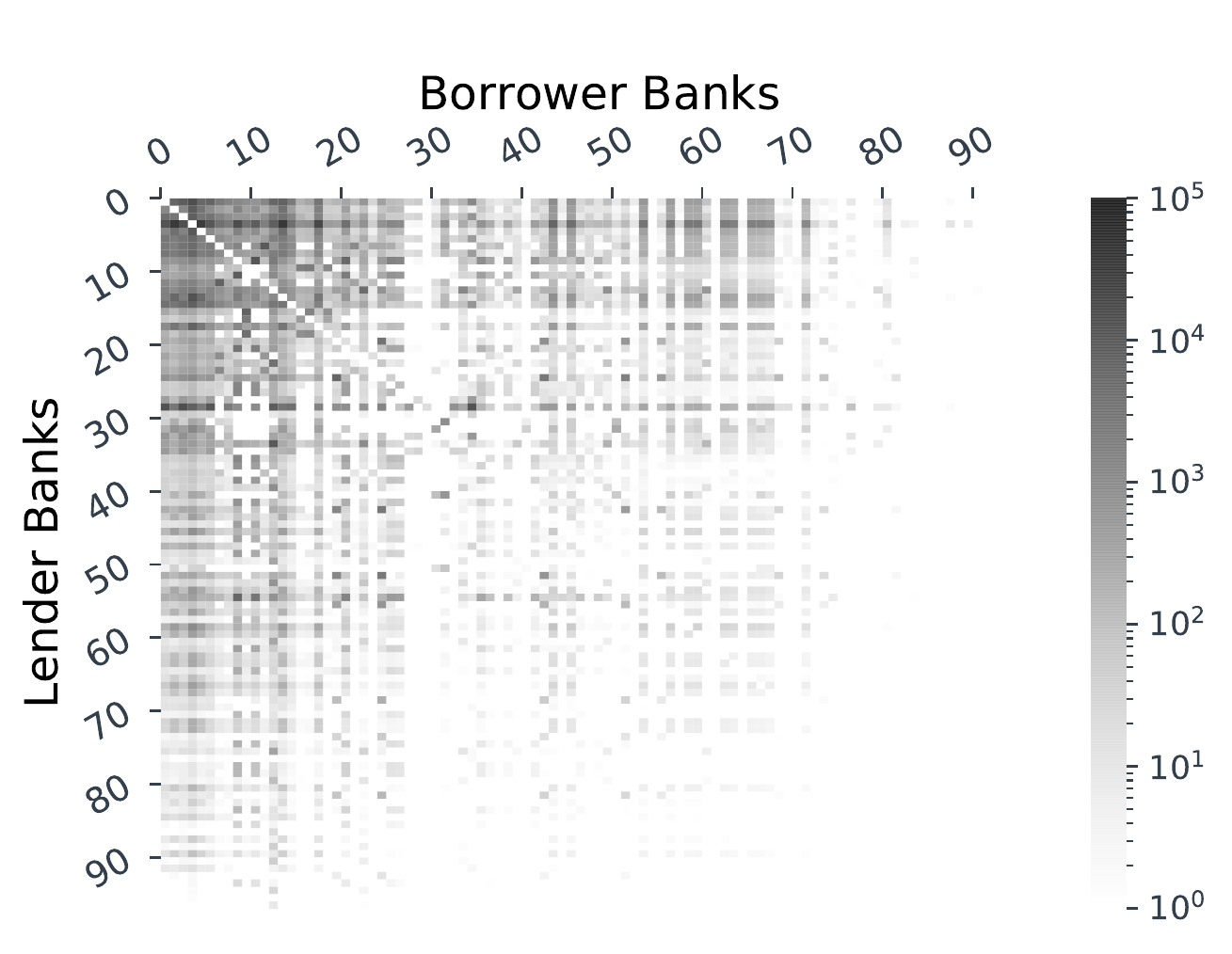}  
		\caption{Fitness model with country-block structure}
		\label{tatweight-noi}
	\end{subfigure}
	\caption{Weight matrix (mil USD), respectively of 2007 and 2012. Mean on the network ensemble. Banks are sorted by total assets of 2013. Colorscale: log10}
	\label{tatweight}
\end{figure}

We have reconstructed the interbank market network using the Fitness model already discussed in section \ref{finessformula}. These reconstruction techniques generate an ensemble of weighted directed networks without self-loops since inter-node transactions are not considered. This reconstruction method requires that the sum of total interbank asset and liabilities is the same $\sum_i A_i=\sum_l L_j$. However, in our dataset, the total amount of interbank liabilities is generally greater than the one of interbank assets so we introduced a "ground" bank. This bank does not participate in the contagion dynamics and acts as an external lender whose assets are given by the difference $A_{gbank}=\sum_j L_j-\sum_i A_i$. In this way, we succeed to generate an ensemble of networks that are consistent with the balance sheet constraints. It is assumed that the density value of the interbank network $\rho_{real}$ does not change over the period considered (2006-2013) and it is equal to $0.3$.\footnote{This density value is reasonable when compared with the density of the European interbank network \citep{anand2018missing}. We repeated the simulations with a lower density parameter and the results were qualitatively identical.} For each year in the period 2006-2013, we have generated an ensemble of 100 weighted directed networks using the equations \ref{pij} and \ref{wijfitness}. Figure \ref{tatweight-noi} reports the mean over the network ensemble of the weight matrices. It is possible to notice from the plot that, as for the standard fitness model reconstruction, the network exhibits a core-periphery structure when the banks are sorted by total asset, similar to the real interbank topology \citep{finger2013network,fricke2015core,craig2014interbank}. 

\begin{figure}[h!]
	\begin{subfigure}{1\textwidth}
		\centering
		
		\includegraphics[width=.48\linewidth]{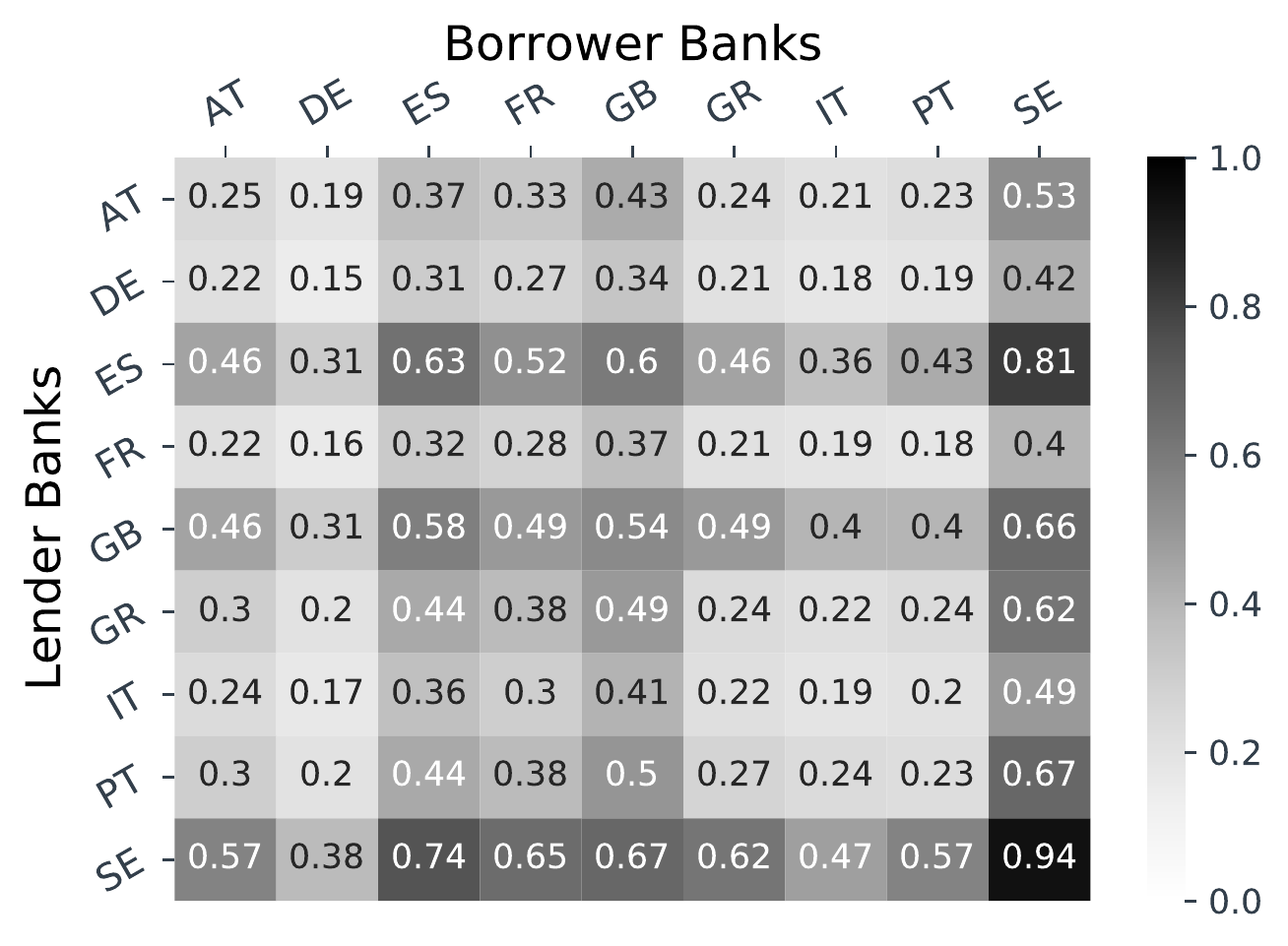}  
		\includegraphics[width=.48\linewidth]{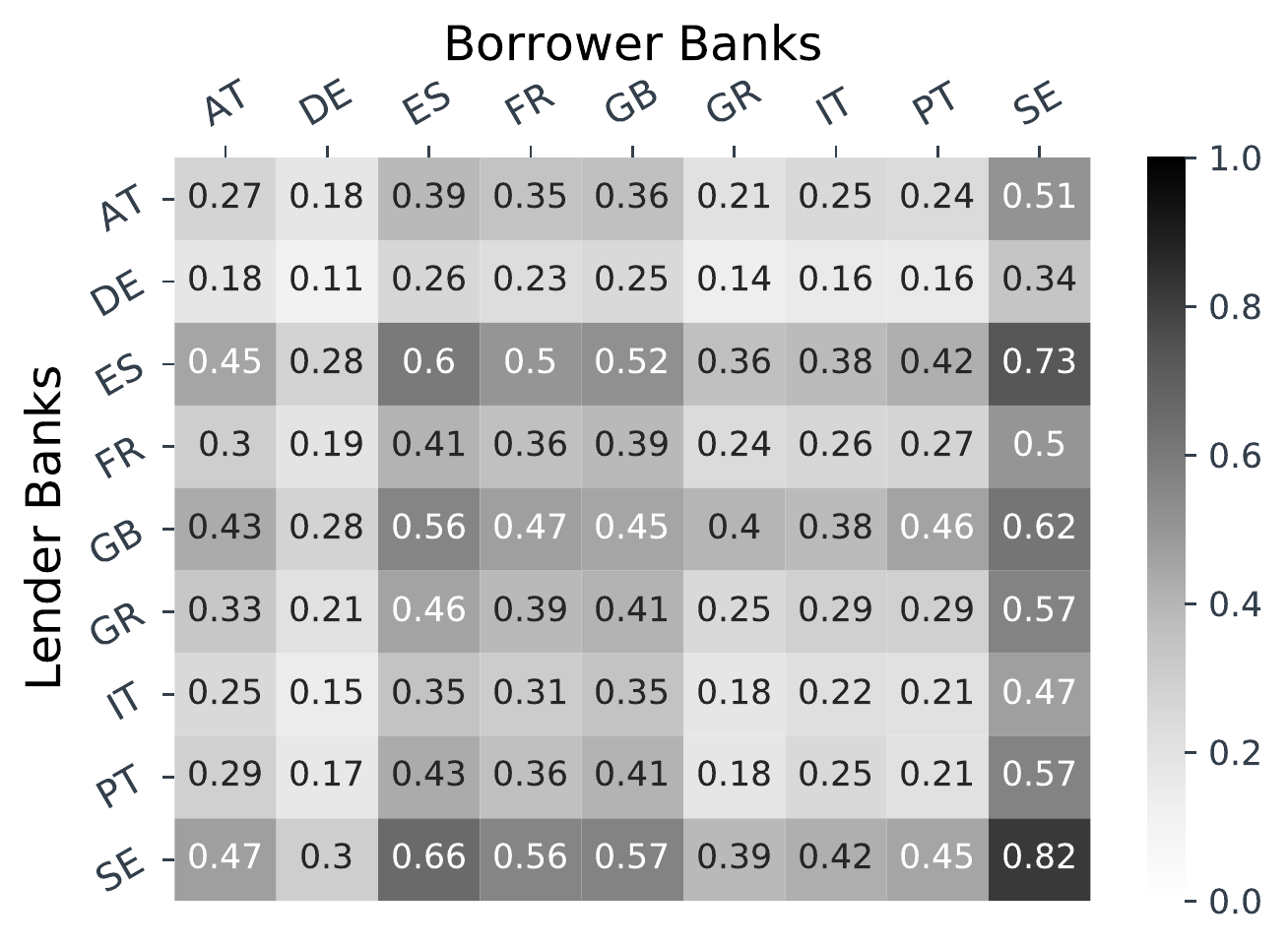}  
		\caption{Fitness model}
		\label{density-giulio}
	\end{subfigure}	
	\begin{subfigure}{1\textwidth}
		\centering
		
		\includegraphics[width=.48\linewidth]{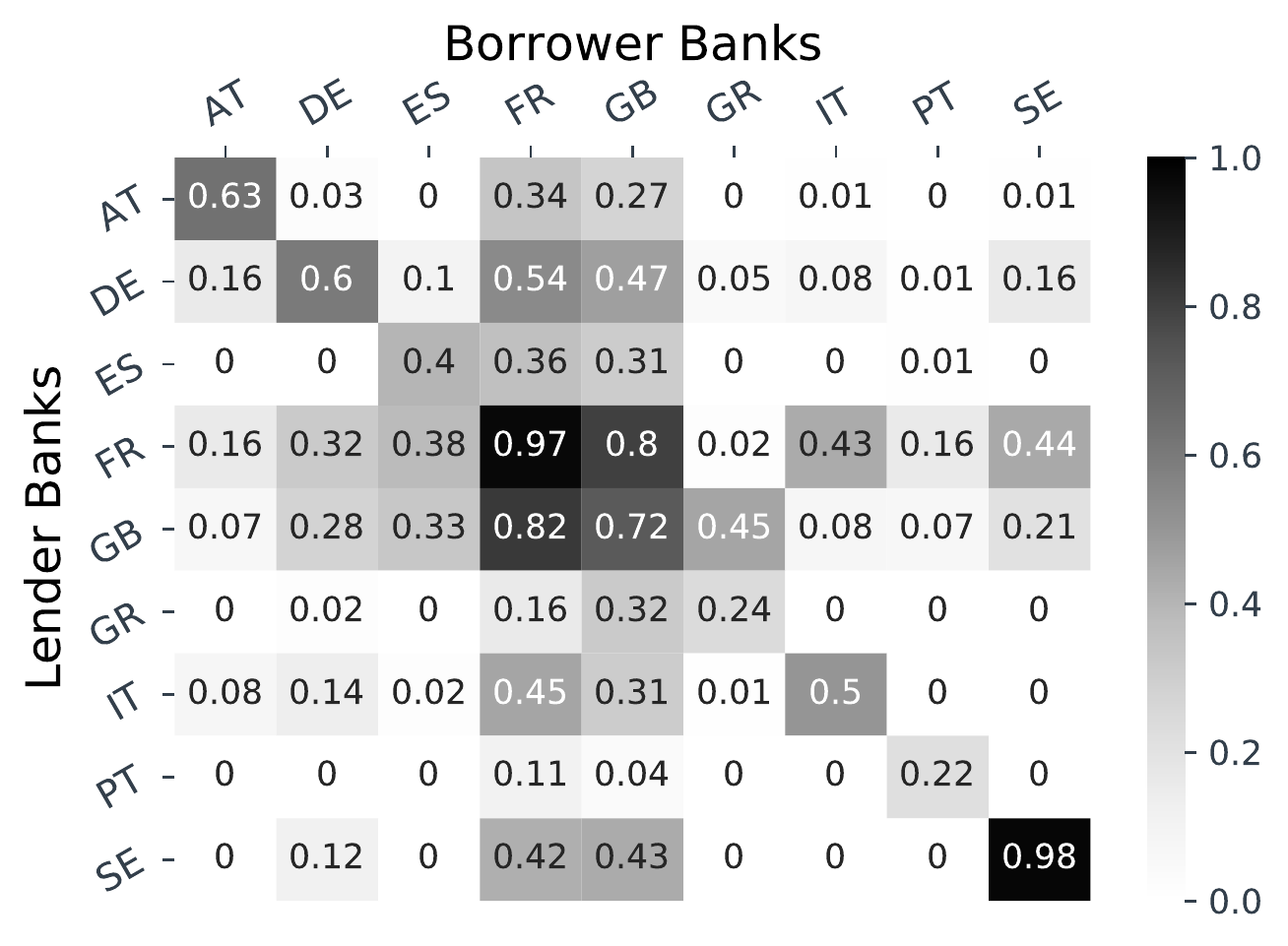}  
		\includegraphics[width=.48\linewidth]{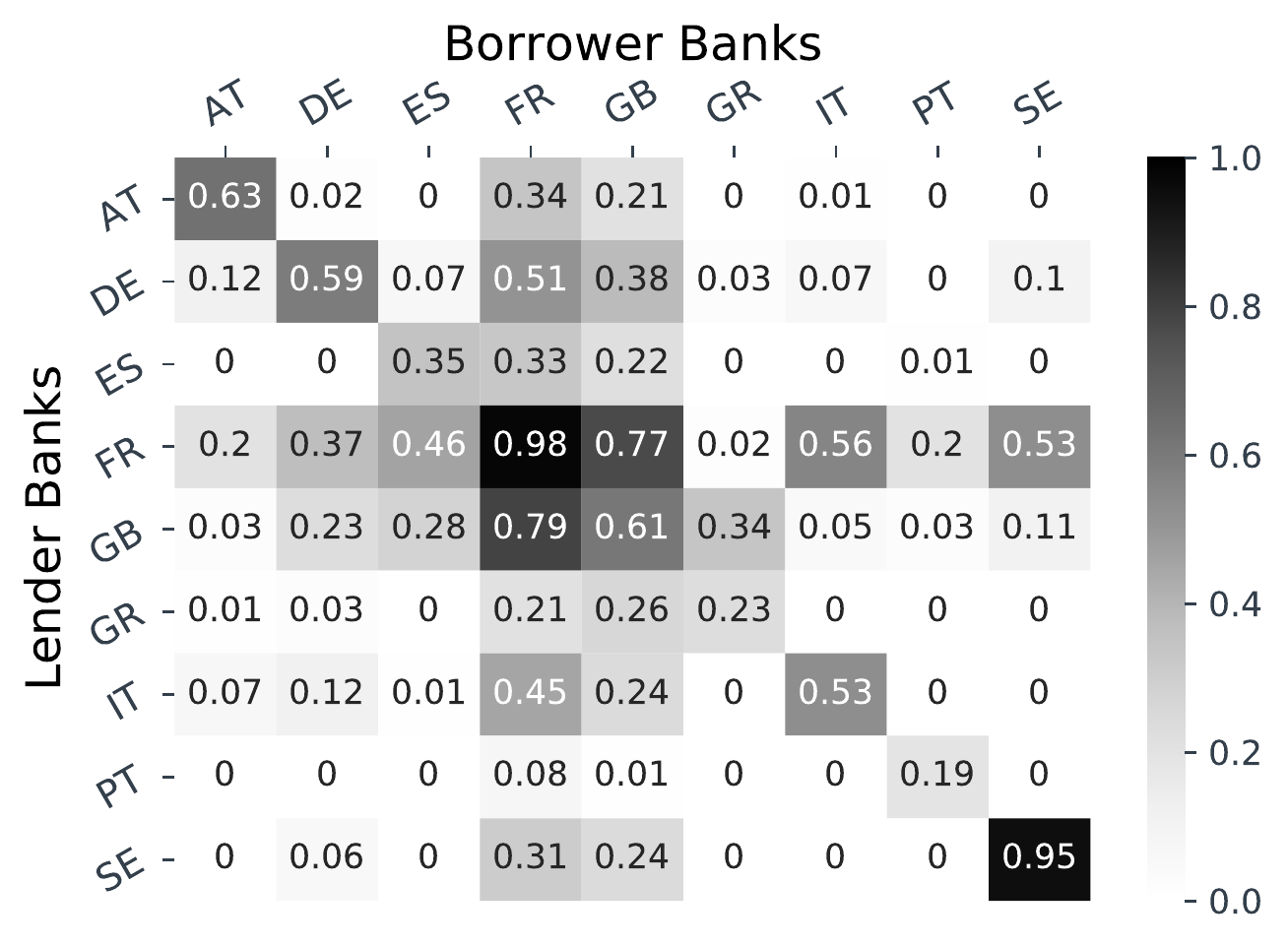}  
		\caption{Fitness model with country-block structure}
		\label{density-noi}
	\end{subfigure}
	\caption{Block density, respectively of 2007 and 2012. Mean on the network ensemble. Banks are sorted by country code. Colorscale: linear }
	\label{ddensity}
\end{figure}

Due to the country-block constraints, the density is not uniform with respect to the banks' country as we could see in the figure \ref{density-noi}.\footnote{We have only generated network topologies with a unique weakly-connected component so there are no isolated nodes.} This result is in line with the concept of preferential lending in the interbank market as we would expect in a real financial system. The comparison with the reconstruction without economic constraint is striking. Without imposing such constraints, the network density in each block is uniform, something difficult to assume from an economic viewpoint. Besides, the average values of the node strength are in line with those ones reported in the balance sheets. In both reconstructions, there are high values of the Sweden block densities. This is because there are only four Swedish banks but very big in term of total assets and interbank assets and liabilities. This implies a higher probability to be connected (eqs. \ref{pijold}, \ref{pij}).

\begin{figure}[h!]
	\begin{subfigure}{1\textwidth}
		\centering
		
		\includegraphics[width=.48\linewidth]{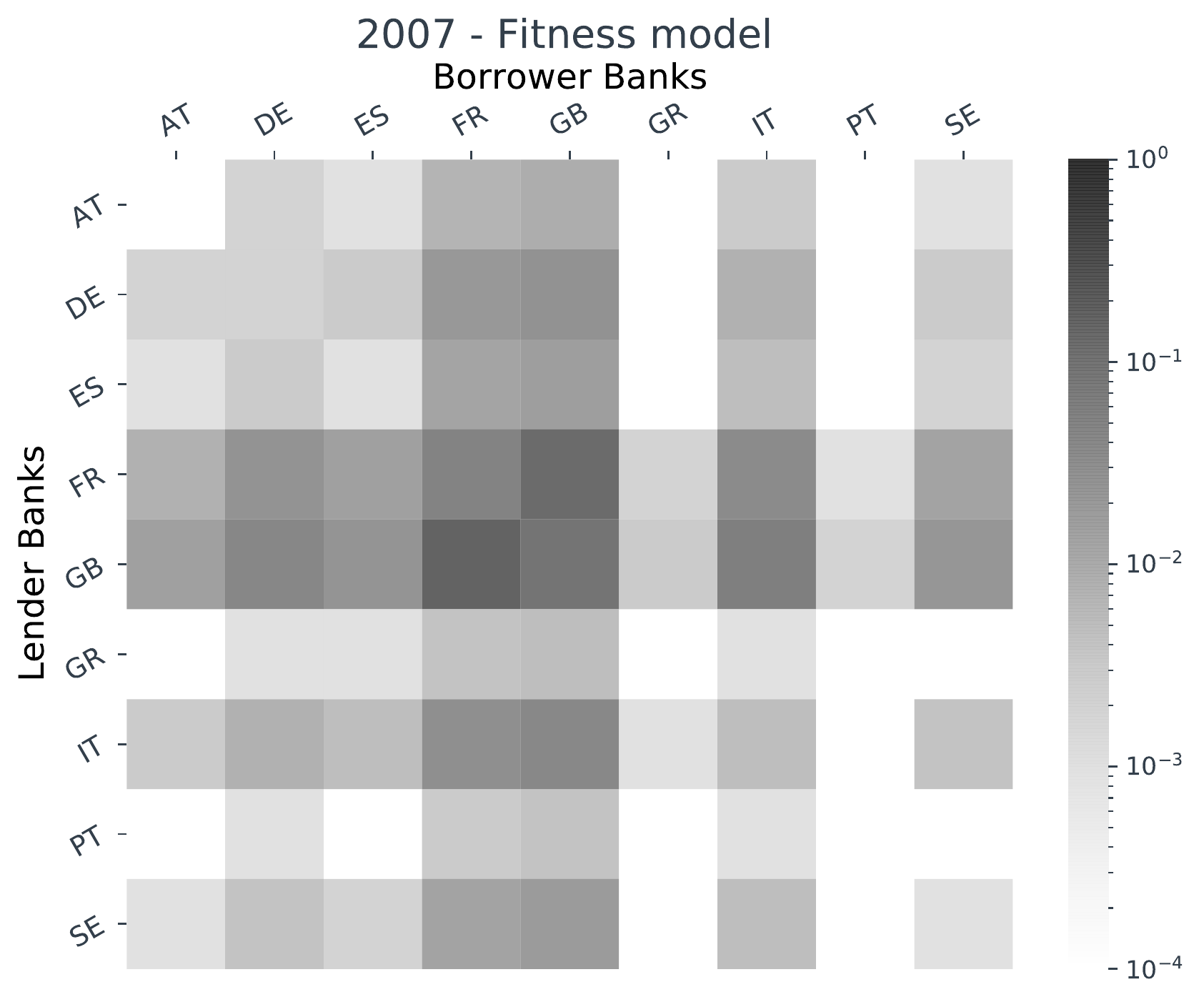}  
		\includegraphics[width=.48\linewidth]{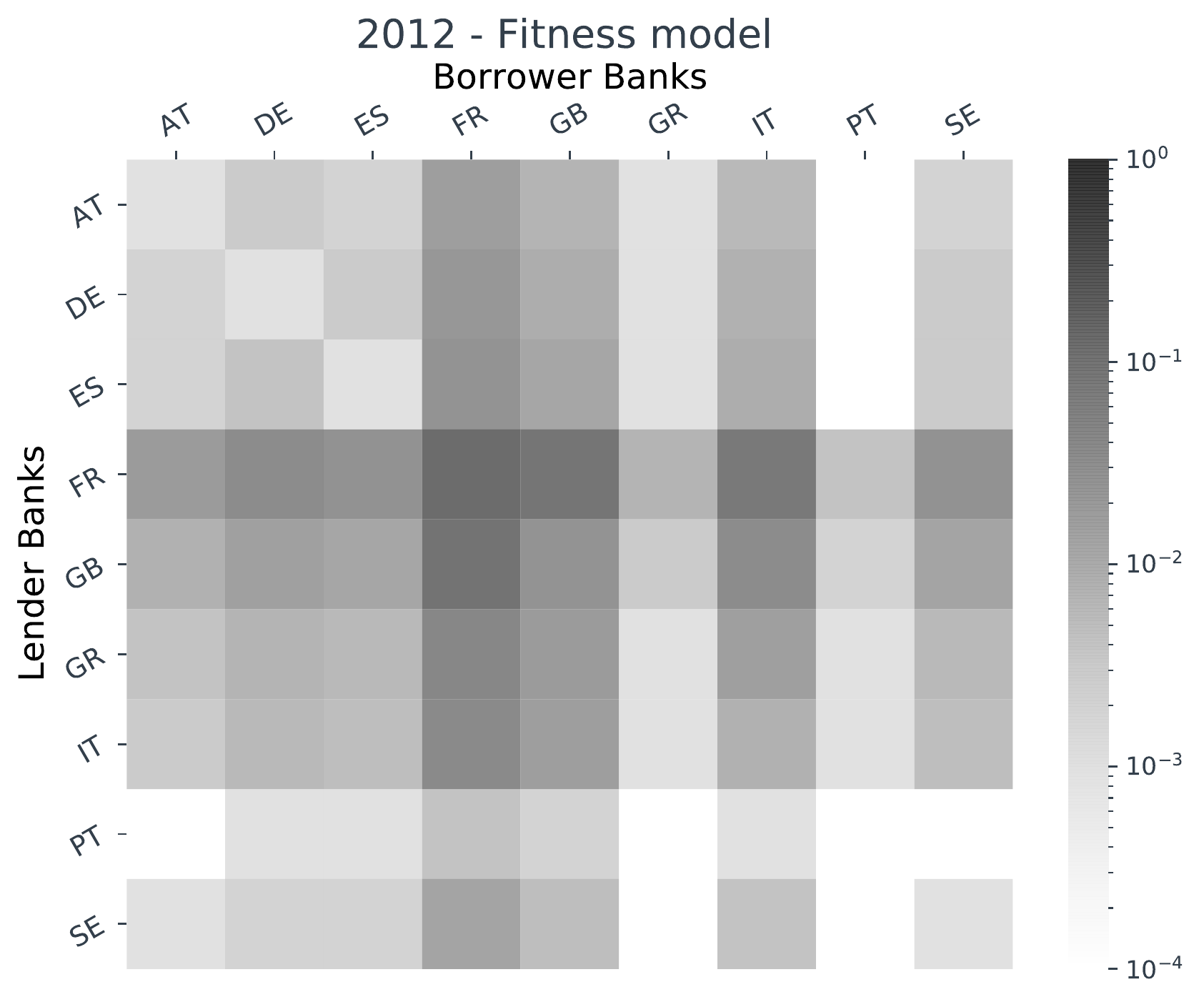} 
		\caption{Fitness model}
		\label{BIS-giulio}
	\end{subfigure}	
	
	\begin{subfigure}{1\textwidth}
		\centering
		
		\includegraphics[width=.48\linewidth]{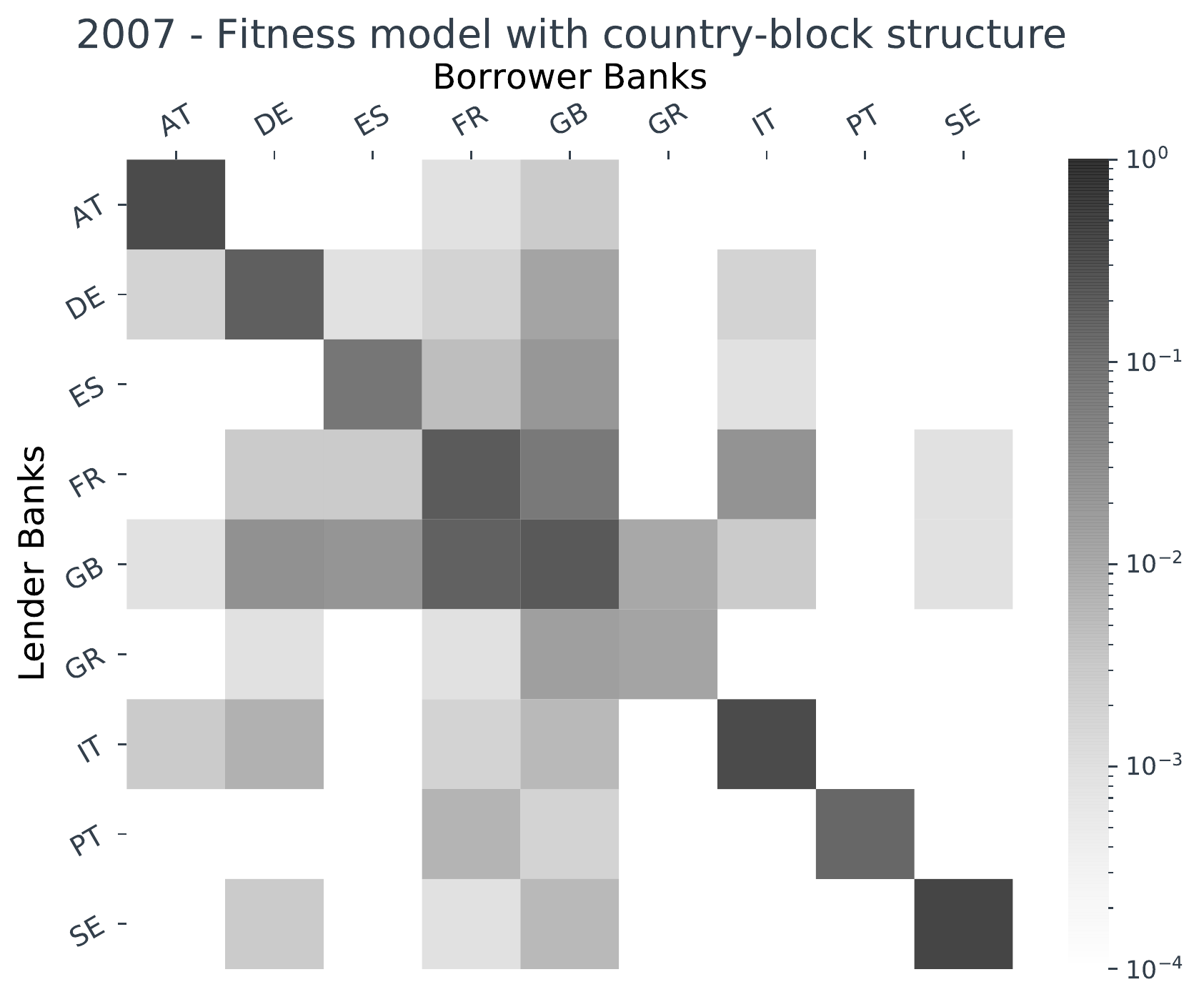}  
		\includegraphics[width=.48\linewidth]{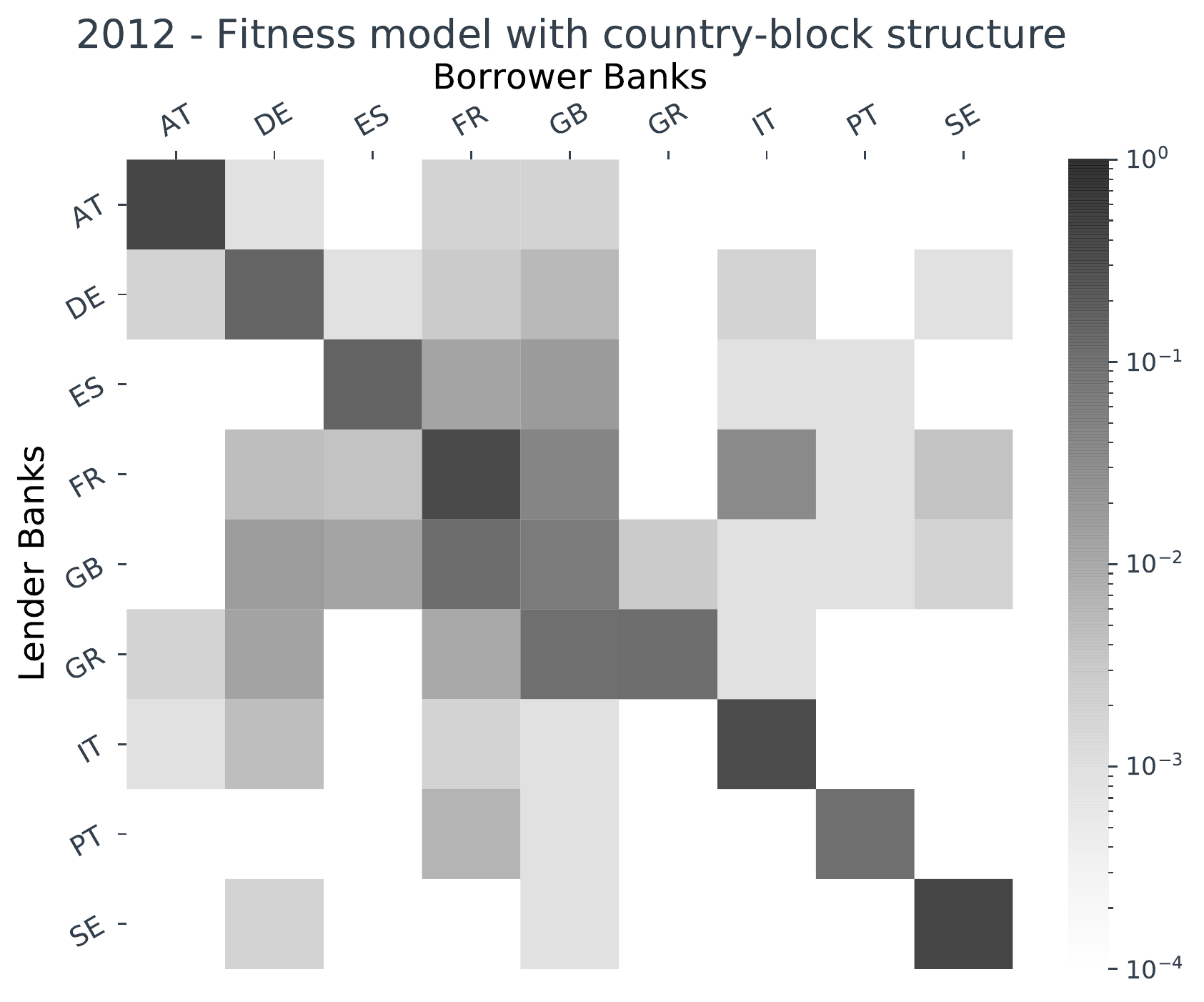}  
		\caption{Fitness model with country-block structure}
		\label{BIS-noi}
	\end{subfigure}
	\caption{Rescaled intra-country lending exposure (eq. \ref{volab}), respectively 2007 and 2012. Mean on the network ensemble. Colorscale: log10 }
	\label{BISblock}
\end{figure}

The contribution given by the block constraints to the reconstruction technique is more evident when we compare the weighted network obtained imposing or not such country requirements, i.e. the volumes of inter and intra-country exposures provided by the BIS database. For the matter of sole graphical analysis, we compare the rescaled inter-country volume exposures (eq.\ref{volab}) of the real and the generated topology, considering the cases in which the block structure is enforced or not. Results (Figs. \ref{bisriscalato}, \ref{BISblock}) confirm a clear similarity between real and reconstructed inter-country exposures. Given the result of this enhanced reconstruction with economic constraints, we use the generated ensemble of networks to run our contagion model to asses systemic risk in the interbank market.

\subsection{Systemic liquidity risk contagion}

\subsubsection{Contagion setting}\label{SIinSI}
As previously explained, we have generated an ensemble of 100 weighted directed networks for each year and on each annual network in the ensemble, we compute 97 stochastic SI contagion simulations for each network of the ensemble. Each of these simulations has a different initial infectious seed, i.e. the infected bank from which the infection starts. Each contagion begins from a single distressed bank and each node is used once as initial seed. We have in total 9700 contagion dynamics for each year. The simulation stops when no more banks are in the distressed state or 50 iterations are reached.

\subsubsection{Contagion without node-term}
\begin{figure}[h!]
	\centering
	\includegraphics[width=0.49\linewidth]{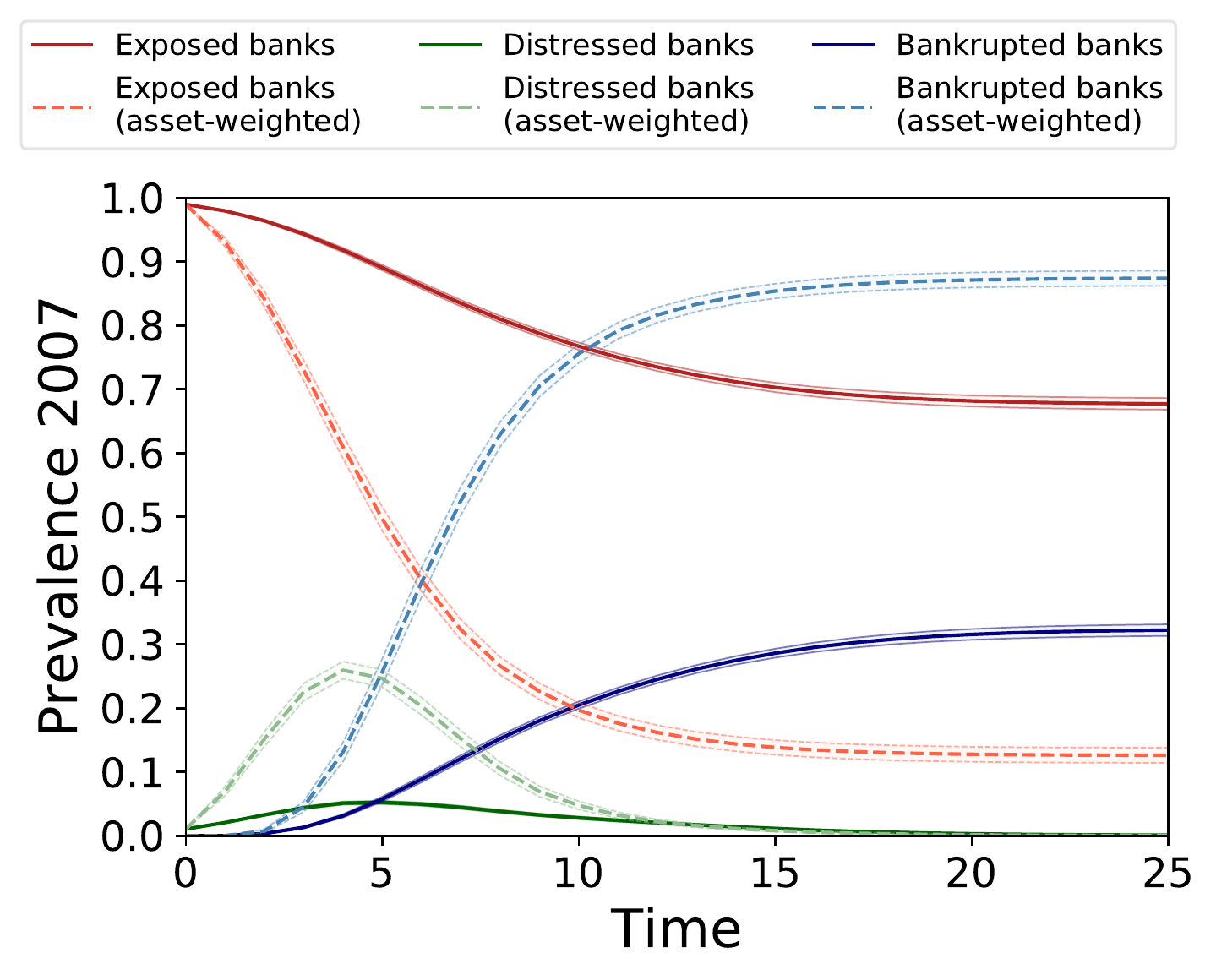}
	\includegraphics[width=0.49\linewidth]{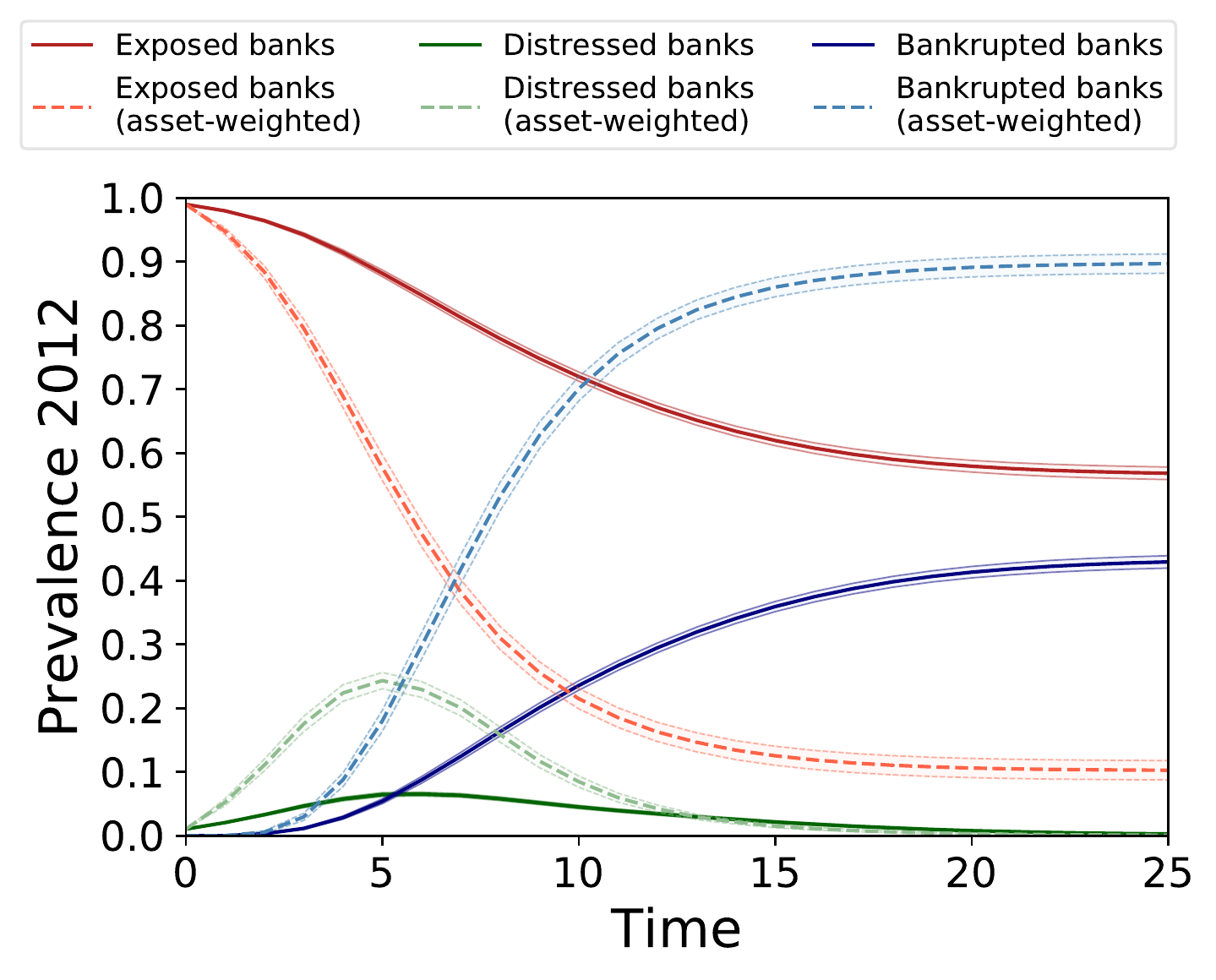}
	\caption{Prevalence dynamics in the 2007 and 2012 for the different bank states. Mean over all the simulations. Contagion rate without node-term}
	\label{prevalencenonodo}
\end{figure}

In this scenario, we consider the functional form of the contagion rate $\lambda$ (eq. \ref{lambdabrandi}) already discussed in \citet{brandi2018epidemics}. In figure \ref{prevalencenonodo}, it is shown the prevalence dynamics in 2007 and 2012, namely the proportion of banks that belong to each compartment at each time-step of the contagion spreading. It is also reported the annual weighted prevalence so each bank is represented by the fraction of total assets that it owns in the considered year. It is possible to notice that although similar, the 2012 contagion spreads slightly faster. The contagion dynamic is also more rapid and sharper when we consider the prevalence weighted by banks' total assets.

\begin{figure}[h!]
	\centering
	\includegraphics[width=0.7\linewidth]{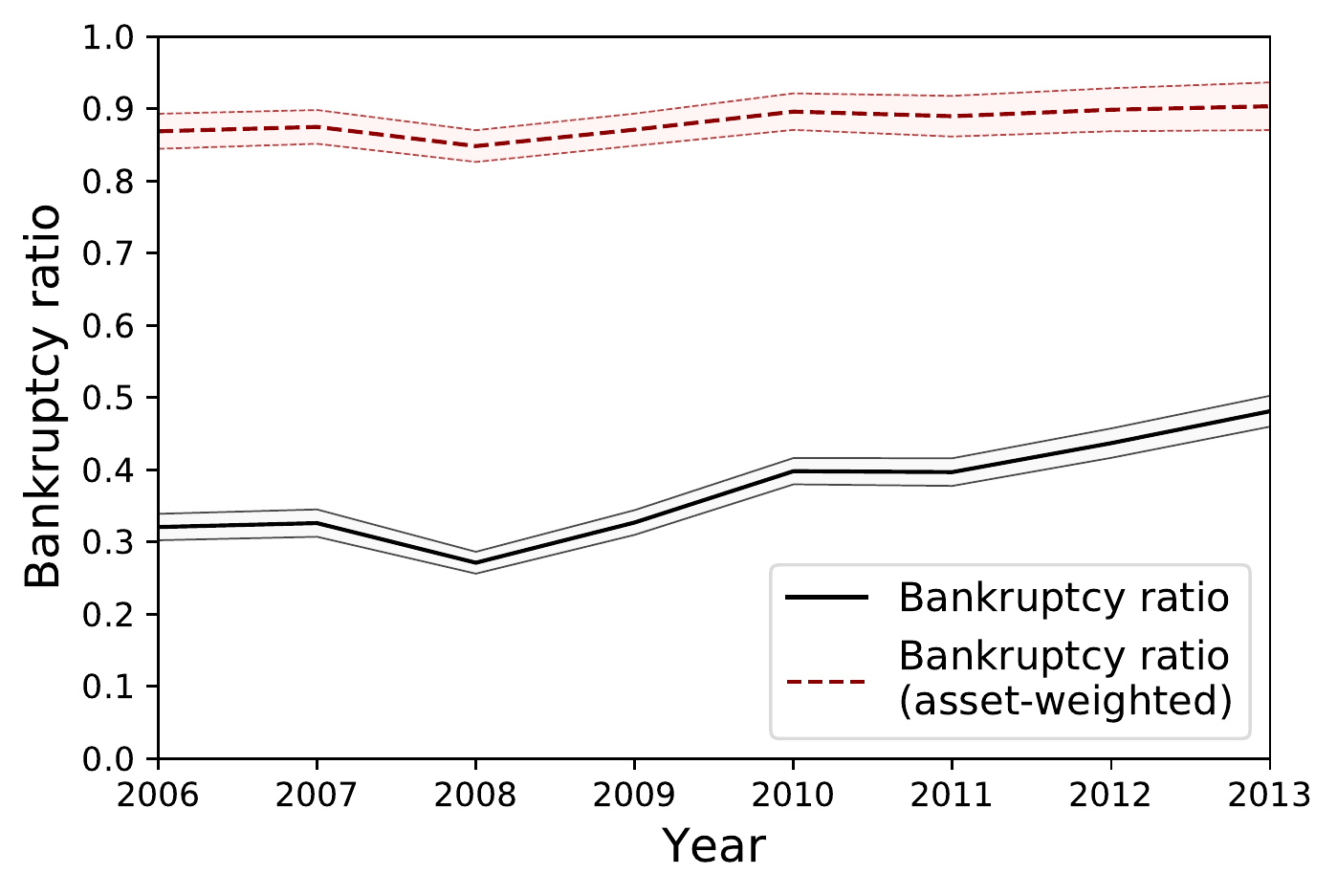}
	\caption{Time evolution of the bankruptcy ratio over the period 2006-2013. Contagion rate without node-term. Mean over the all simulations with 95\% confidence bands}
	\label{trendnonodo}
\end{figure}

\begin{figure}[h!]
	\centering
	\begin{subfigure}{0.49\textwidth}
		\includegraphics[width=1\linewidth]{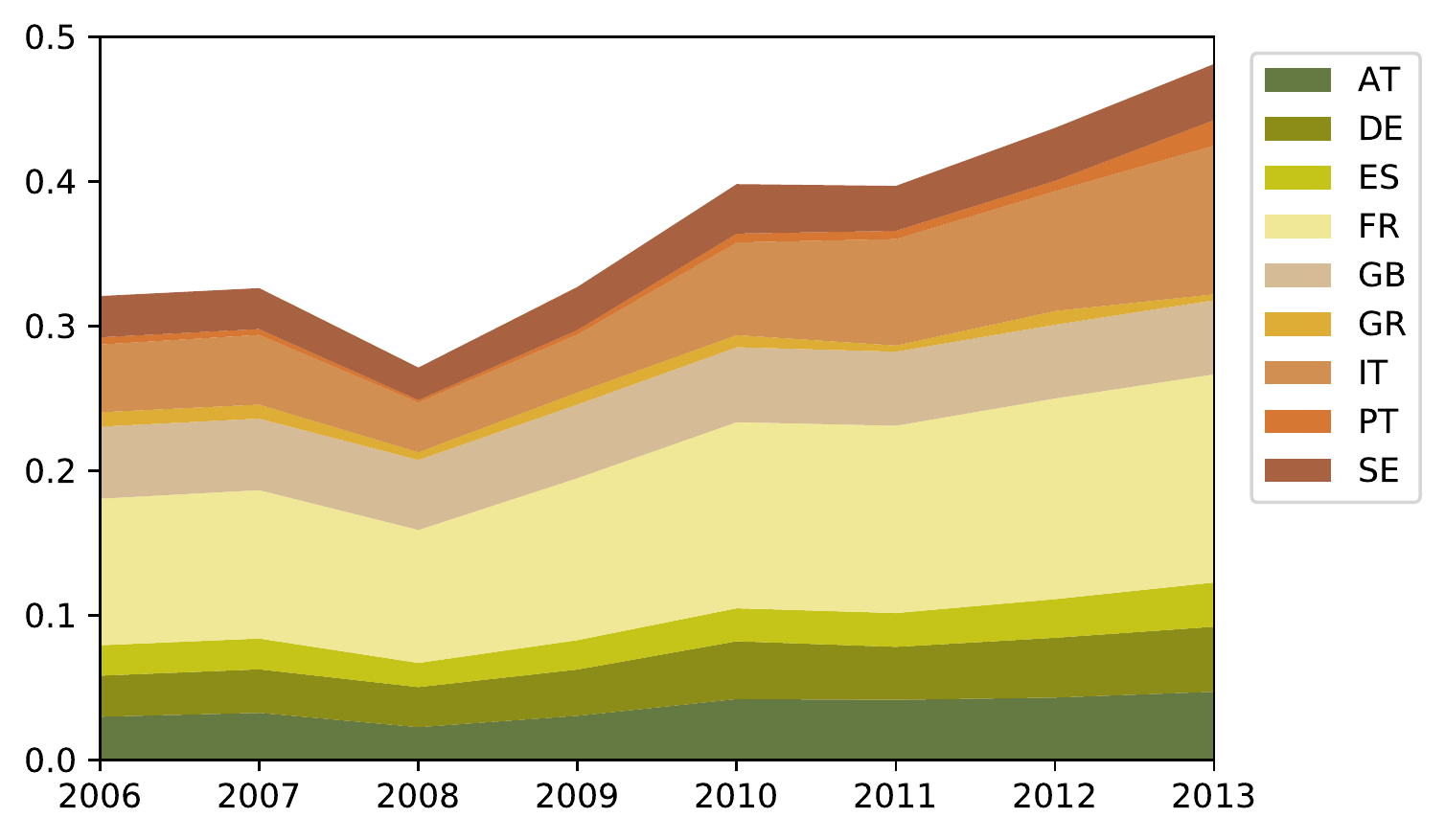}
		\caption{Fraction of defaulted banks}
		\label{fig:decomposition_no_nodeterm1}
	\end{subfigure}
	\begin{subfigure}{0.49\textwidth}
		\includegraphics[width=1\linewidth]{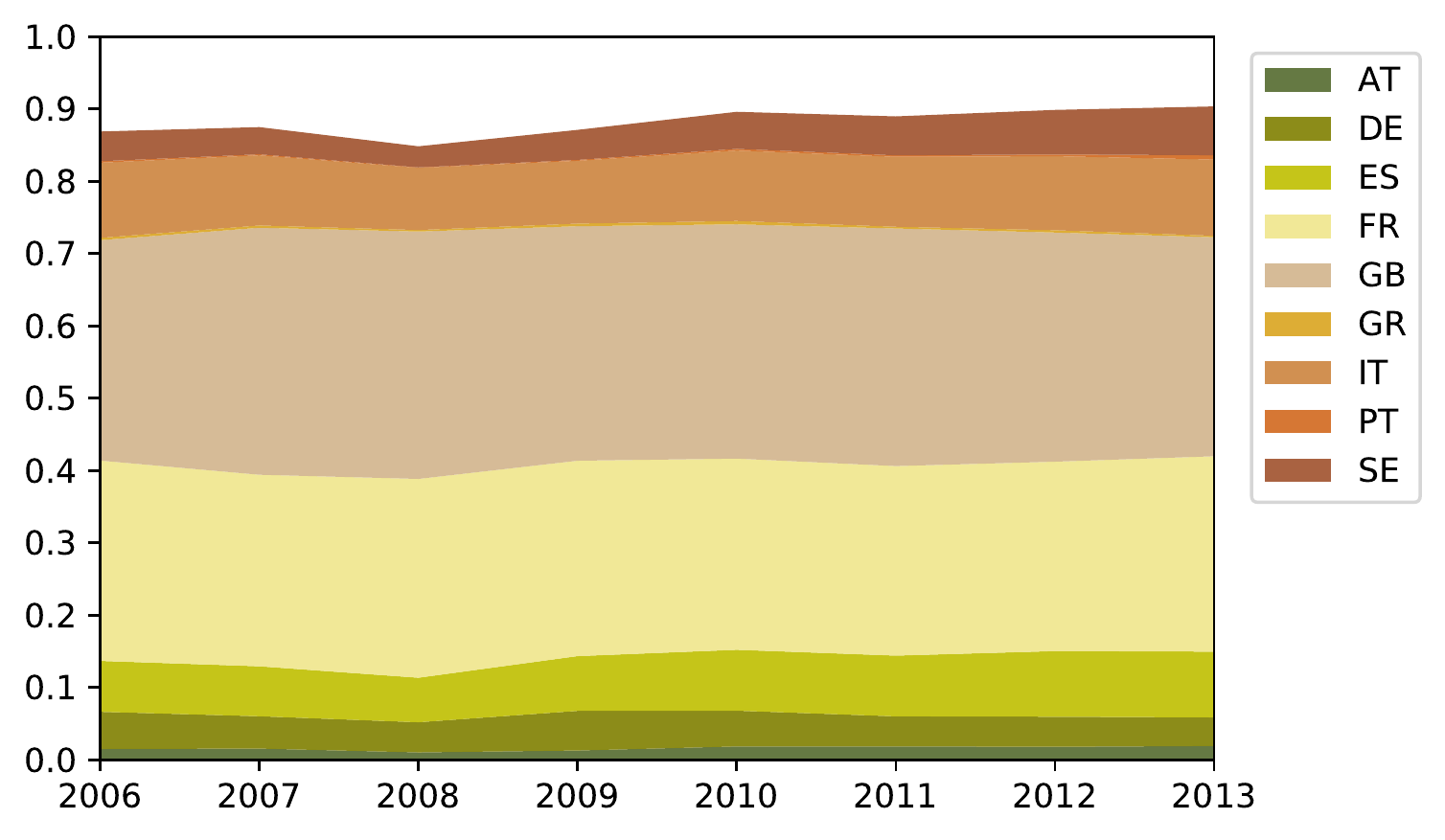}
		\caption{Fraction of assets owned by defaulted banks}
		\label{fig:decomposition_no_nodeterm2}
	\end{subfigure}
	\caption{Countries contribution of the time evolution of the systemic liquidity risk. Contagion rate without node-term}
	\label{fig:decomposition_no_nodeterm}
\end{figure}

To better understand the systemic risk during the years, we plot the bankruptcy ratio dynamics. The ratio is computed as the fraction of bankrupted banks at the end of the infection or as the fraction of total assets owned by bankrupted banks at the end of the infection. It is possible to notice from figure \ref{trendnonodo} that there are differences among years although not as one would expect. If we turn our attention to the assets owned by bankrupted banks, it tends to be around the 90\% of the total market. It means that on average bigger banks are failing more than the smaller ones. This is also due to their higher connectivity (higher in- and out-degree); they are more prone to contagion because they are involved in several contagion paths. It is also true that bigger banks have most of the interbank exposure so their bilateral exposures are also large but the rate of contagion is proportional to the relative exposure, not the absolute one. 
Figure \ref{fig:decomposition_no_nodeterm1} represents the decomposition of the bankruptcy ratio with respect to the country in which banks are based. The graph shows that France, Italy and Great Britain are the ones which contribute the most to systemic risk. It is important to notice that even if some countries are more represented than others, it does not directly reflect in having more banks defaulted. In fact, GB has less than half of the Italian banks and almost 50\% banks less than Germany but its contribution is comparable to the former and higher than the latter. This is because GB banks are more interconnected with foreign banks with respect to other countries' banks. Moreover, figure \ref{fig:decomposition_no_nodeterm2} depicts the decomposition of the asset owned by defaulted banks for each country. In this case, the GB is the main contributor, even if it is not the biggest country in terms of assets and number of banks. This is due to two main factors, i.e. the average bank is big and central in the interbank market. Furthermore, France, even if it has a huge fraction of assets in defaulted banks, has an average figure per each bank smaller than other countries which are less represented in number but with a consistent fraction of assets owned by bankrupted banks, e.g. Spain and Sweden. This result suggests that in some countries big banks are the ones which default more because of their centrality while in other countries the default concerns both central and peripheral banks irrespective of the size. 

\begin{figure}[h!]
	\centering
	\includegraphics[width=0.49\linewidth]{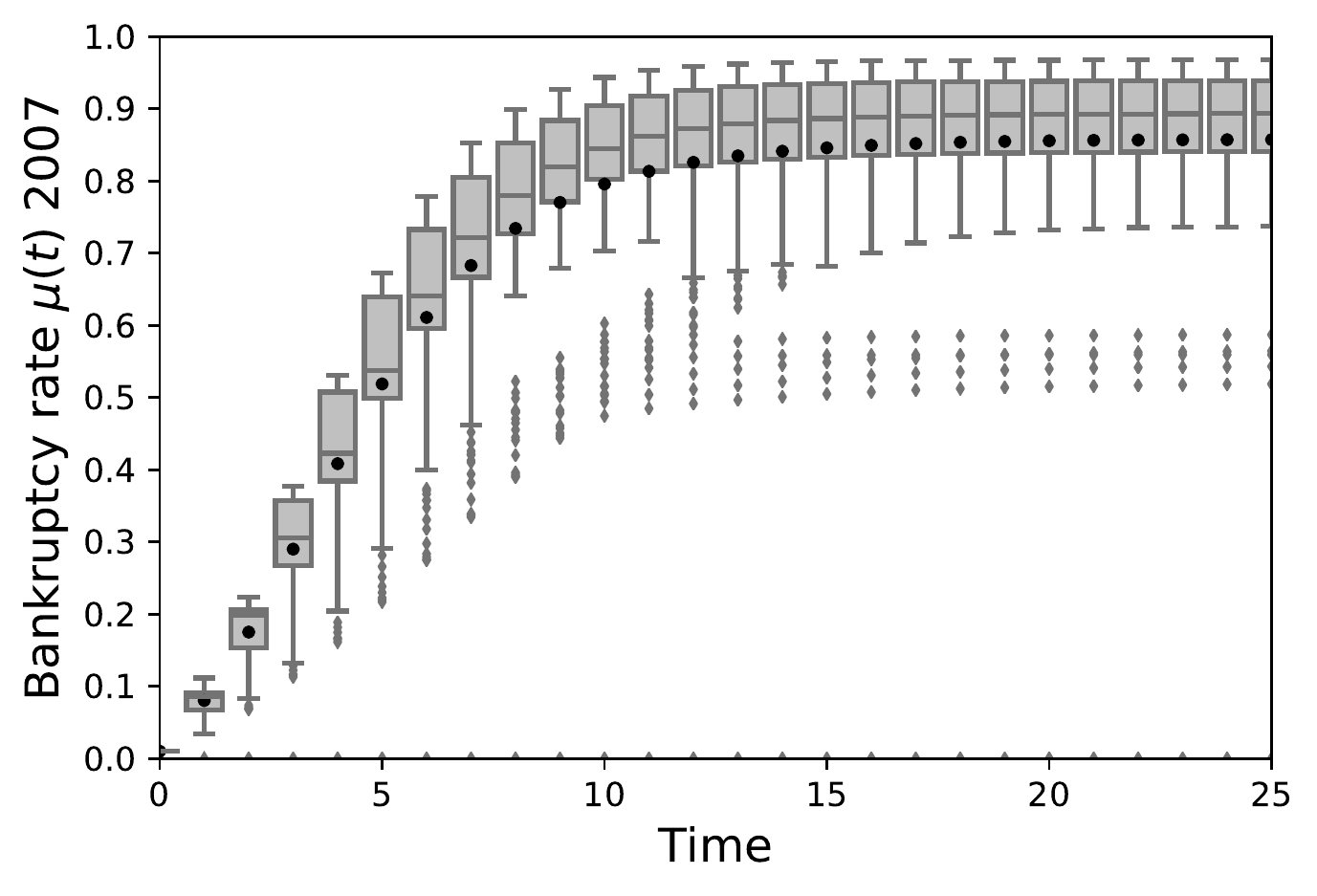}
	\includegraphics[width=0.49\linewidth]{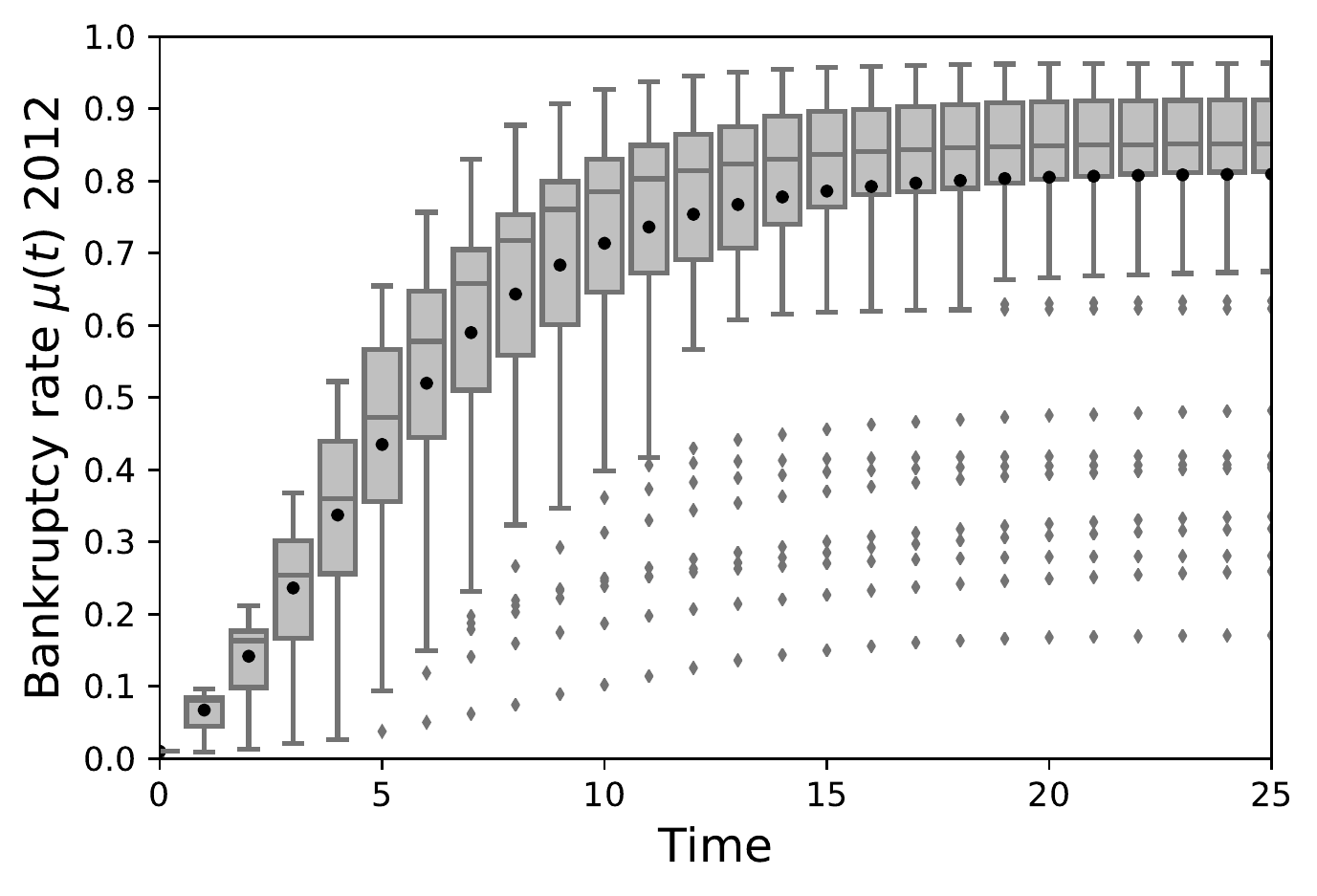}
	\caption{Dynamics of the distribution of the bankruptcy rate $\mu$ (eq.\ref{eq:bankrupt rate}) in 2007 and 2012. Black dots are the mean over the banks. Contagion rate without node-term}
	\label{mudinamicanonodo}
\end{figure}

Figure \ref{mudinamicanonodo} shows the dynamics of the bankruptcy rate $\mu$ (eq.\ref{eq:bankrupt rate}). The slope, the convergence and the mean value change from 2007 and 2012, showing that the contagion dynamic is more severe in 2012. In fact, this rate $\mu$ can be interpreted as a measure of the system health. Banks' bankruptcy rate depends on the real-time state of the banks since it is defined as the fraction of liquidity that a bank needs that were previously lent by infectious banks. The higher the value the greater is the number of the infected lender of a bank and the probability to go bankrupt. Individually, it is a measure of banks' fragility, considering instead the overall dynamic it gives information on the resilience of the whole system.

\subsubsection{Contagion with node-term}
\begin{figure}[h!]
	\centering
	\includegraphics[width=0.49\linewidth]{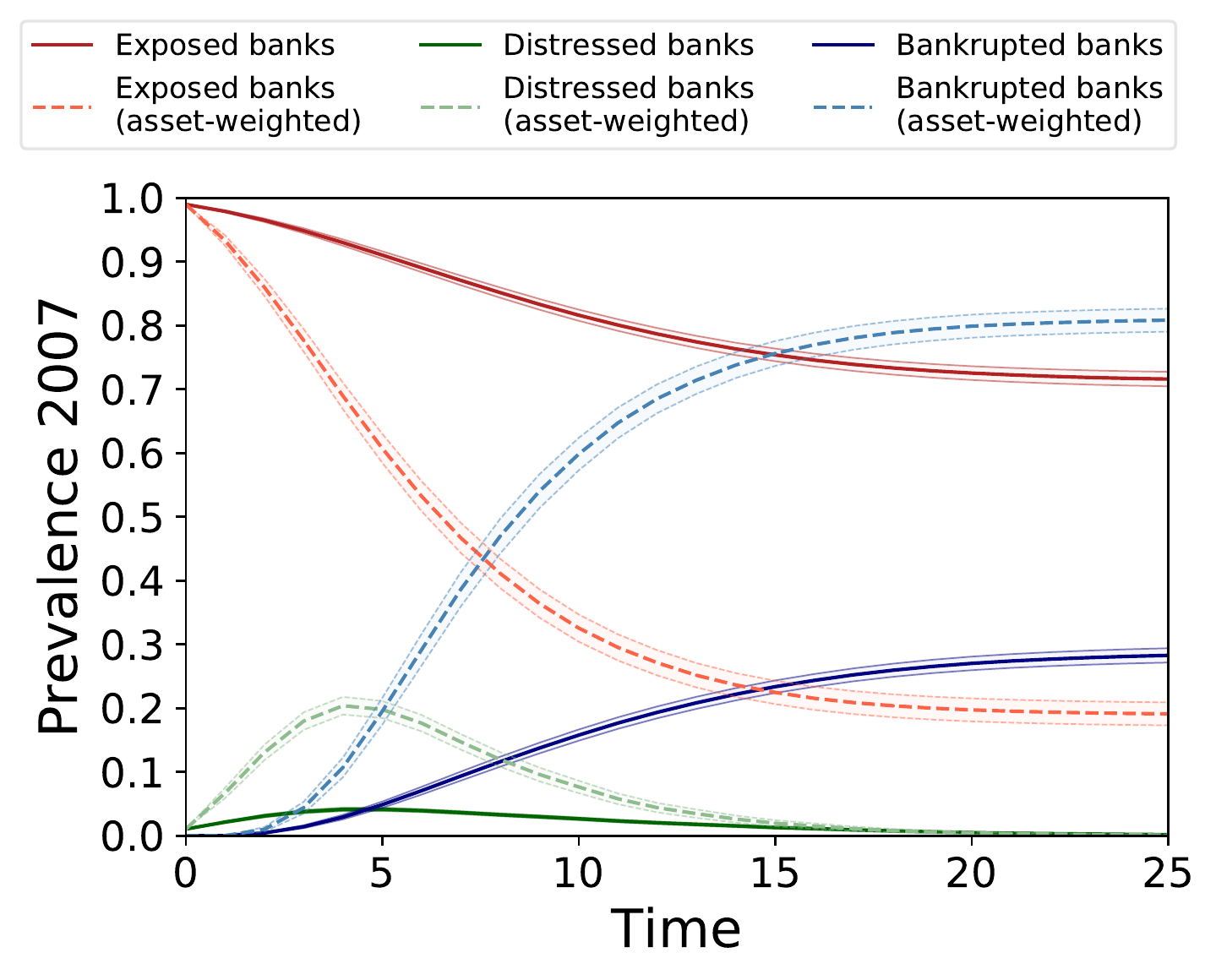}
	\includegraphics[width=0.49\linewidth]{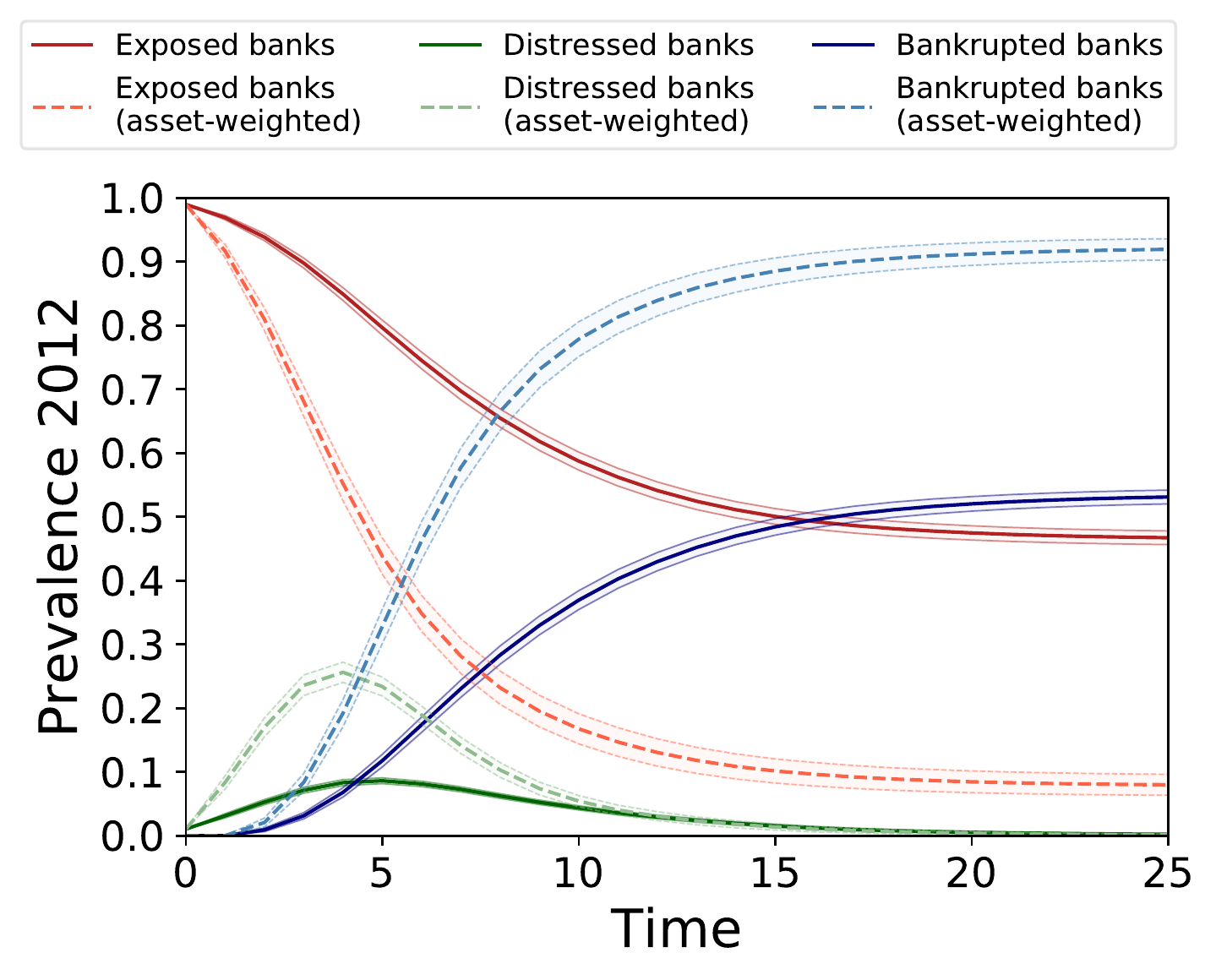}
	\caption{Prevalence dynamics in the 2007 and 2012 for the different bank states. Mean over all the simulations. Contagion rate with node-term}
	\label{prevalence}
\end{figure}

In this scenario, we consider the node-term in the functional form of the contagion rate $\lambda$ (eq. \ref{lambdafinale}).\footnote{All the variable used in the model are transformed using eq. \ref{mapping} over the whole dataset in order to maintain the dynamic of each variable. One could also use the transformation year by year, but this would destroy the dynamic of such variables over time.} This is because we want to take into account banks' contagiousness heterogeneity using balance sheet and country information which account for liquidity risk. Figure \ref{prevalence} shows the prevalence and the weighted prevalence dynamics in 2007 and 2012. As can be noticed, the node-term changed the dynamic of the prevalence, and in particular, the systemic risk in 2012 is much higher and this can be easily noticed by the fact that the fraction of bankrupted banks is higher than the one of healthy banks. 

\begin{figure}[h!]
	\centering
	\includegraphics[width=0.7\linewidth]{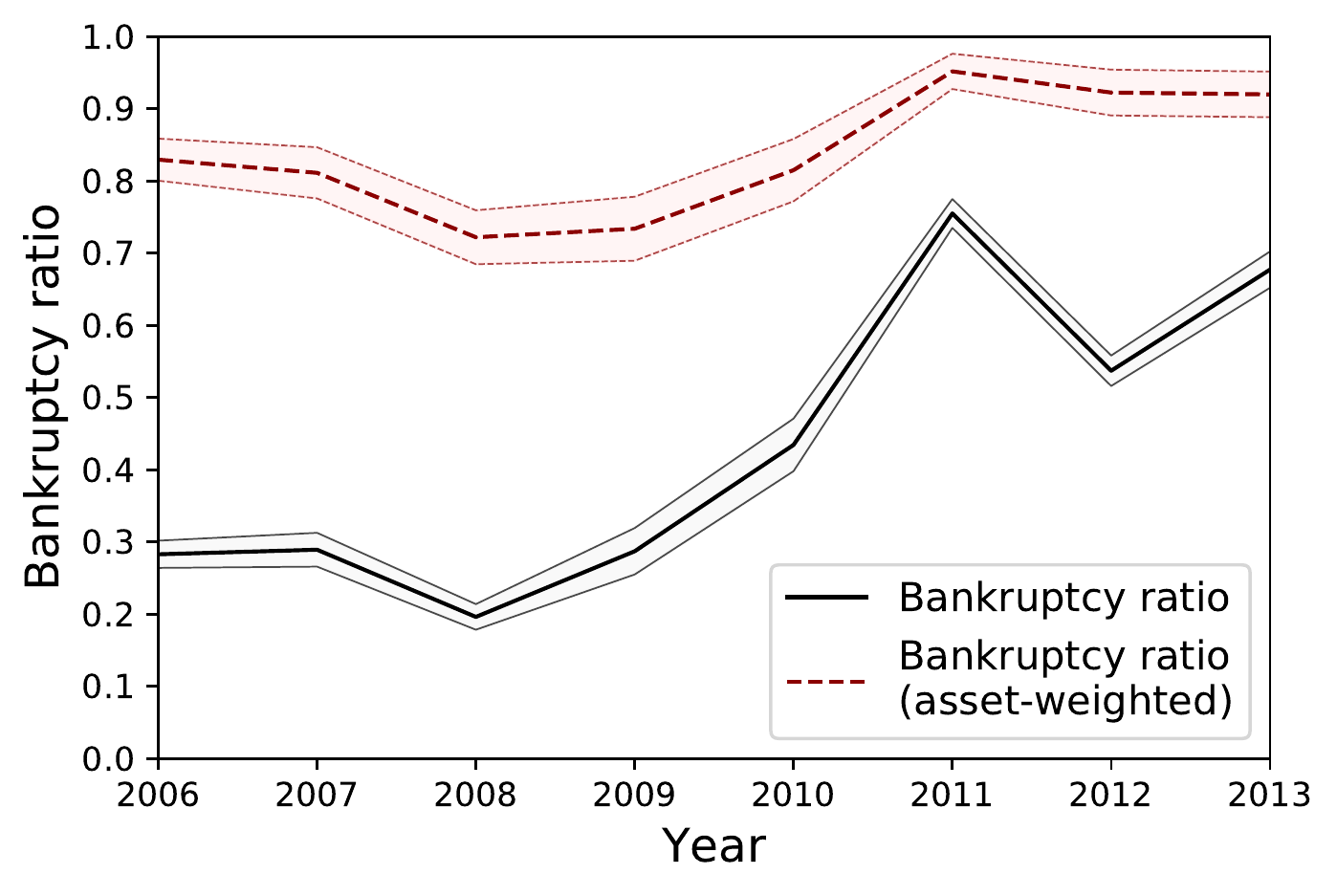}
	\caption{Time evolution of the bankruptcy ratio over the period 2006-2013. Contagion rate with node-term. Mean over the all simulations with 95\% confidence bands}
	\label{trend}
\end{figure}

If we shift the attention to the full dynamics of the bankruptcy ratio, we can notice from figure \ref{trend} that there are significant differences between years and we can observe a clear and evident dynamic on the bankruptcy rate over the years. Moreover, figure \ref{trend} shows as the liquidity systemic risk, i.e. the bankruptcy rate, increases not so much during the global financial crisis but rather during the European sovereign debt crisis. To understand the role played by the node-term in the contagion dynamics we compare the bankruptcy ratio dynamics with node-term in the infection rate $\lambda$ (Fig. \ref{trend}) with the benchmark scenario (Fig. \ref{trendnonodo}). In both cases, we observe an increasing systemic liquidity risk starting from 2008, although the more pronounced when we consider nodes' features. This result strongly suggests that the distribution of the node-term, starting from 2008, increasingly moved upwards to positive values consequently increasing the contagion rate, especially for highly connected banks. As shown, the dynamics of the bankruptcy ratio has a break in 2012, probably due to the ECB forward guidance in 2012 to counteract the European sovereign debt crisis. The contribution of the contagion rate's components is not equivalent. The country term, having strong fluctuations over the years, drives the dynamic of the liquidity risk, while the bank component has an elastic mechanism, magnifying or reducing the liquidity risk dynamics.

\begin{figure}[h!]
	\centering
	\begin{subfigure}{0.49\textwidth}
		\includegraphics[width=1\linewidth]{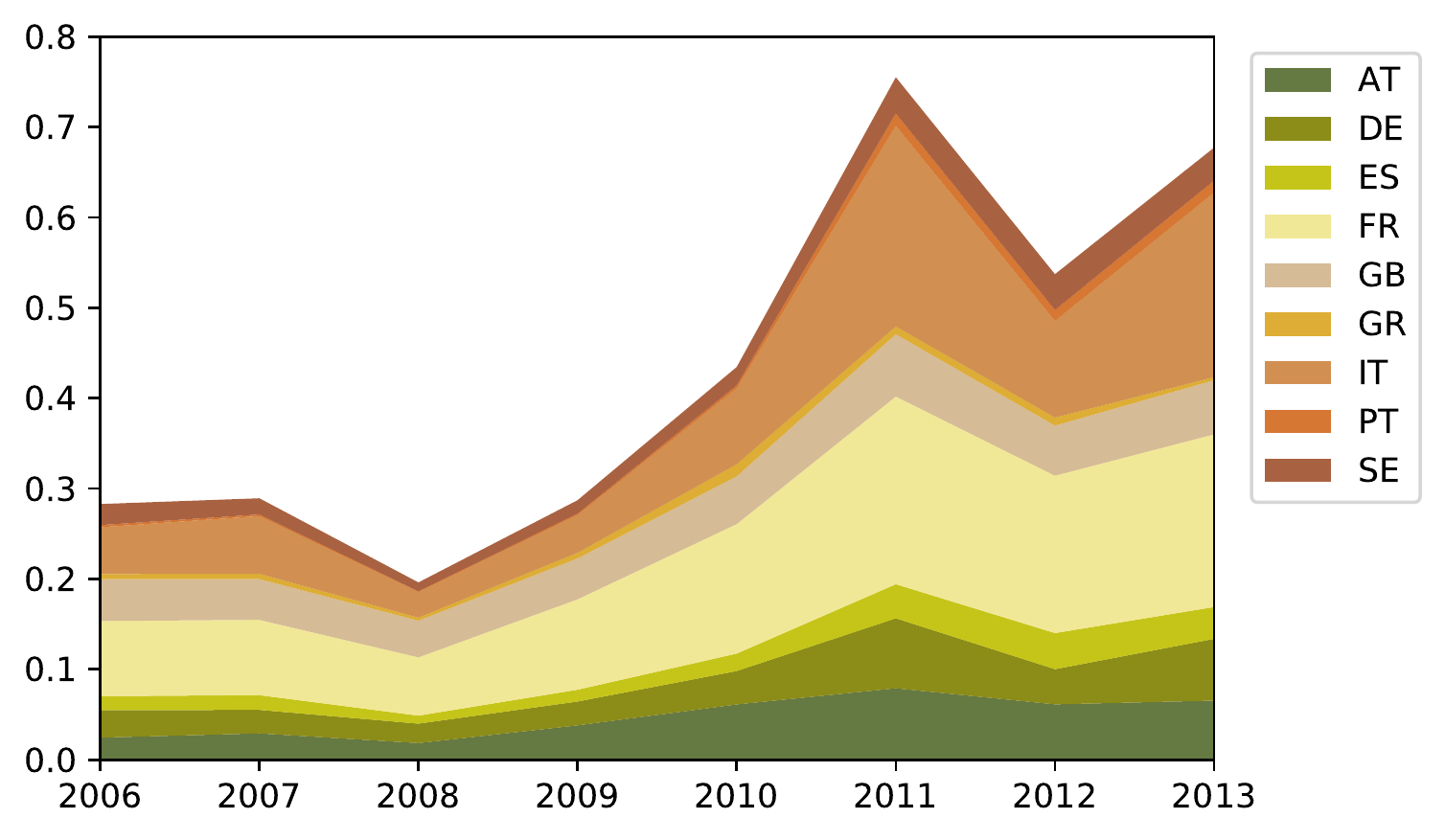}
		\caption{Fraction of defaulted banks}
		\label{decomposition_nodeterm1}
	\end{subfigure}
	\begin{subfigure}{0.49\textwidth}
		\includegraphics[width=1\linewidth]{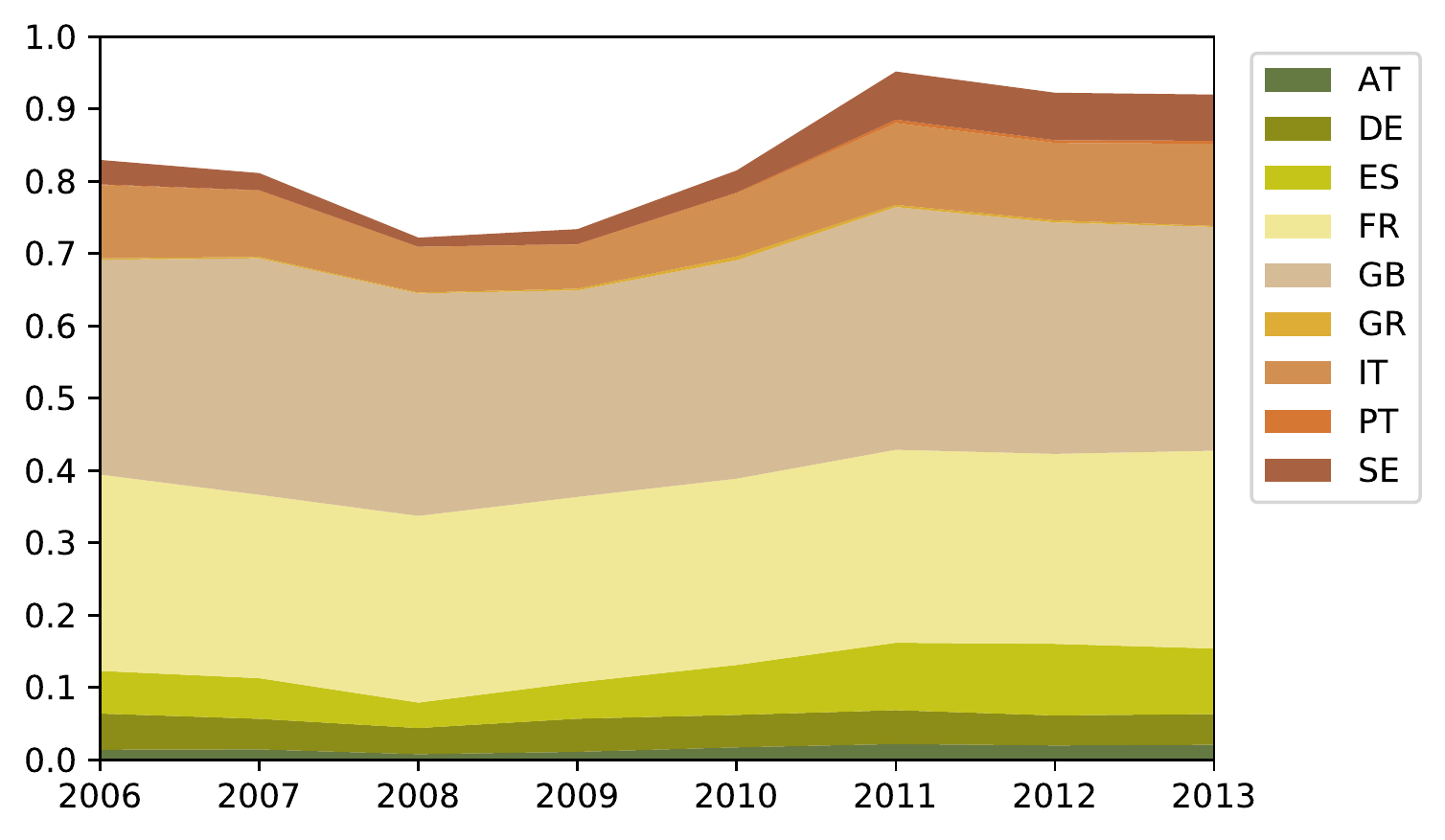}
		\caption{Fraction of assets owned by defaulted banks}
		\label{decomposition_nodeterm2}
	\end{subfigure}
	\caption{Countries contribution of the time evolution of the systemic liquidity risk. Contagion rate with node-term}
	\label{decomposition_nodeterm}
\end{figure}

Going deeper and watching at the systemic liquidity risk contribution by country of figure \ref{decomposition_nodeterm1}, we can see that Italian banks are the one with the highest contribution to the potential systemic contagion. This is due both to the fact that banks operating in this country have a high liquidity risk and a lot of interconnections, both with domestic and foreign banks. In this case, contrary to the benchmark model, we can see that the heterogeneity in liquidity risk incorporated in the node-term, reduced the magnitude of GB banks. This is due to the fact that GB banks are both resilient and with low liquidity risk. Also, France bears a huge share of the systemic risk proxied by the bankruptcy ratio, and this is mostly due to the huge amount of links with risky banks. If we analyse the average risk per-bank contribution in each country (country risk contribution divided by the number of banks in the country), we can notice that Spain and Sweden play a huge role, bearing a non-negligible fraction of systemic risk even if the number of banks is quite small. A similar argument to the benchmark model applies to the asset at default decomposed by country. In fact, figure \ref{decomposition_nodeterm2} shows that GB is the country with the highest fraction of assets at default. This is because, even if not all the GB banks default, the ones which do default, are the biggest ones. The contrary argument applies to France, where even if more banks default, they are of smaller size on average. This highlights that peripheral countries, even with a marginal number of banks in default, can generate a non-negligible impact in both the number of banks and the amount of assets in default. 

\begin{figure}[h!]
	\centering
	\includegraphics[width=0.46\linewidth]{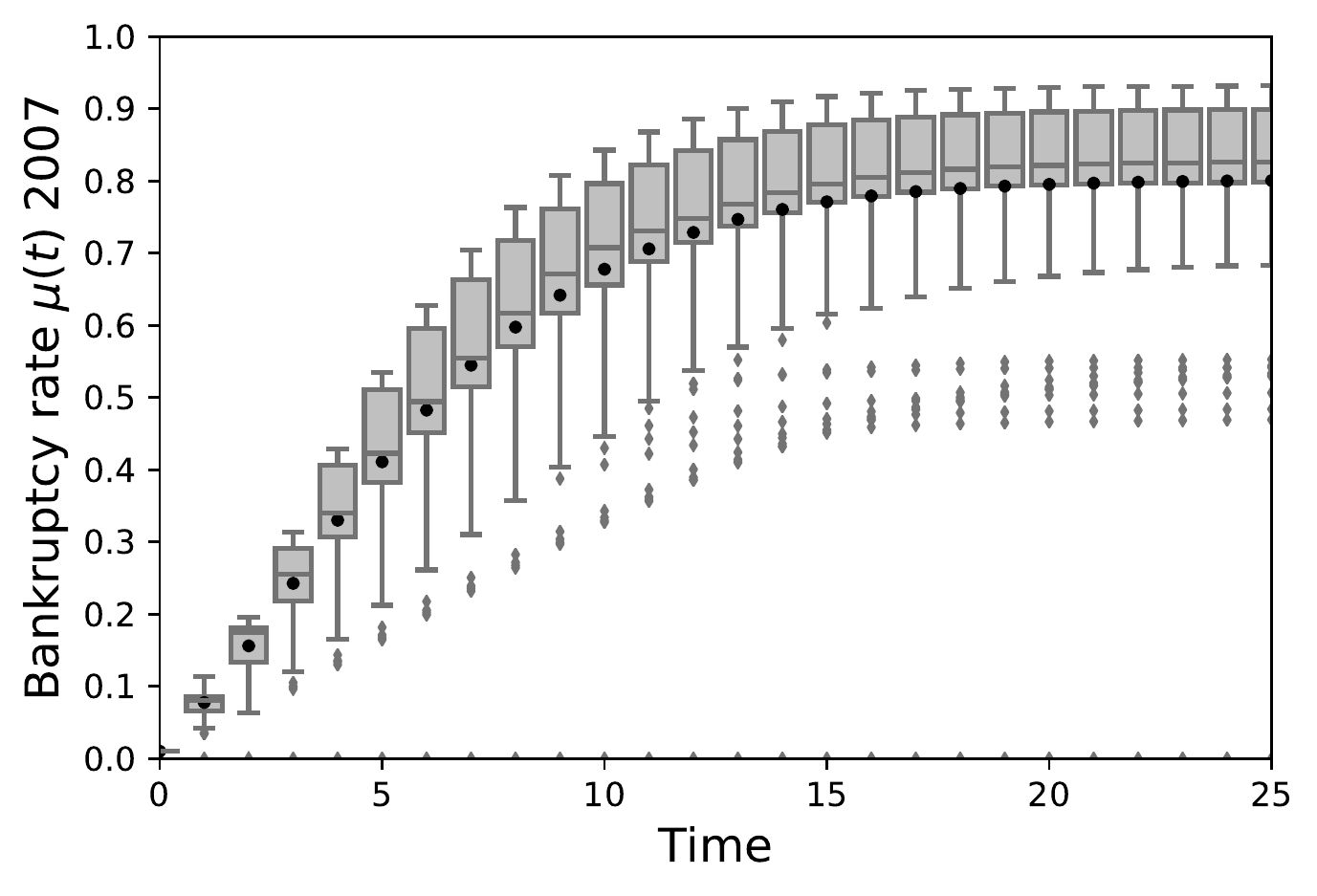}
	\includegraphics[width=0.46\linewidth]{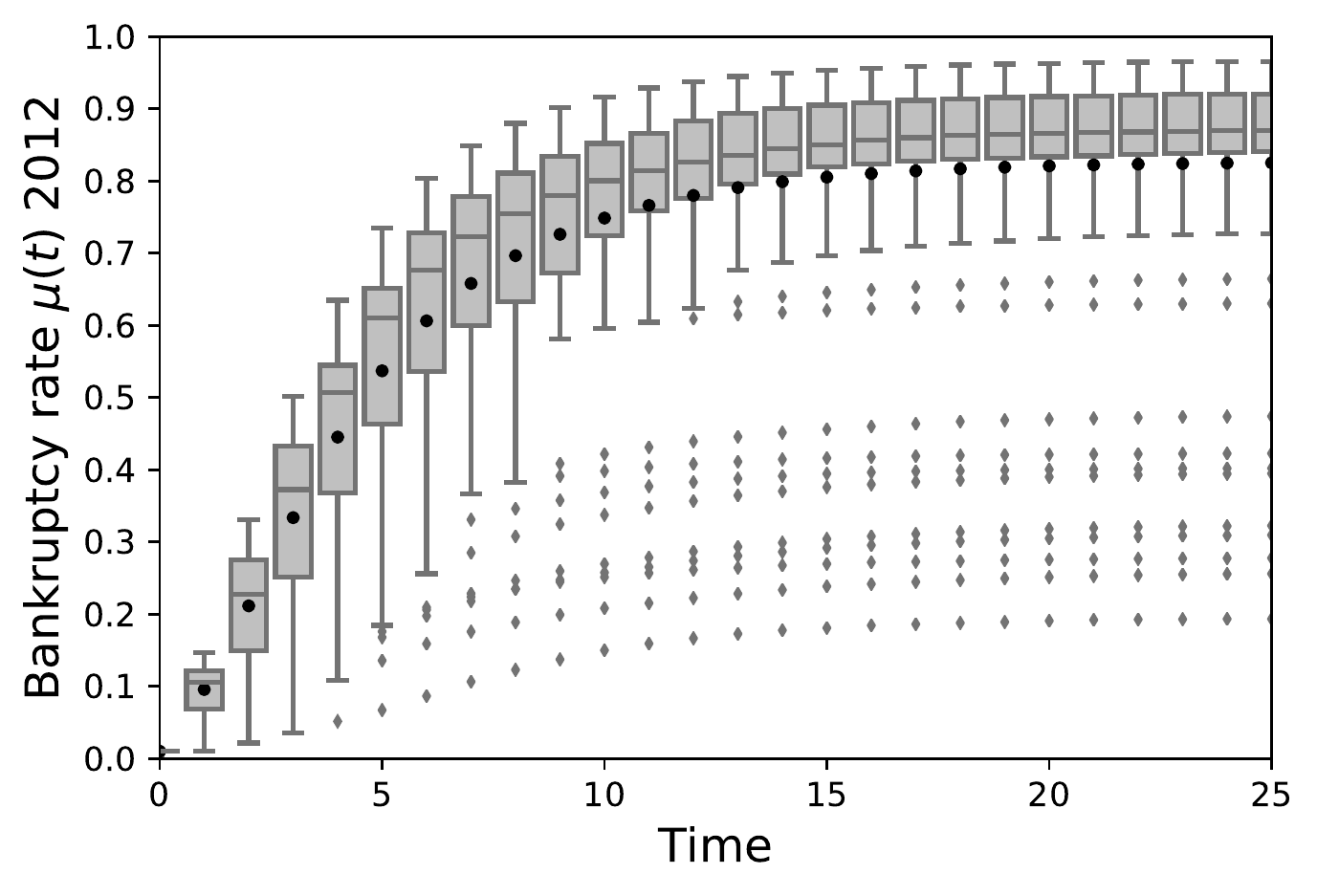}
	\caption{Dynamics of the distribution of the bankruptcy rate $\mu$ (eq.\ref{eq:bankrupt rate}) in 2007 and 2012. Black dots are the mean over the banks. Contagion with node-term}
	\label{mudinamica}
\end{figure}

Figure \ref{mudinamica} shows the dynamics of the bankruptcy rate $\mu$ (eq.\ref{eq:bankrupt rate}). While in 2007 the time evolution of the bankruptcy rate is similar to the benchmark case without node-term (Fig. \ref{mudinamicanonodo}), if we compare the dynamics of 2012 we observe that the node-term speed up the convergence to higher values. This means that most of the banks or at least a huge fraction of central banks are moving toward a riskier domain. This makes evident that the node terms enriches considerably the model, resulting in what we would expect during a banking financial crisis. This result suggests that both topology and country-bank features matters for systemic liquidity risk. 

\subsubsection{Contagion and bankruptcy rates with node-term}
\begin{figure}[h!]
	\centering
	\includegraphics[width=0.49\linewidth]{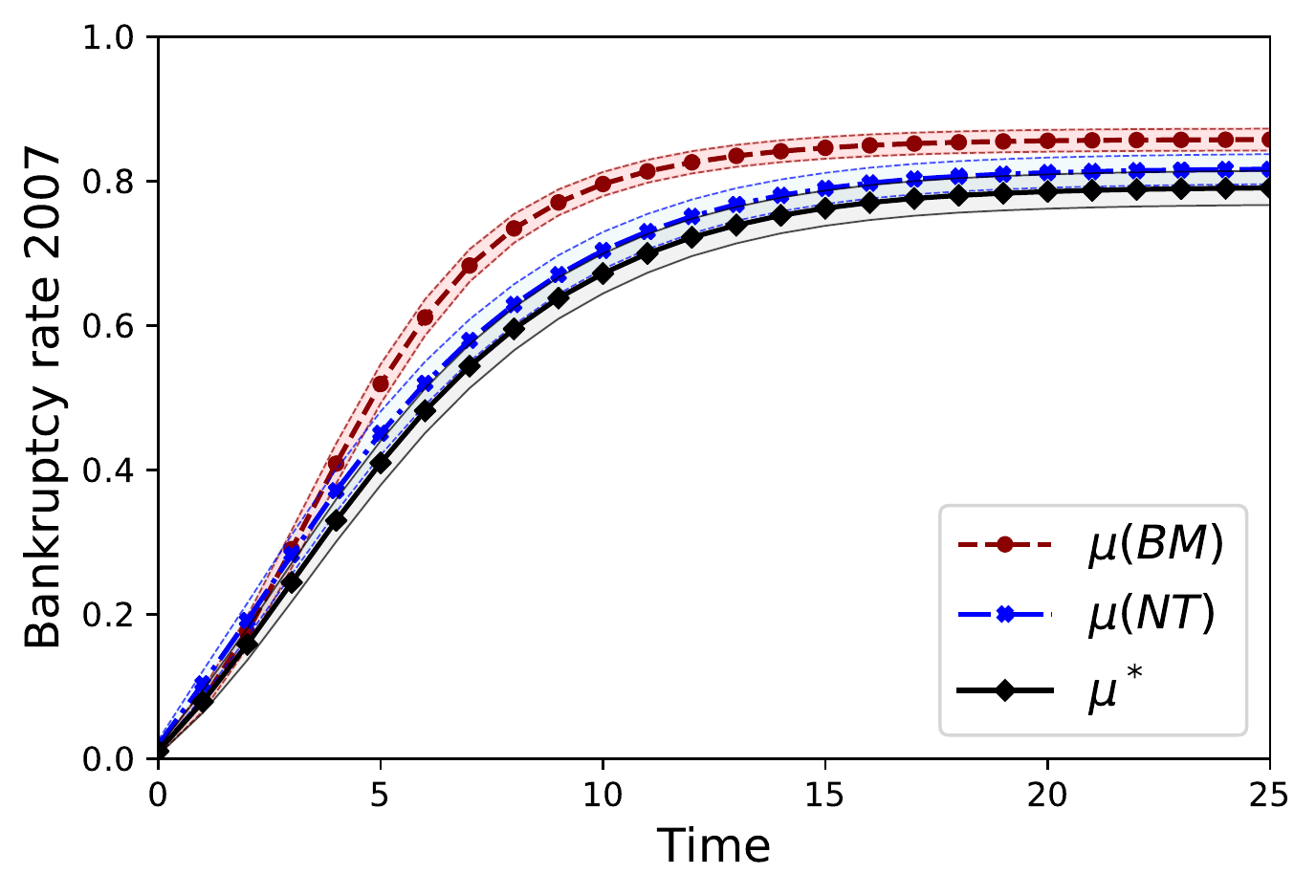}
	\includegraphics[width=0.49\linewidth]{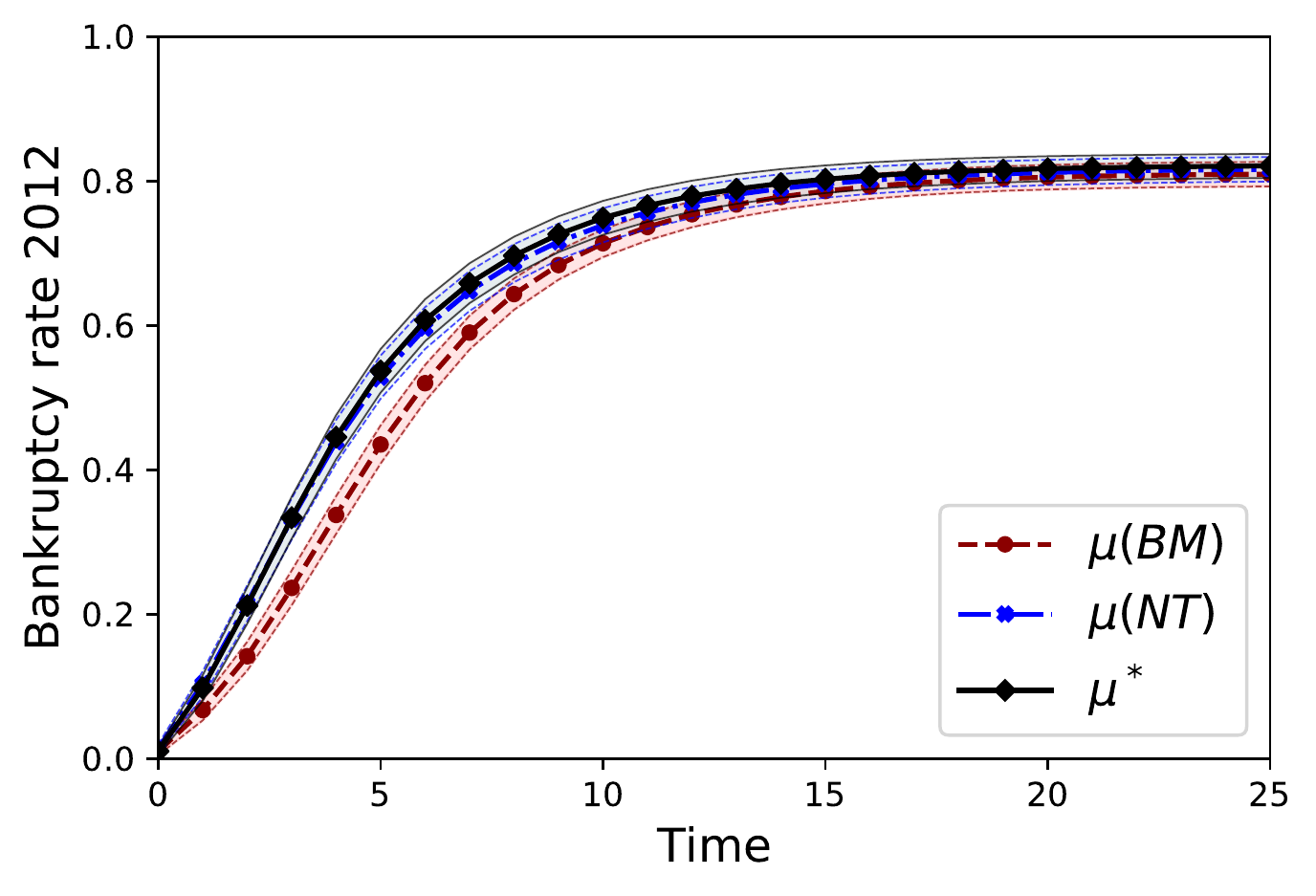}
	\caption{Dynamics of the average bankruptcy rate $\mu$. The dotted line $\mu(BM)$ refers to the benchmark model without node-term, the line-dot line $\mu(NT)$ refers to the model with node-term in the contagion rate while the continuous line is refers to $\mu^*$, the model with liquidity resilience in the bankruptcy rate. Mean and 95\% confidence interval for 2007 and 2012}
	\label{murossoeblustep101}
\end{figure}

In this scenario, we consider the liquidity resilience indicator in the functional form of the bankruptcy rate $\mu$ (eq. \ref{dopomuu}). In this way, the spontaneous transition from the infected $I_1$ to the bankrupted state $I_2$ is also dependent on nodes' feature as banks' interbank liquidity resilience (eq. \ref{liqindicator}). For long enough simulations, the systemic risk would result to be equal since without a recovery mechanism, distressed banks are set to remain in the distress compartment or to default and this node-term would only accelerate or decelerate the bankruptcies. For this reason, showing the systemic liquidity risk would not carry any additional information. What is important though, is not only the final state of the simulation but also its speed to convergence. In fact, if a policymaker is able to monitor a distress situation, it would intervene timely. For this reason, it is of crucial importance to understand if liquidity resilience of some banks (in topologically relevant positions) can indeed slow down the contagion process. If it is the case, a central bank would operate on banks which can help to stop the contagion rather than injecting liquidity directly to more central or more vulnerable banks. To see if we have any effect by the liquidity resilience indicator, we compare the dynamics of the average bankruptcy rate $\mu$ for different model specification. In figure \ref{murossoeblustep101} we represent the different dynamics of the bankruptcy rate over time when the liquidity resilience indicator is considered or not. For the case in which we don't use nodes specific features in the bankruptcy rate, we plot both the benchmark model (BM) and the model with node-term in the contagion rate (NT). As it is possible to notice, in 2007 the three plot are statistically different in mean at 5\% confidence level. In fact, the mean values do not intersect with the 95\% confidence bands of the other specification. In 2012 however, apart from the initial steps of the contagion in which the benchmark model has a statistically lower speed in the bankruptcy rate compared to the other two, the three lines cross, meaning that they are not statistically different at 5\% confidence level. From an economic viewpoint, the 2007 case is more interesting. As it is possible to notice, the introduction of the liquidity resilience indicator lowered the speed of contagion. Even if the difference between the two curves is not huge in absolute terms, it can make the difference when preventing an additional bank to fail that can generate a domino effect is vital.

\subsubsection{Contagion and bankruptcy rates with node-term and dynamic confidence contagion}
\begin{figure}[h!]
	\centering
	\includegraphics[width=0.49\linewidth]{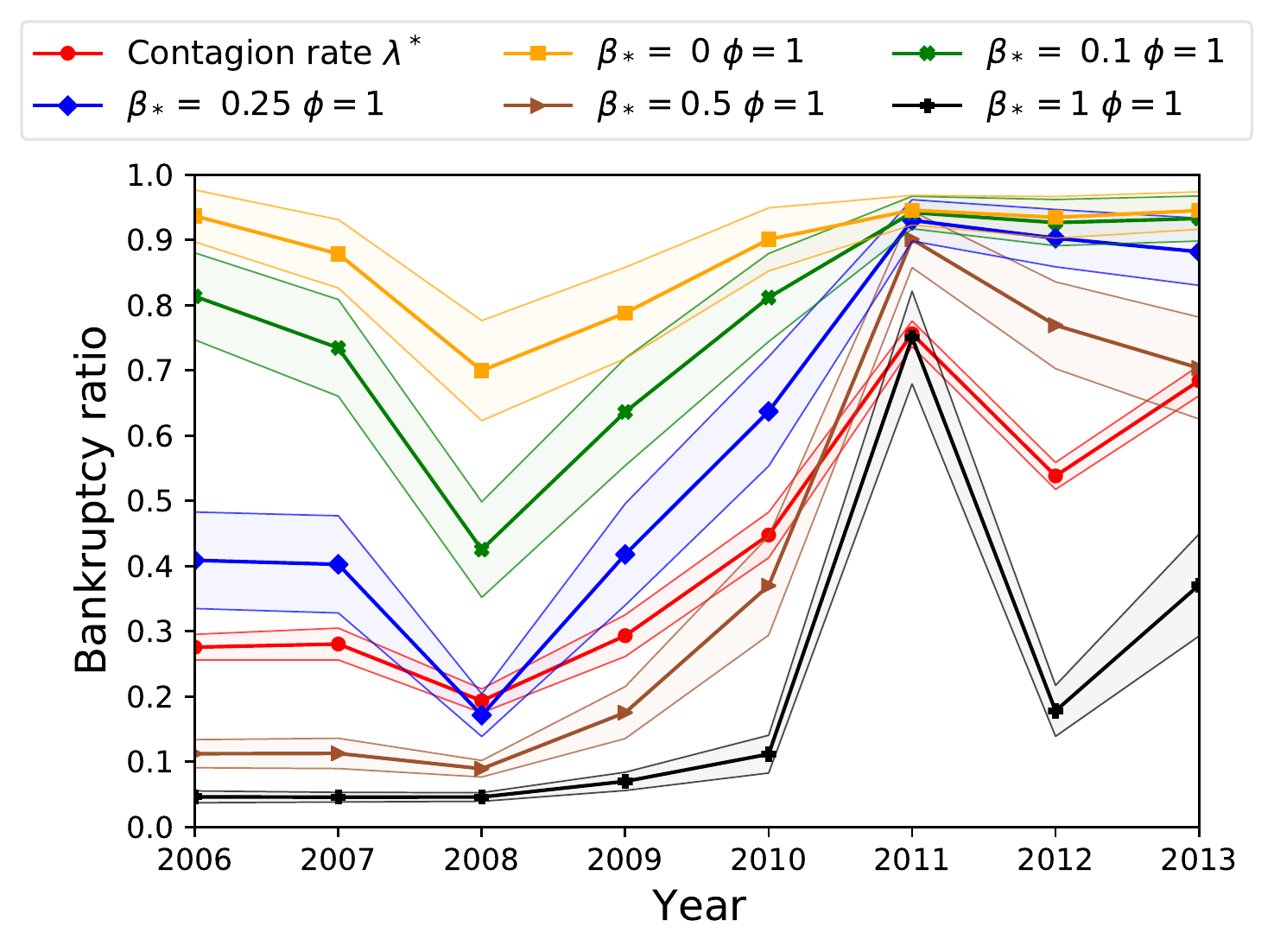}
	\includegraphics[width=0.49\linewidth]{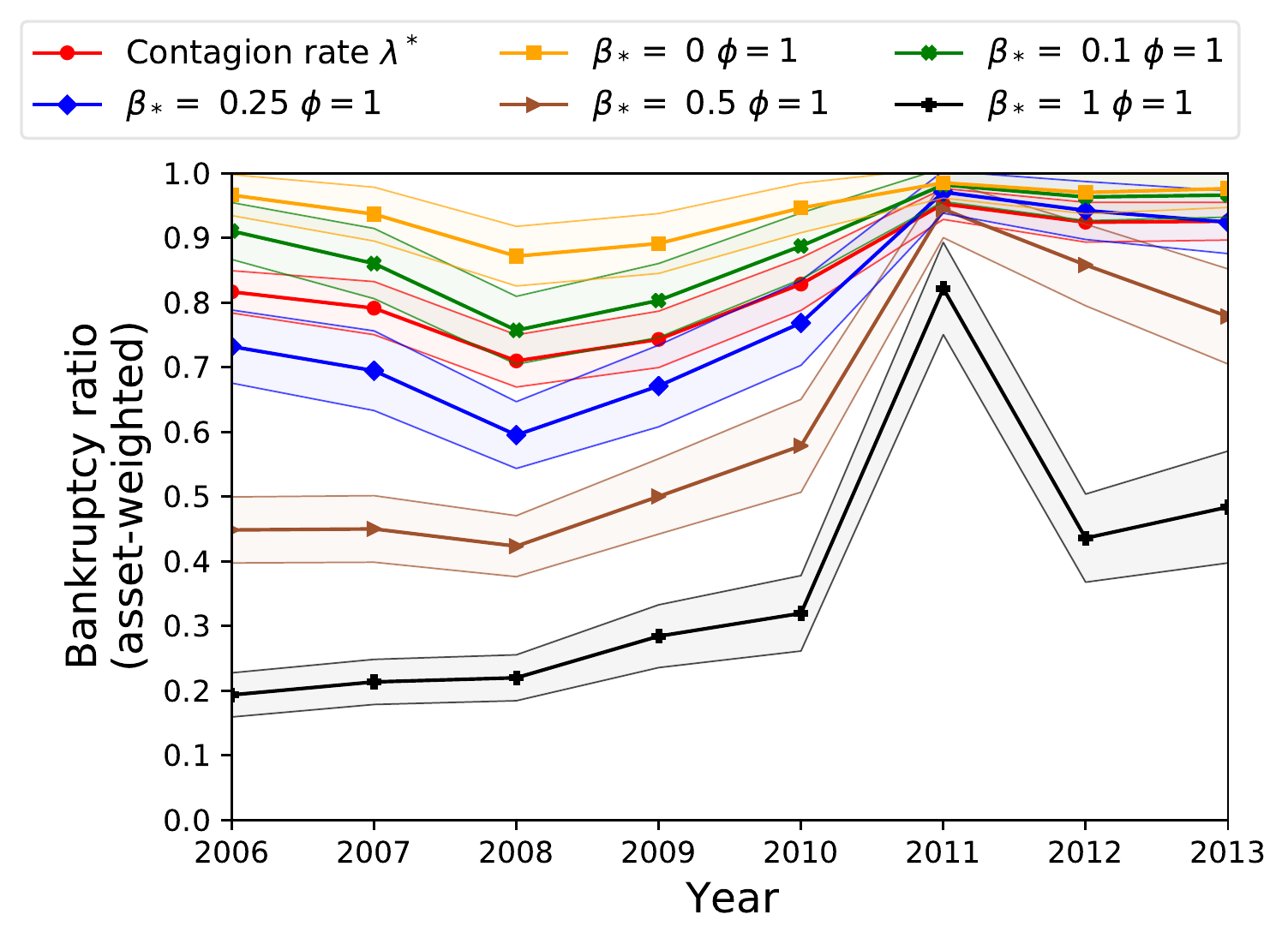}
	\caption{Time evolution of the bankruptcy ratio over the period 2006-2013. Contagion rate with confidence risk dynamics and node-term. Bankruptcy rate with liquidity risk multiplier. Mean over the all simulations with 95\% confidence bands. On the left, the fraction of bankrupted banks at the end of the infection while on the right, the fraction of total assets owned by bankrupted banks at the end of the infection}
	\label{trendstep2}
\end{figure}

In this scenario, we consider a richer model in which we have the node-term in both the contagion and the bankruptcy rate parameters and, in addition, the contagion rate has a dynamic confidence multiplier which makes the contagion rate intrinsically dynamic. In particular, the liquidity resilience indicator enters in the functional form of the bankruptcy rate $\mu^{*}$ (eq. \ref{dopomuu}) while the systemic risk multiplier  $\theta$ enters the dynamic of the contagion rate $\lambda^{+}$ as reported in eq. \ref{lambdadinamicoo2}. The systemic risk multiplier, as described in eq. \ref{labeltheta}, depicts the health of the system at each iteration and it proxies the negative feedback loop and confidence contagion in the interbank market. When a fraction of the banks is failed or in distress, the confidence contagion kicks in since the remaining banks are classified as risky banks. This, in turn, gives rise to a negative spiral effect which magnifies the risk of distress in the market and the freeze of liquidity exchanges. On the contrary, if the market is characterized by few distressed or bankrupted banks, the banks don't see the other banks as risky and continue lending money to them. In practice, there should be a phase transition to the contagion to escalate rapidly. In fact, given the definition of $\theta$ in eq. \ref{labeltheta2}, if more than $s_*=\frac{1}{1+\beta_*}$\footnote{The threshold $s_*$ is computed as the solution of the equation $\theta=(1+\beta_*)s_*=1$.} of the banks are healthy, the contagion rate $\lambda^+$ is slower than $\lambda^*$ while it rapidly increases if less than $s_*$ of the banks are healthy. This means that only if the contagion rate and the liquidity epidemics have enough power, the systemic liquidity risk fully materializes. As defined in eq. \ref{lambdadinamicoo2}, this "psychological" effect is responsible for the reduction or increase of the contagion rate and consequently the spread of the financial distress. Observing the bankruptcy ratio dynamics for different $\theta$ specifications shown in figure \ref{trendstep2}, we can make some comments.\footnote{We show the results for different values of $\beta_*$. Plots related to the non-linear specification are available in the appendix.} First of all, for every specification, a sharp surge in all the curves is observable in 2011 which represents the peak of the sovereign debt crisis. Unlike the previous dynamics (Fig. \ref{trend}), we can notice that the effect played by the parameter $\theta$ is quite important and can be categorized in three main effects: slow-reversal, fast-reversal and a mixed-effect. In fact, for a big value of $\beta_*$ as $1$, the effect is to slow the contagion at every point in time, reaching the same number of bankrupted banks only for 2011. On the contrary, for a very low value of $\beta_*$, the contagion sped up and many more banks default with respect to the $\theta=1$ case (no risk multiplier). However, even if the number of banks defaulted is considerably higher, these are small banks since the assets in default are statistically equivalent. The mixed-effect cases (with intermediate values of $\beta_*$) are the more interesting ones. In fact, these cases are the ones more credible both ex-ante from an economical point of view and ex-post from the results we get. We can notice that the confidence contagion kicks in only in some years and that this has a stronger effect on small (peripheral) banks since even in the years in which the number of banks defaulted is higher, this reflects in an equivalent or lower value of asset defaulted with respect to the $\theta=1$ case.   

\begin{figure}[h!]
	\centering
	\includegraphics[width=0.49\linewidth]{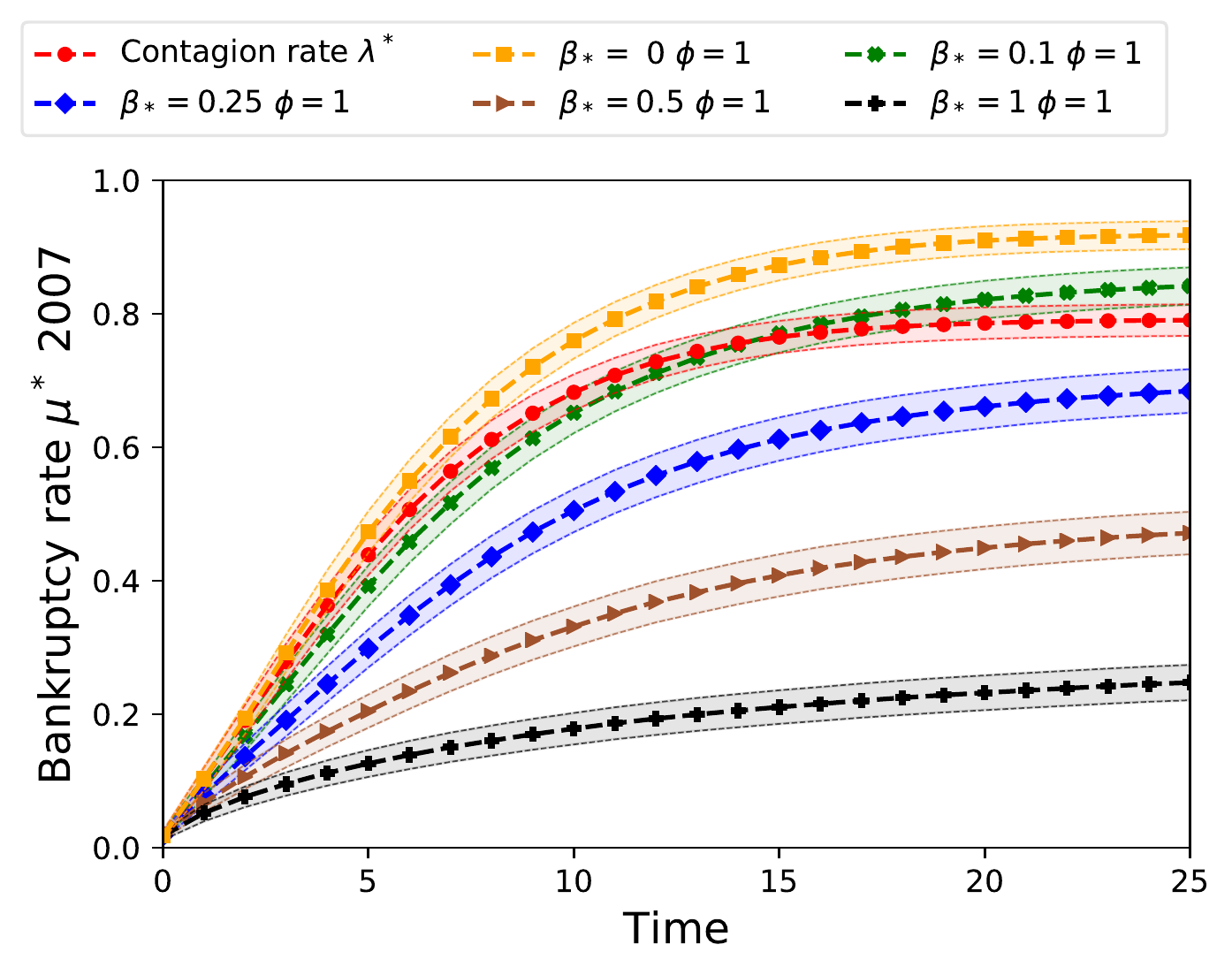}
	\includegraphics[width=0.49\linewidth]{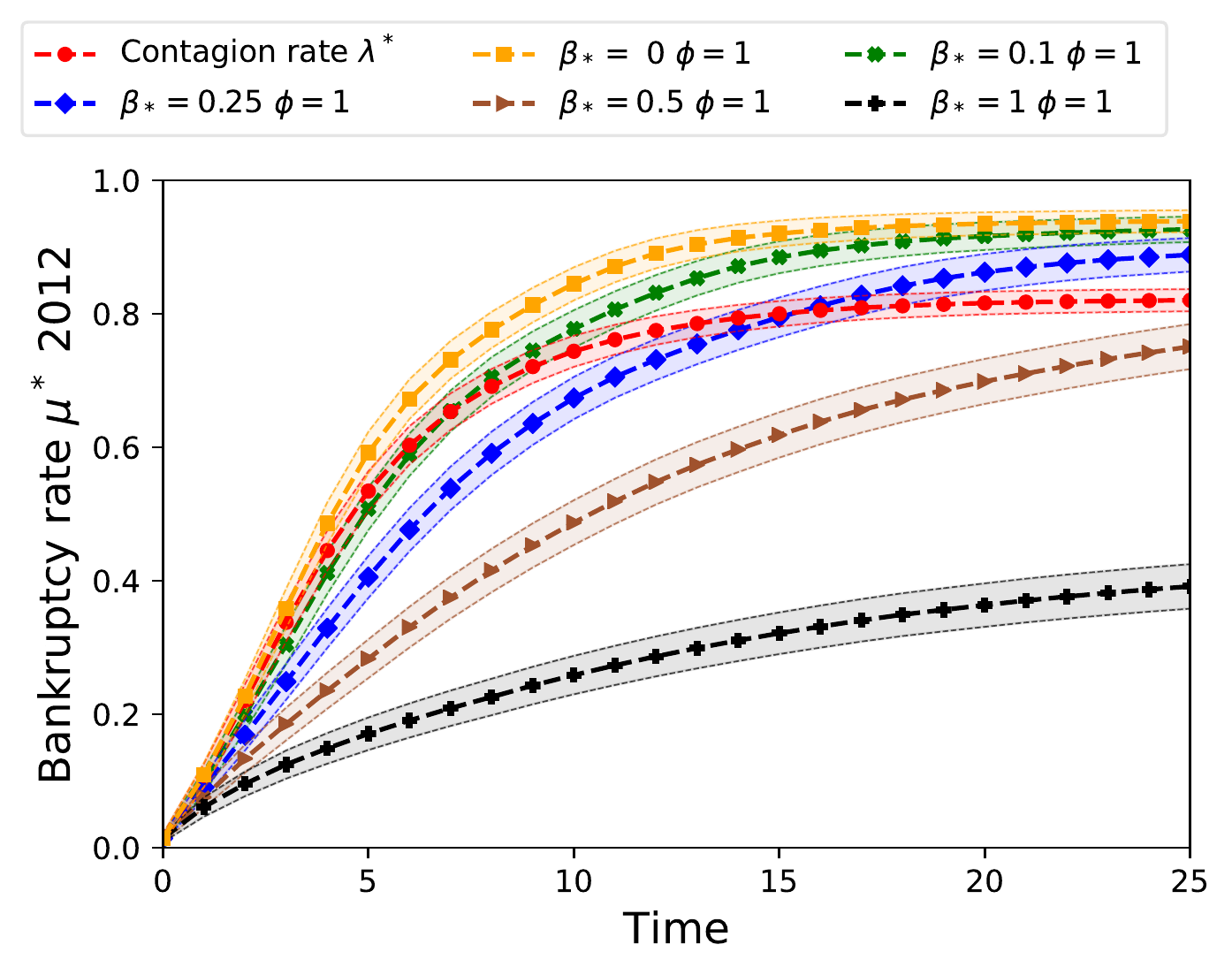}
	\caption{Dynamics of the distribution of the bankruptcy rate $\mu^*$ (eq. \ref{dopomuu}) in 2007 and 2012 for different $\theta$ specifications.  Mean over the all simulations with 95\% confidence bands}
	\label{muconnu}
\end{figure}

We finally show the speed of bankruptcy in this context for different $\theta$ specifications. As it is possible to deduce from the previous plot, if the contagion is lessened for one specific year, also the bankruptcy rate evolution would be lower and vice-versa the opposite. In fact, the bankruptcy rate $\mu^*$ grows as the number of distressed lenders increases, so we also observe a different dynamic of such parameter when the contagion rate is dynamic or static (Fig. \ref{muconnu}). In particular, in 2007 the bankruptcy rate is quite flat for the low-reversal specification while the fast-reversal specification is very similar to the confidence contagion neutral case ($\theta=1$). In this specific year, the mixed-effect specification has a lower speed of default rate and hence a lower or comparable number of banks defaulted with respect to the $\theta=1$ specification. Regarding 2012, the bankruptcy rate for the mixed-effect specifications is quite interesting. In fact, in the first part of the simulation they depict an overall slower default rate, while after some time steps, the opposite is true. This leads to at least $15\%$ more banks defaulted at the end of the contagion process. The behaviour of the contagion due to the psychological effect suggests that if tackled in time, the contagion could be lessened before it degenerates to a systemic collapse.  
This result highlights that an effective macroprudential policy could be to undertake actions in interbank hubs where the contagion and domino effects are more probable rather than inject money in the more central or vulnerable banks. These results make clear that if the psychological effect for bank run is not activated, the contagion spread very slowly and it gives the time to banks and financial institution to appropriately manage the crisis. Financial crisis can be easily generated from financial institutions which are not under the light of regulatory channels, but rather minor players in the financial market.
\section{Conclusions}\label{conclusions}

In this paper, we proposed a stochastic epidemic-like model for systemic (funding) liquidity risk in the European interbank market. We built the model enriching the Exposed-Distressed-Bankrupted (EDB) contagion model recently proposed in \citet{brandi2018epidemics} adding node-specific features related to liquidity risk. In particular, we used the bid-ask spread and a banking liquidity indicator to amplify (or contract) the contagion dynamic on the interbank network. We also augmented the model with a bank liquidity resilience indicator to take into account the heterogeneous probability of default of a bank hit by a liquidity shock. We finally introduced the confidence contagion via a systemic parameter which proxies the health of the banking system. We simulated the model using an agent-based simulation since the result is not achievable in closed form. We run the model on the European interbank market reconstructed with an economic enhanced method that uses external economic information to anchor the classic fitness-based reconstruction model. In particular, we used the BIS data on inter-countries banking exchanges to constraint the reconstruction model and we could appreciate a huge improvement with respect to the unconstrained version. The results of the contagion process are promising. We could appreciate that using node features in the contagion mechanism they enriched considerably its dynamic providing good information for systemic risk assessment. We could notice that both the market topology and the contagion dynamics play a non-trivial role. In fact, however, Greece has the highest (country) market liquidity indicator, their risk is attenuated on the final result because of a lower amount of interconnections with foreign banks with respect to other countries as Italy, which other than having a high liquidity risk, is strongly connected with foreign banks. These results show that if a specialized model cannot be used, this model is easy still flexible to reproduce different contagion dynamics.    
A limitation is represented by the limitedness of the data about banks' exposures. In fact, other features of the lender and the borrower could be considered, e.g. the volume of liquid and illiquid assets, but, as the first step, we consider only the interbank market as the possible channel of contagion. Such a simplification, as well as adopting an epidemiological model, in fact, allows to reduce the complexity of the spreading dynamics and it is possible to run the model when most of the information is missing, as in this case. Another comment regarding the reconstruction must be made. In fact, the results must be taken at face value since the external information, the BIS data, refers to Q4 2013, while a more robust result would have been achieved using the corresponding year of the BIS from 2006 to 2013. In fact, it is possible to argue that this information is not fully representative of the countries linkages in previous years, especially pre-crisis years. However, this is the only information we could rely on. If banks would not be anonymous, other information could be used to anchor the reconstruction method. 
A future step in the use of epidemic-like models for financial contagion is to introduce the credit risk channel in the funding liquidity risk dynamics as prescribed by the Bank of England \citep{kapadia2012liquidity}. This would require a multilayer network with spillover effects between layers devoted to liquidity and credit risk. The model can also be easily extended to more parameters to enter the node-term $\gamma$ and $\nu$ or to change the functional form of the contagion and bankruptcy rate to mimic more some specific financial mechanism. In conclusion, this modelling approach could be used when relying on more sophisticated models is not possible but using only topological measure would be too simplistic. 

\section*{Acknowledgements}
	The authors thank Marco Bardoscia for useful discussions.

%
%

\bibliographystyle{spbasic}      
\bibliography{bibTEMPLATE}   

%
%
\begin{appendices}
	\section{Generalization of the systemic risk multiplier formulation}
	We generalize the systemic risk multiplier formulation $\theta$ (eq. \ref{labeltheta}) as follows:
	\begin{equation}
		\theta^+(t)=h(\theta, \phi)
		\label{thetanl}
	\end{equation}
	where $h(\theta,\phi)$ is a function which increases or decreases the speed of the systemic risk multiplier. In this work, we adopt the following function:
	\begin{equation}
		h(x,\phi)=\frac{x^\phi}{max(x)^{\phi-1}}
		\label{h_equation}
	\end{equation}
	where $\phi$ is the speed parameter and $max(x)$ at the denominator is used to ensure that the boundaries are not changed by the transformation. It could be noticed that in the case of $\phi=1$, we restore as a particular case the previous formulation of $\theta$ in eq. \ref{labeltheta}. For example, if a policymaker doesn't think that the psychological behaviour over banks systemic risk follows a linear pattern in time but rather a quadratic one, this would imply a $\phi=2$. This is a free parameter that must be set by expert judgement or anchored to market expectations. The idea is to reproduce the increasing pressure on the interbank market when a fraction of the banks are infected or defaulted. By definition of the contagion rate $\lambda_{ij}^+$ (eq. \ref{lambdadinamicoo2}), when the systemic risk multiplier crosses the critical value $\theta^+(t^*)=1$, the produced effect on the propagation switches from deceleration to acceleration.
	\section{Results for non-linear systemic risk multiplier}
\begin{figure}[h!]	 
		\centering
		\includegraphics[width=0.49\linewidth]{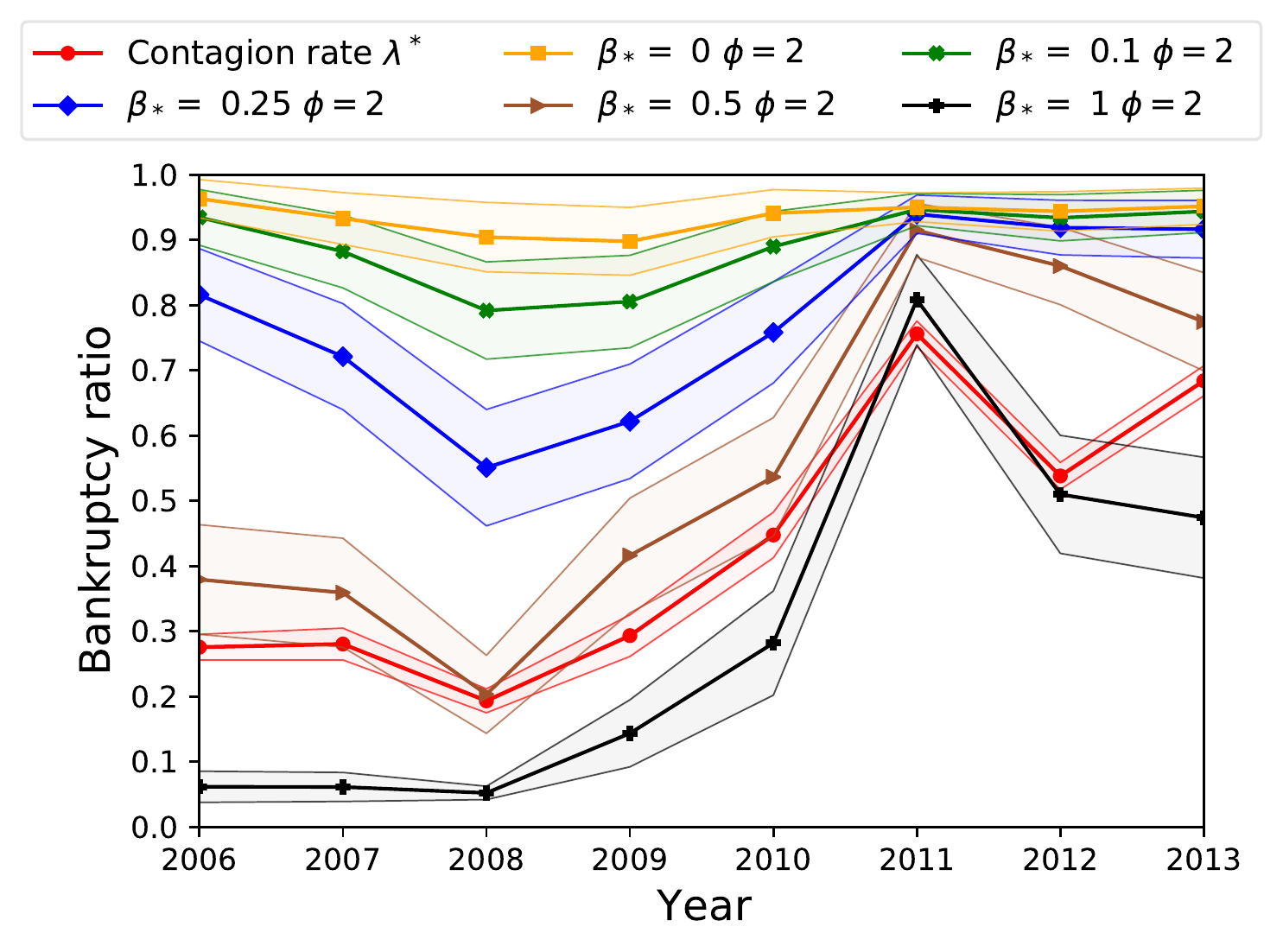}
		\includegraphics[width=0.49\linewidth]{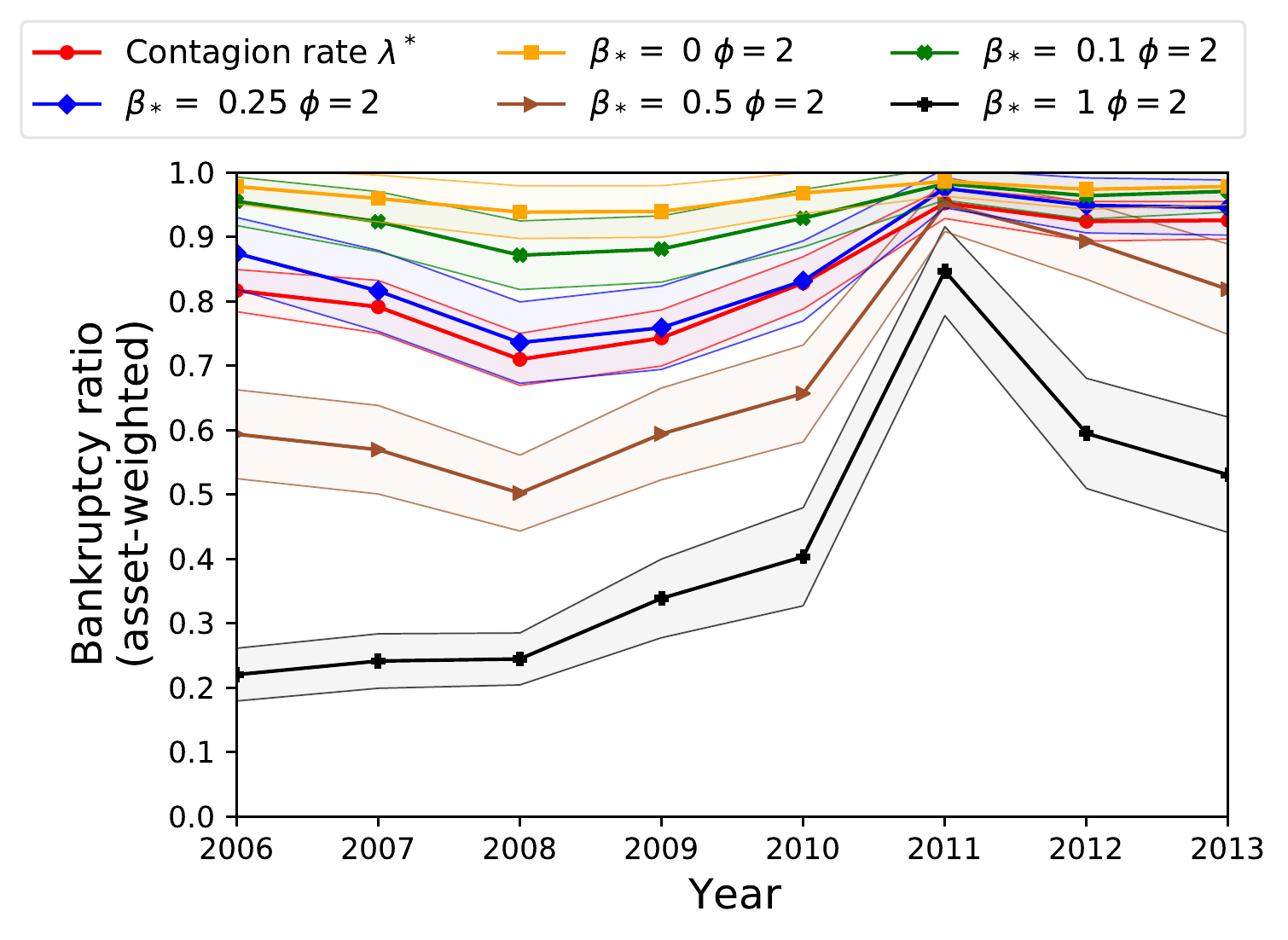}
		\caption{Time evolution of the bankruptcy ratio over the period 2006-2013. Contagion rate with confidence risk dynamics and contagion and bankruptcy node-term. Systemic risk multiplier given by $\theta^+=h(\theta,2)$. Mean over the all simulations with 95\% confidence bands. On the left, the fraction of bankrupted banks at the end of the infection while on the right, the fraction of total assets owned by bankrupted banks at the end of the infection}
		\label{fig22}
	\end{figure}

	We report here the results of the systemic liquidity risk when we have a non-linear systemic risk multiplier. Figure \ref{fig22} depicts the results of the time evolution of the systemic risk in the period 2006-2013 when we consider $\theta^+=h(\theta,2)$ as the systemic risk multiplier in the contagion rate. We can notice that when a stronger behavioural effect is implemented, the confidence contagion kicks in in all the considered scenarios apart from the extreme case of $\beta_*=1$.

\begin{figure}[h!]
		\begin{subfigure}{1\textwidth}
			\centering
			
			\includegraphics[width=.48\linewidth]{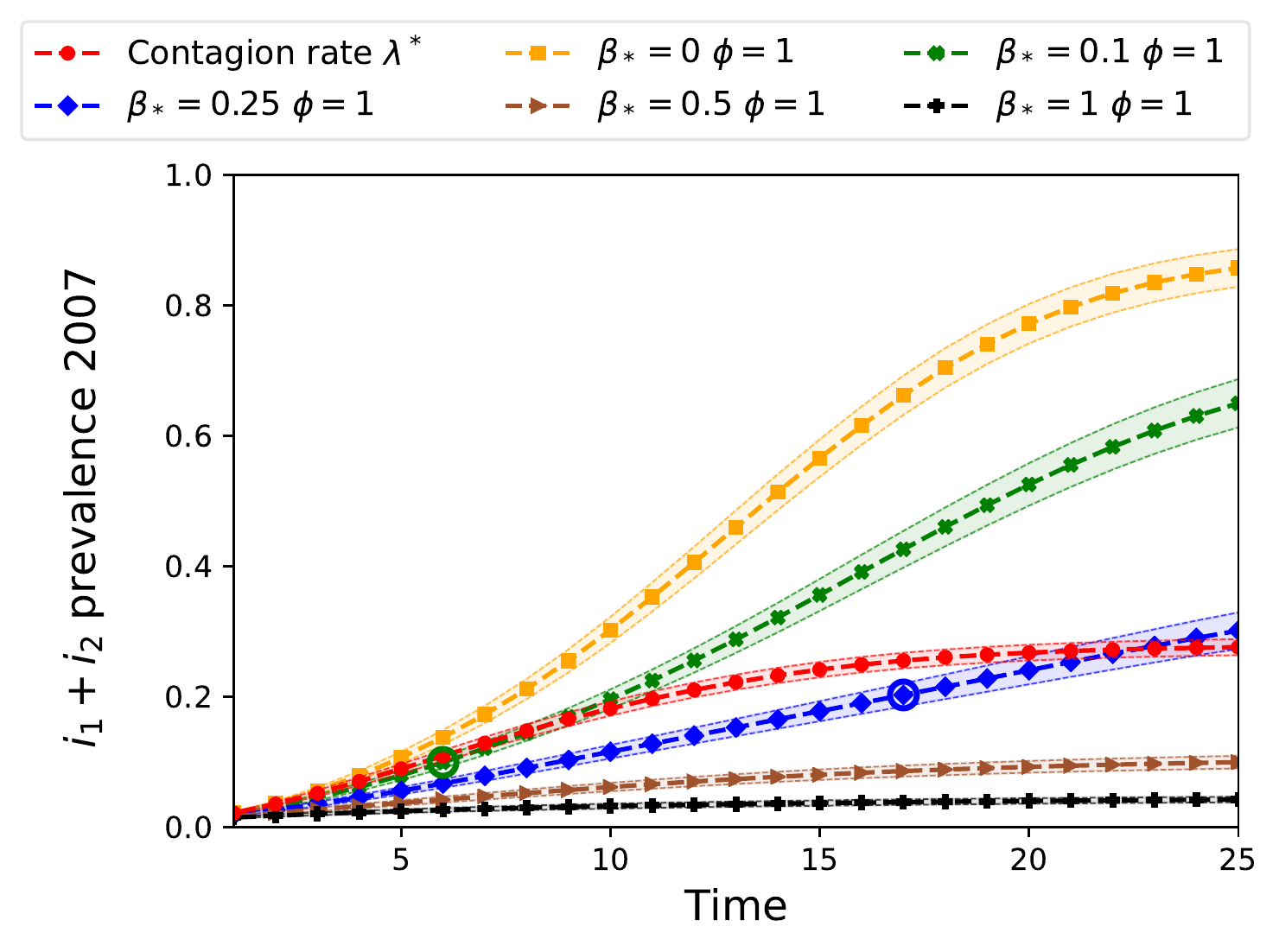}  
			\includegraphics[width=.48\linewidth]{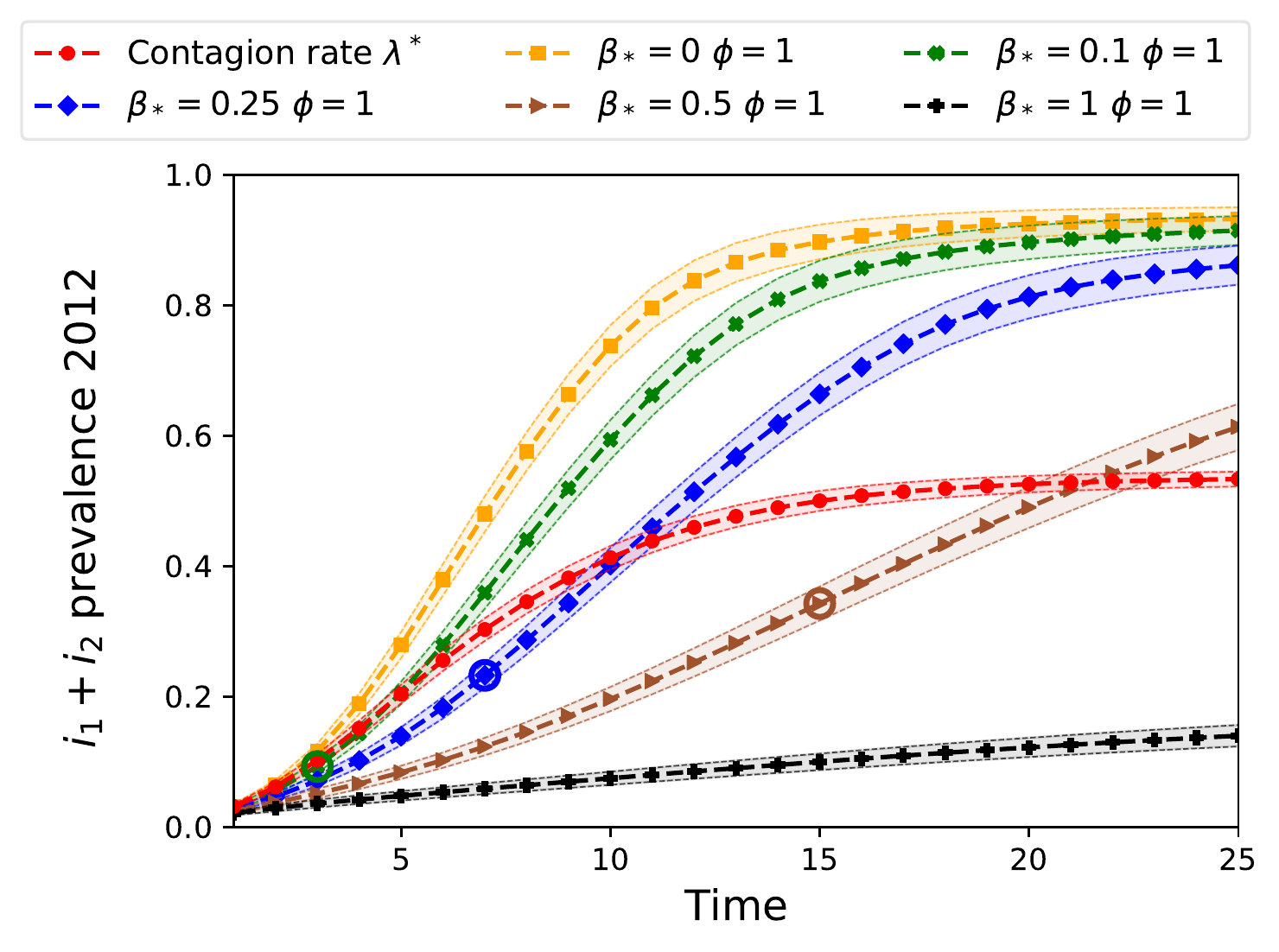} 
			\caption{$\phi=1$}
			\label{23a}
		\end{subfigure}	
		
		\begin{subfigure}{1\textwidth}
			\centering
			
			\includegraphics[width=.48\linewidth]{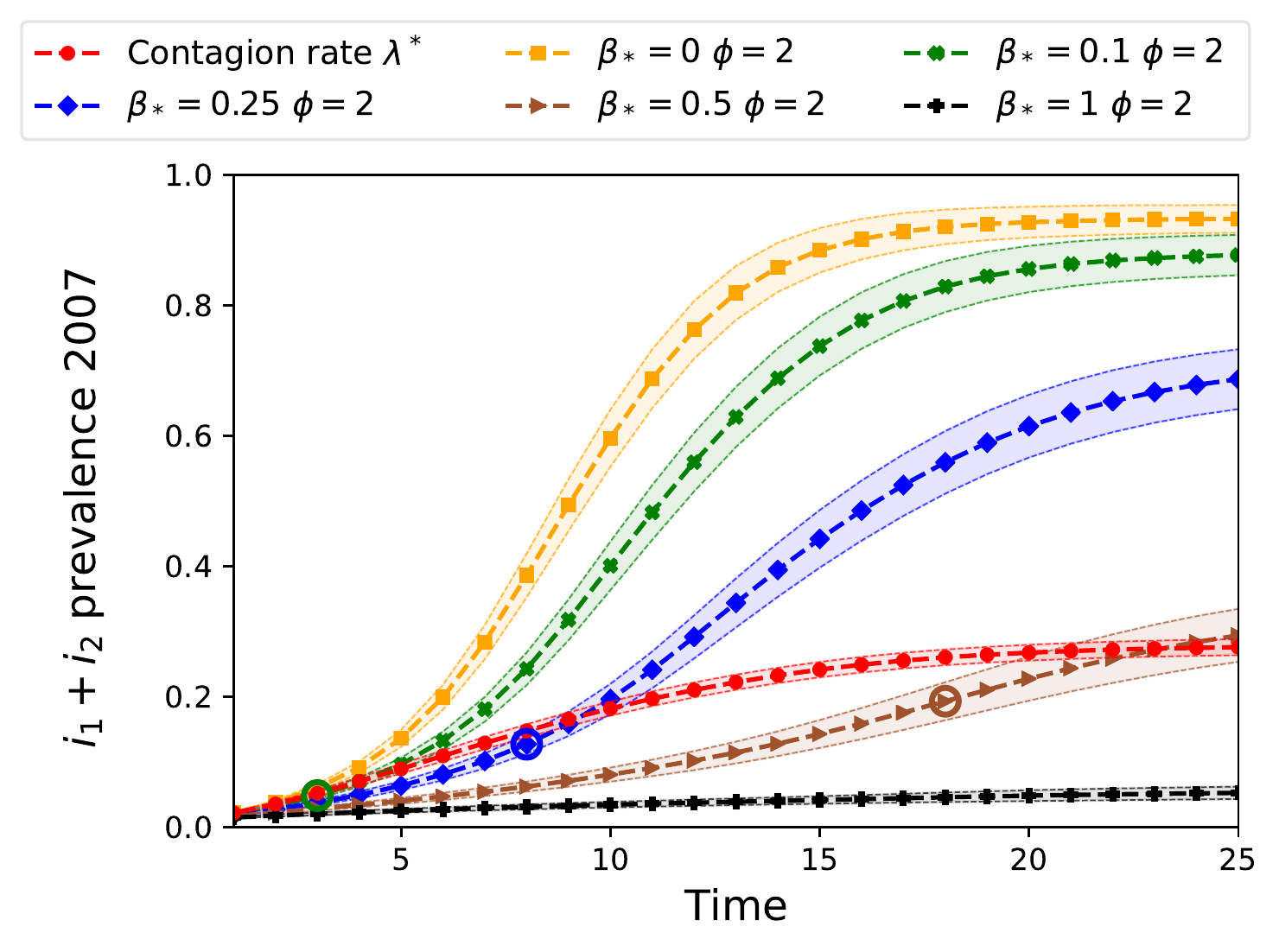}  
			\includegraphics[width=.48\linewidth]{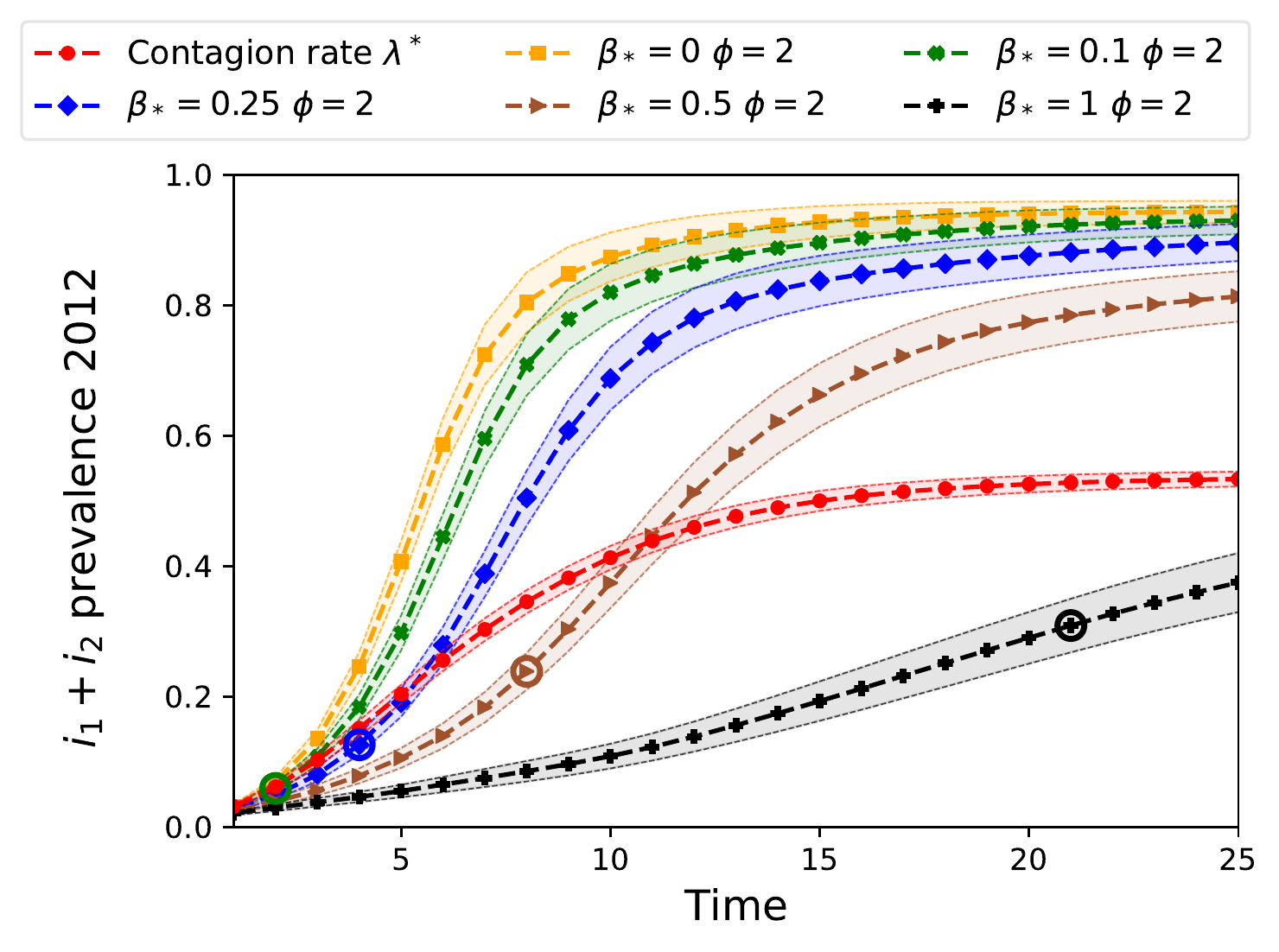}  
			\caption{$\phi=2$}
			\label{23b}
		\end{subfigure}
		\caption{Dynamics of the fraction of infected or bankrupted banks in 2007 and 2012. Empty circles represent the values of $i_1+i_2$ that determine the critical value $\theta^+=1$, namely the transition from a deceleration to an acceleration of the propagation}
		\label{fig23}
	\end{figure}

	Figure \ref{fig23} depicts the time of reversal speed of the contagion for both the linear and non-linear scenarios, i.e. $\phi=1$ and $\phi=2$. In particular, empty dots underline the critical time $t^*$ such that $\theta^+=1$. It is possible to notice from both graphs that, as expected, all the scenarios start from deceleration and, depending on the parameters $\beta_*$ and $\phi$, the time of reversal speed is reached at a higher or lower pace. In particular, for low values of $\beta_*$, the critical time of reversal speed is reached quite soon while for very high values of $\beta_*$, it is not reached at the end of the contagion simulation.

\begin{figure} \centering
		\centering
		
		\includegraphics[width=.48\linewidth]{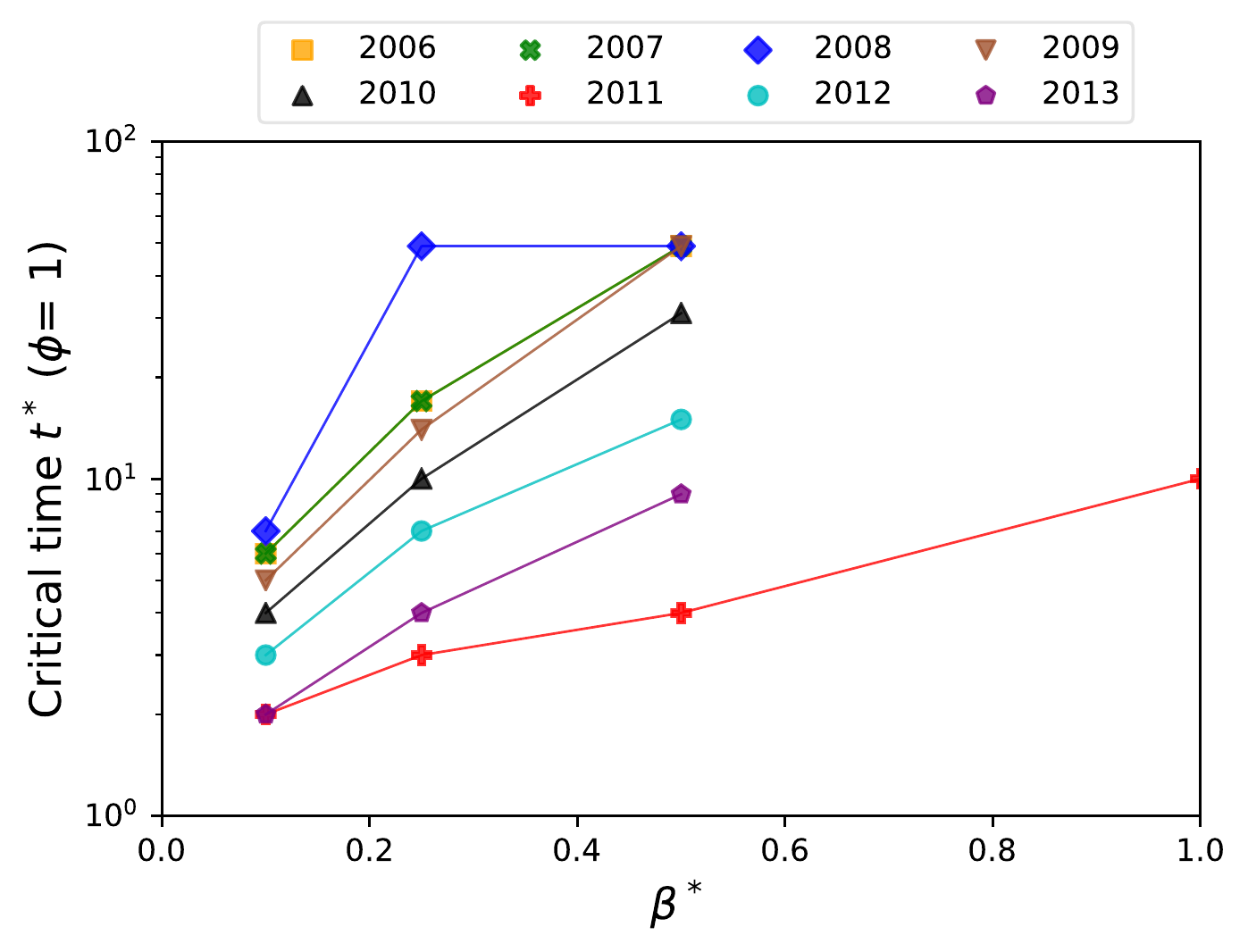}  
		\includegraphics[width=.48\linewidth]{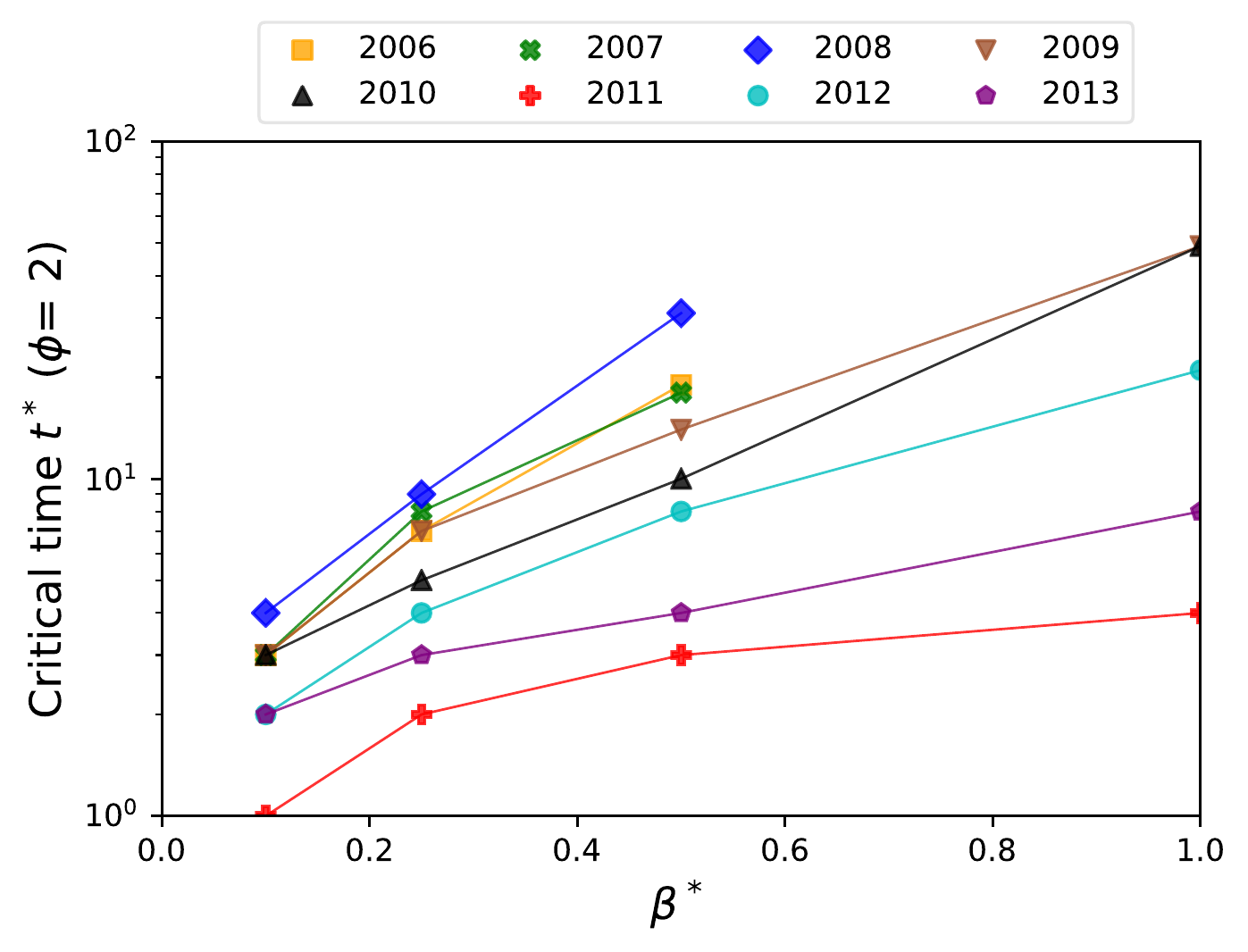}  
		\caption{Critical time values $t^*$ over the period 2006-2013 considering different values of $\phi$ and $\beta^*$. The simulation stops after 50 iterations}
		\label{fig24}
	\end{figure}

	Finally, figure \ref{fig24} shows that the time reversal with respect of each year for different values of $\beta_*$ and $\phi$. It is clear from those plots that (both in the linear and non-linear scenarios) 2011 is the year with a lower critical time of reversal speed and that the slope is less steep, meaning that the node-term in the contagion rate has had a huge role in the speed of time reversal.

\end{appendices}
\end{document}